\documentclass[%
twocolumn,
reprint,
nofootinbib,
 amsmath,amssymb,
pra,
10pt]{revtex4-2}
\usepackage{graphicx,color}
\usepackage{bm}
\usepackage{braket}
\usepackage{ulem}
\usepackage{hyperref}
\usepackage{adjustbox}
\usepackage{bm}
\usepackage{booktabs}
\usepackage{tabularx}

\definecolor{battleshipgrey}{rgb}{0.52, 0.52, 0.51}
\definecolor{cadet}{rgb}{0.33, 0.41, 0.47}
\definecolor{charcoal}{rgb}{0.21, 0.27, 0.31}

\hypersetup{
	colorlinks   = true, 
	urlcolor     = battleshipgrey, 
	linkcolor    = cadet, 
	citecolor   = charcoal 
}

\begin{document}

\preprint{APS/123-QED}

\title{Detangling the quantum tapestry of intra-channel interference in below-threshold nonsequential double ionization with few-cycle laser pulses}
\author{S. Hashim$^1$}
\author{R. Tenney$^{1,2}$}
\author{C. Figueira de Morisson Faria$^1$}%
\email{c.faria@ucl.ac.uk}
\affiliation{%
$^1$Department of Physics \& Astronomy, University College London \\Gower Street London  WC1E 6BT, United Kingdom\\
$^2$Department of Mathematics, City, University of London,\\ Northampton Square,
London EC1V 0HB, UK
}%

\date{\today}

\begin{abstract}

We perform a systematic analysis of single-channel quantum interference  in laser-induced nonsequential double ionization with few-cycle pulses, using the strong-field approximation. We focus on a below-threshold intensity for which the recollision-excitation with subsequent ionization (RESI) mechanism is prevalent. We derive and classify several analytic interference conditions for single-channel RESI in arbitrary driving fields, and address specific issues for few-cycle pulses.  Since the cycles in a short pulse are no longer equivalent, there are several events whose dominance varies. We quantify this dominance for single excitation channels by proposing a dominance parameter. Moreover, there will be many more types of superimposed interference fringes that must be taken into consideration. We find an intricate tapestry of patterns arising from phase differences related to symmetrization, temporal shifts and a combination of exchange and event interference. 

\end{abstract}

\maketitle
\section{Introduction}

The archetypal example of electron-electron correlation in intense laser fields is laser-induced nonsequential double ionization (NSDI)\cite{FigueiradeMorissonFaria2011,Becker2012}. Thereby, a returning electron, upon recollision, gives enough kinetic energy to its parent ion to release another electron \cite{Corkum1993}. While a large amount of studies interpret this correlation as classical, since the past few years quantum effects in NSDI have gained increased attention. This interest started with the theoretical findings that quantum interference is more robust than initially anticipated \cite{Hao2014,Maxwell2015,Maxwell2016}, and, depending on the circumstances, may survive integration over several degrees of freedom \cite{Hao2014,Maxwell2015} and even focal averaging \cite{Maxwell2016}, followed by experiments \cite{Quan2017}.  Furthermore, recent work has shown that entanglement may be unambiguously present in NSDI \cite{Maxwell2022}, by looking at the correlation in the orbital angular momenta of the freed electrons. In \cite{Lewenstein2022}, this has been discussed as an example of entanglement in a Zerfall process. 

The unexpected robustness of quantum interference was the first evidence that NSDI is not classical. Nonetheless, before those findings, quantum effects in NSDI have remained largely unexplored (for a few studies see, however, \cite{Wang2012,Jin2018,Yang2021,Bai2023}). 
This indifference may be justified by the huge success of classical models in reproducing the key experimental findings in NSDI for three decades (for reviews see, e.g., \cite{FigueiradeMorissonFaria2011,Becker2012}). Furthermore, early studies of quantum-classical correspondence have shown that quantum interference gets washed out upon integration over the electron momentum components perpendicular to the laser-field polarization \cite{FigueiradeMorissonFaria2004a,Faria2004,Faria2004b}, 
which is the typical scenario in experiments. Still, imprints of the type of electron-electron interaction \cite{FigueiradeMorissonFaria2004a,Faria2004b}, and of electronic bound states \cite{Faria2005,Shaaran2010,Shaaran2011}
are present. Under particular circumstances, two-center interference in molecules can also be embedded in classical NSDI models \cite{FigueiradeMorissonFaria2008}.

Classical approaches also comprise the overwhelming majority of NSDI studies in tailored fields \cite{Liu2004,Quan2009,Bergues2012,Yu2012,Fu2012,Zhang2014, Mancuso2016,Huang2016,Kubel2016,Ma2017,Chen2017,Huang2018,Song2018,Pang2020,Liu2024,Chen2020}, with good qualitative agreement with experiments. Furthermore, whenever quantum methods such as the strong-field approximation (SFA) \cite{Shaaran2010,Shaaran2011,Shaaran2012,Faria2012,Shaaran2018,Shaaran2019} and the quantitative rescattering theory (QRS) \cite{Chen2010,Chen2019,Chen2021,Chen2022}, or the full numerical solution of the time-dependent Sch\"odinger equation (TDSE) \cite{Lein2000,Parker2006,Baier2006,Baier2007,Liao2008,Eckhardt2010} have been used, the emphasis was on the shapes of the electron momentum distributions due to the type of electron-electron interaction \cite{Lein2000,Parker2006} or the field shape \cite{Baier2006,Baier2007}, and the physical mechanisms behind them. These features can be explained classically and have been reproduced using classical models \cite{Liu2004,Liu2004b}. For examples associated with the type of electron-electron interaction and features specific to a few-cycle pulse, such as asymmetric distributions, see \cite{Parker2006,Baier2006,Baier2007} and \cite{Liao2008,Eckhardt2010}, respectively. Another likely reason why quantum interference has not been paid attention to in TDSE computations for NSDI is possibly the widespread use of reduced-dimensionality models \cite{Lein2000,Liao2008,Eckhardt2010}. In these models, the degrees of freedom perpendicular to the laser-field polarization are absent, which will cause quantum interference to be overestimated. A reasonable extrapolation to a real-life scenario is to neglect this interference since transverse momenta would be integrated over in an experiment.  However, quantum interference and its connection to symmetry \cite{Busuladvic2017,Neufeld2019,Yue2020,Habibovich2021} have been extensively studied for high-order harmonic generation  \cite{Lein2007,Augstein2012,Milos2015} and strong-field ionization \cite{Busuladvic2017,Kang2021,Maxwell2021}, and photoelectron holography \cite{Faria2020,Rook2022}. 

The apparent contradiction regarding the quantum or classical nature of NSDI lies in the different parameter ranges employed in each of the two sets of studies. Classical behavior has been identified for driving-field intensities in which the first electron, upon return, gives enough energy to the core so that a second electron can be immediately released in the continuum by overcoming the ionization potential of its parent ion. This physical mechanism is known as electron-impact ionization (EI). On the other hand, if the kinetic energy of the first electron is not sufficient to trigger EI, the second electron is promoted to an excited state, and it is freed to the continuum after a time delay. This mechanism is known as recollision-excitation with subsequent ionization (RESI) and occurs for driving-field intensities in the below-threshold regime.  Although classical studies have recovered some features specific to RESI, in this regime there is more room for quantum interference. 

In a quantum mechanical framework, transition amplitudes associated with different excitation channels leading to the same final electron momenta will interfere. Furthermore, even if a single channel is involved, quantum interference may occur for events separated by specific time intervals, or transition amplitudes stemming from the two electrons being indistinguishable. 
The first paper in which quantum interference was studied in RESI focused on inter-channel interference \cite{Hao2014}. Using the strong-field approximation (SFA), it established that a fourfold symmetry that exists for correlated electron momentum distributions in the  $p_{1\parallel}p_{2\parallel}$ plane spanned by the electron momentum components parallel to the laser-field polarization, could be broken by choosing appropriate coherent superpositions of excitation pathways. These superpositions were employed to reproduce distributions occupying the second and fourth quadrants of the parallel momentum plane. 

These findings were substantially extended in our previous publications, in which we have identified and classified intra- and interchannel two-electron quantum interference in RESI with a linearly polarized monochromatic field \cite{Maxwell2015}. We have shown that, for a single excitation channel, quantum interference manifests itself as hyperbolic fringes, and bright fringes at both diagonals $p_{1\parallel}=\pm p_{2\parallel}$ of the $p_{1\parallel}p_{2\parallel}$ plane.  For inter-channel interference, we have also identified hyperbolic structures, although obtaining analytic expressions proved challenging. Besides the excellent agreement of these expressions with our numerical calculations, 
in \cite{Maxwell2016}, we specifically linked features observed in experiments with the quantum interference in RESI predicted by our model.  Our results have also shown that additional phase shifts arise from the bound-state wavefunction from which the excited electron tunnels. This builds upon previous work,  in which we have found that RESI has potential for imaging, as the excited state wavefunction strongly influences the shape of the resulting RESI electron momentum distributions \cite{Shaaran2010,Shaaran2011}. 

Nonetheless, several simplifications have been made in the abovementioned computations. One of them is to assume that the driving field is monochromatic \cite{Maxwell2015,Maxwell2016}. This is a good approximation for long enough pulses, and a reasonable assumption for short pulses if the carrier-envelope phase (CEP) is averaged over. However, in the latter case, one must compensate for the different frequency widths. To that end, in \cite{Maxwell2016} we have added extra amplitudes and phases, whose existence has been justified using qualitative arguments and averages in specific momentum ranges. These arguments backed the findings that, for longer pulses, $d$ intermediate states are the most important, while for few-cycle pulses $s$ states prevail.

Still, one must bear in mind that, for ultrashort pulses, the physics is markedly different. First, the frequency and intensity widths introduced by a pulse are expected to affect the dominant excitation channels. Second, the field cycles will no longer be equivalent, which may hinder the fringe contrast if different events within the pulse interfere or add to the problem's complexity. Third, the carrier-envelope phase (CEP) will play a major role in determining the shapes of the distributions, which will be linked to dominant events within a pulse. Fourth, there will be broken symmetries for fixed CEPS, which will influence the electron-momentum distributions and the quantum phase differences leading to interference patterns. Evidence that the fourfold symmetry is broken is provided in previous work \cite{Shaaran2012,Faria2012}, where we have studied the influence of the CEP on single-channel RESI distributions, and in experiments \cite{Bergues2012,Kubel2014}. Therein, quantum interference was not included.

In the present work, we discuss quantum interference in RESI with few-cycle pulses. We will focus on single-channel interference effects, as they depend more critically on the driving field, while the target plays a more significant role when more than one excitation channel is present.  When dealing with a pulse, a central question is what events dominate the underlying RESI and which ones can be neglected. In \cite{Shaaran2012,Faria2012}, we have assessed the dominance of specific events in an intuitive, ad hoc way, but it is of interest to seek a more systematic strategy. Furthermore, we generalize the intra-channel interference conditions derived in \cite{Maxwell2015} to a field of arbitrary shape.  We show that relaxing the assumptions stemming from the symmetries specific to the monochromatic field adds complexity to the problem and gives rise to a rich tapestry of phase shifts and quantum interference conditions. 
Throughout, we employ the modified version of the strong-field approximation (SFA) developed in \cite{Shaaran2010,Shaaran2010a}, in which excitation and electron-electron correlation have been incorporated. Although the SFA makes significant approximations, such as neglecting the residual binding potential in the electron's continuum propagation, it is useful for disentangling different types of quantum interference. First, it can be constructed focusing on the specific RESI process, without the presence of electron-impact ionization. Second, if constructed in conjunction with saddle-point methods, the SFA is an orbit-based approach that incorporates tunneling and quantum interference. Thus, specific quantum pathways can be switched on and off at will. In contrast, TDSE computations can be used as a benchmark for semi-analytic methods, but the different physical mechanisms are difficult to disentangle. For detailed discussions on the strengths and weaknesses of \textit{ab-initio} and analytical methods see the perspective article \cite{Armstrong2021}. Detailed reviews of theoretical methods in NSDI and of the strong-field approximation are provided in \cite{Faria2011} and \cite{Symphony}, respectively. 

The paper is organized as follows. In Sec.~\ref{sec:backgd}, we revisit the theoretical background necessary to understand the subsequent discussions. Thereafter, in Sec.~\ref{sec:interfcondition}, we bring the interference conditions generalized to an arbitrary field.  In Sec.~\ref{sec:pulseshape}, we investigate the pulse in detail, by systematically finding dominant events (Sec.~\ref{sec:dominance}) and their influence on target-related features (Sec.~\ref{sec:prefmapping}). Sec.~\ref{sec:PMDpulses} is devoted to calculating electron-momentum distributions, bringing the studies of the previous sections together with particular emphasis on quantum interference. Finally, in Sec.~\ref{sec:conclusions} we state our main conclusions. We will use atomic units throughout.

\section{Background}
\label{sec:backgd}
\subsection{General expressions}
In the SFA, the RESI transition amplitude related to the $\mathcal{C}$-the excitation channel is given by
\begin{eqnarray}
&&M^{(\mathcal{C})}(\mathbf{p}_{1},\mathbf{p}_{2})=\hspace*{-0.2cm}\int_{-\infty }^{\infty
}dt\int_{-\infty }^{t}dt^{^{\prime }}\int_{-\infty }^{t^{\prime
}}dt^{^{\prime \prime }}\int d^{3}k  \notag \\
&&\times V^{(\mathcal{C})}_{\mathbf{p}_{2}e}V^{(\mathcal{C})}_{\mathbf{p}_{1}e,\mathbf{k}g}V^{(\mathcal{C})}_{\mathbf{k}%
	g}\exp [iS^{(\mathcal{C})}(\mathbf{p}_{1},\mathbf{p}_{2},\mathbf{k},t,t^{\prime },t^{\prime
	\prime })],  \label{eq:Mp}
\end{eqnarray}
where the action
\begin{eqnarray}
&&S^{(\mathcal{C})}(\mathbf{p}_{1},\mathbf{p}_{2},\mathbf{k},t,t^{\prime },t^{\prime \prime
})=  \notag \\
&&\quad E^{(\mathcal{C})}_{\mathrm{1g}}t^{\prime \prime }+E^{(\mathcal{C})}_{\mathrm{2g}}t^{\prime
}+E^{(\mathcal{C})}_{\mathrm{2e}}(t-t^{\prime })-\int_{t^{\prime \prime }}^{t^{\prime }}%
\hspace{-0.1cm}\frac{[\mathbf{k}+\mathbf{A}(\tau )]^{2}}{2}d\tau  \notag \\
&&\quad -\int_{t^{\prime }}^{\infty }\hspace{-0.1cm}\frac{[\mathbf{p}_{1}+%
	\mathbf{A}(\tau )]^{2}}{2}d\tau -\int_{t}^{\infty }\hspace{-0.1cm}\frac{[%
	\mathbf{p}_{2}+\mathbf{A}(\tau )]^{2}}{2}d\tau  \label{eq:singlecS}
\end{eqnarray} describes the process in which an electron, initially at a bound state of energy $-E^{(\mathcal{C})}_{1g}$, is released at $t^{\prime\prime}$, returns to its parent ion at $t^{\prime}$ with intermediate momentum $\mathbf{k}$ and excites a second electron from a state with energy  $-E^{(\mathcal{C})}_{2g}$ to a state with energy  $-E^{(\mathcal{C})}_{2e}$. Upon rescattering, the first electron acquires the final momentum $\mathbf{p}_1$, while the second electron is freed at a later time $t$ with final momentum $\mathbf{p}_2$. The prefactors $V^{(\mathcal{C})}_{\mathbf{k}g}$, $V^{(\mathcal{C})}_{\bm{p}_1e,\mathbf{k}g}$ and $V^{(\mathcal{C})}_{\bm{p}_2e}$ are associated to the ionization of the first electron, the recollision-excitation process and the tunnel ionization of the second electron, respectively. Within the SFA, they contain all information about the electronic bound states, and about all the interactions \cite{Shaaran2010,Shaaran2010a}. The expressions stated above are general and the superscript $(\mathcal{C})$ makes them easily adaptable to coherent superpositions of channels and bound states. Nonetheless, in the present paper, we assume that the initial state of the system will be the same, so that it will be dropped subsequently (see Sec.~\ref{sec:target}).  

The ionization prefactor for the first electron is explicitly given by 
\begin{equation}\label{eq:pre3}
V^{(\mathcal{C})}_{\mathbf{k}g} = \bra{\mathbf{k}}V\ket{\psi_{1g}^{(\mathcal{C})}} = \frac{1}{(2\pi)^{3/2}}\int d^3r_1e^{-i\mathbf{k} \cdot\mathbf{r}_1}V(\mathbf{r}_1)\psi_{1g}^{(\mathcal{C})}(\mathbf{r}_1),
\end{equation}
where $V(\mathbf{r}_1)$ is the neutral atom's binding potential, typically assumed to be of Coulomb-type, $V_{\mathrm{ion}}$ the potential for the singly ionized target and $\psi_{1g}^{(\mathcal{C})}$ is the ground-state wave function for the first electron.
The excitation prefactor for the second electron reads
\begin{eqnarray}
\label{eq:Vp1ekg}V^{(\mathcal{C})}_{\mathbf{p}_1e,\mathbf{k}g}\hspace*{-0.15cm}&=& \hspace*{-0.15cm} \bra{\mathbf{p}_1,\psi_{2e}^{(\mathcal{C})}}V_{12}
\ket{\mathbf{k},\psi_{2g}^{(\mathcal{C})}}
    \\
    &=&\hspace*{-0.15cm} \frac{V_{12}(\mathbf{p}_1-\mathbf{k})}{(2\pi)^{3/2}}\hspace*{-0.15cm}\int \hspace*{-0.1cm}d^3r_2e^{-i(\mathbf{p}_1-\mathbf{k})\cdot \mathbf{r}_2}\psi_{2e}^{*(\mathcal{C})}(\mathbf{r}_2)\psi_{2g}^{(\mathcal{C})}(\mathbf{r}_2),  \notag 
\end{eqnarray}
where $\psi_{2e}^{(\mathcal{C})}(\mathbf{r}_2)$ and $\psi_{2g}^{(\mathcal{C})}$ are the excited and ground states wave functions for the second electron. 
The electron-electron interaction reads
\begin{equation}
    V_{12}(\mathbf{p}_1 - \mathbf{k}) =  \frac{1}{(2\pi)^{3/2}}\int d^3rV_{12}(\mathbf{r})\exp[-i(\mathbf{p}_1-\mathbf{k})\cdot\mathbf{r}]
\end{equation}
where \(\mathbf{r} = \mathbf{r}_1-\mathbf{r}_2\), and it is taken to be of contact-type, as in \cite{Faria2012,Maxwell2016}. 
 
 Finally, the ionization prefactor associated with the second electron is given by
\begin{eqnarray}
    V^{(\mathcal{C})}_{\mathbf{p}_2e} &=& \bra{\mathbf{p}}V_{\mathrm{ion}}\ket{\psi_{2e}^{(\mathcal{C})}}  \\ &=& \frac{1}{(2\pi)^{3/2}}\int d^3r_2V_{\mathrm{ion}}(\mathbf{r}_2)e^{-i\mathbf{p_2}\cdot\mathbf{r}_2}\psi_{2e}^{(\mathcal{C})}(\mathbf{r}_2). \notag
    \label{eq:Vp2e}
\end{eqnarray}

In the SFA, the target structure is incorporated via the prefactors $V^{(\mathcal{C})}_{\mathbf{k}g}$, $V^{(\mathcal{C})}_{\bm{p}_1e,\mathbf{k}g}$ and $V^{(\mathcal{C})}_{\bm{p}_2e}$. In our calculations, we employ hydrogenic wave functions $\psi_{nlm}(\mathbf{r})=R_{nl}(r)Y_l^m (\theta, \phi)$ for the electronic bound states. The explicit expressions for these prefactors are  given in  the appendix, and detailed derivations are given in our previous publications \cite{Shaaran2010,Maxwell2015}. In all examples studied in this paper, we have taken the magnetic quantum number $m=0$ to facilitate a comparison with our previous work \cite{Maxwell2015,Maxwell2016} and with the existing results in the literature \cite{Hao2014}.  Furthermore, we consider the ionization prefactors in the velocity gauge to avoid bound-state singularities. Length-gauge prefactors contain additional momentum shifts, which will not play an important role for the ionization prefactors and will cancel out for the excitation prefactor. A detailed discussion has been provided in \cite{Shaaran2010}. 

We calculate the multiple integrals in Eq.~(\ref{eq:Mp}) using the steepest descent method.
This leads to the saddle-point equations
\begin{equation}\label{eq:sp1}
[\mathbf{k}+\mathbf{A}(t'')]^2 = -2E^{(\mathcal{C})}_{1g},
\end{equation}
\begin{equation}\label{eq:sp2}
\mathbf{k} = -\frac{1}{t'-t''}\int_{t''}^{t'}d\tau \mathbf{A}(\tau)
\end{equation}
\begin{equation}\label{eq:sp3}
[p_{1\parallel}+A(t')]^2+[\mathbf{p}_{1\perp}]^2 = [\mathbf{k}+\mathbf{A}(t')]^2 - 2(E^{(\mathcal{C})}_{2g}-E^{(\mathcal{C})}_{2e})
\end{equation}
and 
\begin{equation}\label{eq:sp4}
[p_{2\parallel} + A(t)]^2+[\mathbf{p}_{2\perp}]^2 = -2E^{(\mathcal{C})}_{2e}.
\end{equation}
Eqs.~(\ref{eq:sp1}) and (\ref{eq:sp4}) give the conservation of energy upon tunnel ionization for the first and second electrons, respectively. Eq.~(\ref{eq:sp2})  constrains the intermediate momentum so that the first electron returns to the site of its release, and Eq.~(\ref{eq:sp3}) gives the rescattering event in which the first electron gives part of the kinetic energy $E_{k}(t',t'')=[\mathbf{k}+\mathbf{A}(t')]^2/2$ to excite the second electron.  
We denote $p_{n\parallel}$ and $\mathbf{p}_{n\perp}$, $(n=1,2)$, the momentum components parallel and perpendicular to the laser-field polarization, respectively. One should note that  $\mathbf{p}_{n\perp}$ are two-dimensional vectors spanning the plane perpendicular to the driving-field polarization. 
The real parts of the solutions of the saddle-point equations are directly related to the classical recollision and ionization times. 
The integrals are then approximated by sums over these stationary variables. For the second electron, the saddles are well separated in all momentum regions, while, for the first electron, pairs of saddles must be considered collectively using the uniform approximation in \cite{Faria2002}. 

Eq.~\eqref{eq:sp3} relates the kinetic energy $E_{k}(t',t'')$ of the returning electron to the energy difference $\Delta E^{(\mathcal{C})}= E^{(\mathcal{C})}_{2g}-E^{(\mathcal{C})}_{2e}$ between the ground and excited state of the second electron. With regard to the space spanned by the momenta $p_{1\parallel},\mathbf{p}_{1\perp}$ of the first electron, this saddle-point equation represents a sphere, whose radius is $\sqrt{2(E_{k}(t',t'')-\Delta  E^{(\mathcal{C})})}$. If  $E_{k}(t',t'')> \Delta  E^{(\mathcal{C})}$, this radius is real and Eq.~\eqref{eq:sp3} may have a classical counterpart.  If we are interested in $p_{1\parallel}$ only, we can assume that $\mathbf{p}_{1\perp}^2 /2$ is an additional energy shift which will effectively alter the ionization threshold. Therefore, assuming $\mathbf{p}_{1\perp}=\mathbf{0}$ helps to define an upper bound for  Eq.~\eqref{eq:sp3}  in terms of the momentum range $p_{1\parallel}$ which may be occupied, should rescattering have a classical counterpart. We call this momentum range the \textit{classically allowed region} (CAR). This concept has been introduced in our previous publications \cite{Shaaran2010,Shaaran2010a}.  

Furthermore, the saddle-point equations state that RESI may be viewed as two time-ordered processes similar to above-threshold ionization (ATI) \cite{Shaaran2010,Shaaran2010a}. The first electron behaves as if it were undergoing rescattered ATI, with the difference that, for RESI, the process is inelastic. This means that the maximal final kinetic energy of the first electron can approach the rescattered ATI cutoff of $10U_p$, where $U_p$ is the ponderomotive energy. Moreover, Eq.~\eqref{eq:sp4} resembles the saddle-point equation obtained by direct ATI, for which the maximum classical kinetic energy, known as the direct ATI cutoff, is 2$U_p$. This leads to the equation of a sphere in terms of $p_{2\parallel}$, $\mathbf{p}_{2\perp}$, whose radius is $2\sqrt{U_p}$  \cite{Becker2002Review}. The ponderomotive energy is the time-averaged kinetic energy acquired by an electron from the field. For a linearly polarized monochromatic wave, it is related to the amplitude of the vector potential by $A_0 = 2\sqrt{U_p}$.

\subsection{Two-electron probability density}
\label{sec:probdensities}
The quantity of interest is the correlated two-electron probability density as a function of the momentum components $p_{n\parallel}$ $n=1,2$ parallel to the driving-field polarization. This is given by 
\begin{align}
	\mathcal{P}(p_{1\parallel},p_{2\parallel})= \int\int d^2 p_{1\perp}d^2 p_{2\perp}\mathcal{P}(\mathbf{p}_{1},\mathbf{p}_{2}), \label{Eq:Channels}
\end{align}
where $\mathcal{P}(\mathbf{p}_{1},\mathbf{p}_{2})$ is the fully resolved two-electron momentum probability density, and the transverse momentum components have been integrated over. 
When calculating this probability density, several issues must be taken into consideration.  First, both electrons are indistinguishable, which means that Eq.~(\ref{eq:Mp}) must be symmetrized upon $\mathbf{p}_1 \leftrightarrow \mathbf{p}_2$, that is, electron exchange. Second, in a real target, there will be several excitation channels, each of which will be associated with the transition amplitude Eq.~(\ref{eq:Mp}). Third, there will be several events within the pulse for the first and second electrons. Thus, within the saddle-point approximation, the overall RESI amplitude must contain sums over (i) transitions involving different excitation channels; (ii) the symmetrization related to electron indistinguishability, which will occur for each pair of excitation and ionization times; (iii) the events within a pulse. Quantum mechanically, all these contributions add up coherently.

The fully coherent sum over events, channels and symmetrization reads
\begin{equation}
\mathcal{P}_{(\mathrm{ccc})}(\mathbf{p}_{1},\mathbf{p}_{2})=\left|\sum_{\varepsilon}\sum_{\mathcal{C}}\left[M_{\varepsilon}^{(\mathcal{C})}(\mathbf{p}_1,\mathbf{p}_2)+M_{\varepsilon}^{(\mathcal{C})}(\mathbf{p}_2,\mathbf{p}_1)\right]\right|^2, 
\label{eq:fullcoherent}
\end{equation}
where the symbols $\varepsilon$ and $\mathcal{C}$ denote event and channel, respectively.  In the present work, we will focus on single-channel interference, so that sums over channels will not be taken into consideration. 

For a single channel, a coherent sum over the events $\varepsilon$ in the pulse and symmetrization leads to
\begin{equation}
\mathcal{P}^{(\mathcal{C})}_{(\mathrm{cc})}(\mathbf{p}_{1},\mathbf{p}_{2})=\left|\sum_{\varepsilon}\left[M_{\varepsilon}^{(\mathcal{C})}(\mathbf{p}_1,\mathbf{p}_2)+M_{\varepsilon}^{(\mathcal{C})}(\mathbf{p}_2,\mathbf{p}_1)\right]\right|^2,
\label{eq:1coherent}
\end{equation}
while the incoherent counterpart of Eq.~(\ref{eq:1coherent}) is given by
\begin{equation}
\mathcal{P}^{(\mathcal{C})}_{(\mathrm{ii})}(\mathbf{p}_{1},\mathbf{p}_{2})=\sum_{\varepsilon}\left[\left|M_{\varepsilon}^{(\mathcal{C})}(\mathbf{p}_1,\mathbf{p}_2)\right|^2\hspace*{-0.2cm}+\hspace*{-0.1cm}\left|M_{\varepsilon}^{(\mathcal{C})}(\mathbf{p}_2,\mathbf{p}_1)\right|^2\right]. 
\label{eq:1ii}
\end{equation}

In the present work, we are interested in disentangling several types of quantum interference. Therefore, we construct the two-electron probability density in several ways, depending on the question we wish to address. 

Throughout, we will use the notation $\mathcal{P}^{(\mathcal{C})}_{(S\varepsilon )}$  where the indices $S$, $\varepsilon,$ and $\mathcal{C}$ relate to the electron symmetrization, the event, and the channel, respectively.  The indices $c$ and $i$ on the left-hand side stand for coherent and incoherent, respectively. Thus, $\mathcal{P}^{(\mathcal{C})}_{(c c)}$ states that the sum considered in the single-channel probability density is coherent over the pulse events and symmetrization, and has been calculated for the $\mathcal{C}$-th channel. 
If the symmetrization is done coherently, but the events are summed over incoherently, this gives 
 \begin{equation}
\mathcal{P}^{(\mathcal{C})}_{(\mathrm{ci})}(\mathbf{p}_{1},\mathbf{p}_{2})=\sum_{\varepsilon}\left|M_{\varepsilon}^{(\mathcal{C})}(\mathbf{p}_1,\mathbf{p}_2)+M_{\varepsilon}^{(\mathcal{C})}(\mathbf{p}_2,\mathbf{p}_1)\right|^2. 
\label{eq:1ci}
\end{equation}
Alternatively, one could sum the events within a pulse coherently and perform the symmetrization incoherently. This yields
 \begin{equation}
\mathcal{P}^{(\mathcal{C})}_{(\mathrm{ic})}(\mathbf{p}_{1},\mathbf{p}_{2})=\left|\sum_{\varepsilon}M_{\varepsilon}^{(\mathcal{C})}(\mathbf{p}_1,\mathbf{p}_2)\right|^2\hspace*{-0.2cm}+\hspace*{-0.1cm}\left|\sum_{\varepsilon}M_{\varepsilon}^{(\mathcal{C})}(\mathbf{p}_2,\mathbf{p}_1)\right|^2. 
\label{eq:1ic}
\end{equation}

This notation differs from that employed in our previous work \cite{Maxwell2015,Maxwell2016}, for which the interference due to electron symmetrization was called ``event interference". Apart from the combinations given above, given the wealth of interference patterns, it may also be useful to sum transition amplitudes pairwise. A discussion of the phase shifts that occur in these pairwise sums is given in Sec.~\ref{sec:interfcondition}.

Finally, under some circumstances, it is useful to compute partial momentum distributions for each electron, given by 
\begin{equation}
   M^{(n)}(p_{n\parallel})=\int d^2 p_{n \perp}|M^{(n)}(\mathbf{p}_n)|^2,
   \label{eq:partialdistr}
\end{equation}
with $n=1,2$.

\subsection{Target considerations}
\label{sec:target}
 As a target, we will consider Argon, for which there are six main excitation channels.  The absolute values of ground-state energies associated with $3s$ and $3p$ are taken to be $E^{(1)}_{2g}= E^{(\mathcal{C})}_{2g}= 1.016$ a.u., $\mathcal{C}=2$ to $6$, and the first ionization potential is $E^{(\mathcal{C})}_{1g}=0.58$ a.u. for all channels. Thus, for simplicity, we are dropping the superscript for $E_{1g}$. One should note that the bound-state geometry differs for $3s$ and $3p$, which will influence the excitation prefactor. For clarity, these channels are provided in Table \ref{tab:channels}. The influence of the target occurs via the bound-state energies and the prefactors. According to the saddle-point equations (\ref{eq:sp1})-(\ref{eq:sp4}),  the bound-state energies will affect the ionization probability of the first and second electron, and the classically allowed region in momentum space for the rescattered electron. The more tightly bound the electrons are, the lower their tunneling probability will be. Furthermore, large (small) energy gaps $\Delta E^{(\mathcal{C})}= (E^{(\mathcal{C})}_{2g}-E^{(\mathcal{C})}_{2e})$ stemming from the second electron being excited will potentially favor small (large) classically allowed regions for the first electron upon rescattering. 

The bound-state geometry will introduce an additional momentum bias in the problem via the prefactors. The shapes of the electron-momentum distributions will mostly be influenced by the prefactor $V_{\mathbf{p}_2e}$ \cite{Shaaran2010,Maxwell2015}. Before integration over the transverse degrees of freedom, this prefactor will exhibit radial and, for $l \neq 0$, angular nodes. The other prefactors will have less influence in the shapes of the electron momentum distribution, due to the time dependence of the intermediate momentum $\mathbf{k}$ (see saddle-point equation \eqref{eq:sp2}). This time dependence will cause a blurring in their nodes.  The radial and angular nodes of the prefactors may lead to additional phase shifts, which will influence the overall interference patterns. A detailed study of these features has been performed in \cite{Maxwell2015} for a monochromatic field, but some of these issues must be revisited for a pulse. 

\begin{table}[] 
\begin{tabular}{ccc}
\hline \hline
 Channel & Excited-state configuration  &  $E^{(\mathcal{C})}_{2e}$ (a.u.)  \vspace*{0.1cm} \\ \hline
 1& $3s3p^6 (3s \rightarrow 3p)$ &  0.52  \\
 2& $3p^53d (3p \rightarrow 3d)$ &  0.41  \\
 3& $3p^54d (3p \rightarrow 4d)$ &  0.18  \\
 4&  $3p^54s (3p \rightarrow 4s)$  &  0.40\\
 5& $3p^54p (3p \rightarrow 4p) $& 0.31 \\
 6& $3p^55s (3p \rightarrow 5s)$ & 0.19 \vspace*{0.1cm} \\ \hline \hline
\end{tabular}
\caption{Relevant excitation channels for $\mathrm{Ar}^+$, in order of increasing principal and orbital quantum numbers. From left to right, the first column gives the number associated with the channel, the second column states the electronic configuration and the excitation pathway, and the third column provides the excited-state energy in atomic units. \label{tab:channels}}
\end{table}

\subsection{Momentum constraints for a pulse}
\label{sec:eventinterf}
 
 The saddle-point Eqs.~(\ref{eq:sp3}) and (\ref{eq:sp4}) can be used to determine constraints in momentum space for each rescattering event within the pulse. The first electron is released such that, for $\mathrm{Re}[t'']$, the electric field is close to an extremum. For a specific event, the first electron's most prominent return, at $\mathrm{Re}[t']$, is near a field zero crossing displaced by approximately three-quarters of a field cycle. The second electron may then tunnel in any of the subsequent times $t$, for which its real part is near a field extremum.  However, due to bound-state depletion, the prevailing ionization time occurs around the extremum immediately after the zero crossing. Similarly, the first electron may return following longer orbits, but such returns are expected to be suppressed due to wave-packet spreading.
 
 Using this information, we estimate that the final momentum of both electrons will be located around  $(p_{ 1\parallel},p_{ 2\parallel})=( -A(t'),-A(t))$. For times $\mathrm{Re}[t']$ sufficiently close to the peaks of the field, $-A(t')\simeq \pm 2\sqrt{U_p}$ and $-A(t)\simeq 0$. 
 The extension of the momentum region occupied will depend on the electron's kinetic energy upon return at each rescattering event. This region may be large, small, or even have no classical counterpart. Electron indistinguishability dictates that there will also be events whose amplitudes are centered at $(p_{ 1\parallel},p_{2\parallel})=( -A(t),-A(t'))$. The final kinetic energy for the first electron may extend to almost $10U_p$, while that of the second electron may reach up to $2U_p$. Taking the perpendicular momentum components to vanish provides an upper bound for the regions occupied in the $p_{1\parallel}p_{2\parallel}$ plane \cite{Shaaran2010,Shaaran2010a}.

Similar to the procedure in \cite{Maxwell2015} for a monochromatic wave, we consider a single channel, neighboring events within the pulse, displaced by approximately half a cycle, and those present due to the electron exchange symmetry of the system.  However, there are key differences for a few-cycle pulse: (i) the half-cycle symmetry $\textbf{A}(t)= \pm \textbf{A} (t \pm T/2)$, where $T$ is a field cycle, is broken; (ii) the cycles are not equivalent due to the presence of the pulse envelope; (iii) the assumptions made upon $A(t')$ and $A(t)$ are approximate, and work better close to the center of the pulse. 

\begin{figure}
    \centering
    \includegraphics[width=\columnwidth]{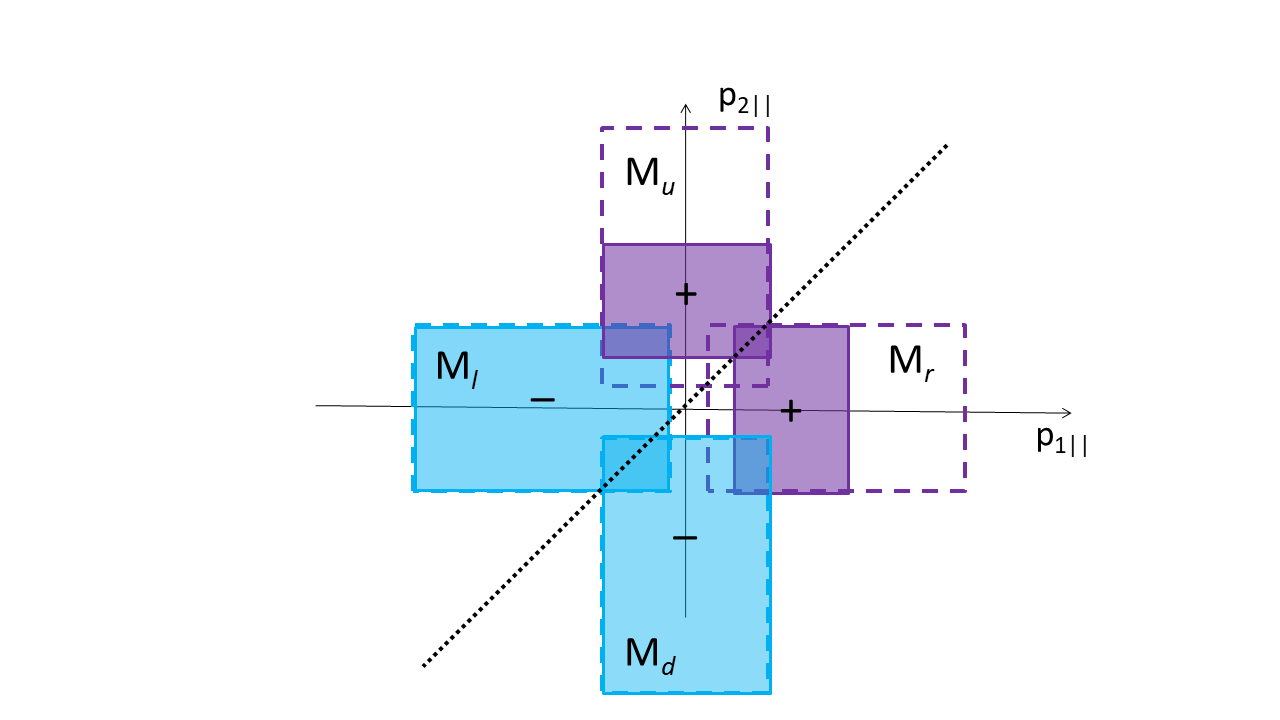}
    \caption{Schematic representation of the momentum-space regions occupied by the RESI transition amplitudes $M_l$, $M_u$, $M_r$ and $M_d$. The dashed rectangles indicate the momentum constraints that would hold for a monochromatic driving field, while the shaded areas show their counterparts for few-cycle pulses. The overlapping areas indicate momentum regions for which quantum interference will be significant. We have used the same color for  $M_l$ and $M_d$ (blue), and $M_u$ and $M_r$ (purple), to highlight the property $M_l(\mathbf{p}_1,\mathbf{p}_2)=M_d(\mathbf{p}_2,\mathbf{p}_1)$ and $M_r(\mathbf{p}_1,\mathbf{p}_2)=M_u(\mathbf{p}_2,\mathbf{p}_1)$, which gives a reflection symmetry about the diagonal $p_{1\parallel}=p_{2\parallel}$. This diagonal is indicated by the dotted black line. The negative (positive) signs in the center of the blue (purple) shaded regions indicate the most probable momenta.   }
    \label{fig:schematic1}
\end{figure}

This will lead to the 
transition amplitudes $M_l$, $M_u$, $M_r$, and $M_d$, with $M_l(\mathbf{p}_1,\mathbf{p}_2)=M_d(\mathbf{p}_2,\mathbf{p}_1)$ and $M_r(\mathbf{p}_1,\mathbf{p}_2)=M_u(\mathbf{p}_2,\mathbf{p}_1)$, where we have dropped the superscript $(\mathcal{C})$ for simplicity. In Fig.~\ref{fig:schematic1}, we provide a schematic representation of the momentum constraints associated with each of these transition amplitudes as the shaded regions.  Outside these constraints, the probability density is strongly suppressed, as it has no classical counterpart.  
Additionally, for a monochromatic wave, the half-cycle symmetry implies that $M_l(\mathbf{p}_1,\mathbf{p}_2)=M_r(-\mathbf{p}_1,-\mathbf{p}_2)$ and $M_u(\mathbf{p}_1,\mathbf{p}_2)=M_d(-\mathbf{p}_1,-\mathbf{p}_2)$. This leads to a fourfold symmetry in the momentum-region constraints, 
as outlined by the dashed rectangles in the figure, which, in the absence of further momentum bias, causes the correlated two-electron distribution to be cross-shaped \cite{Shaaran2010,Shaaran2010a}. For a few-cycle pulse, the half-cycle symmetry is broken, but the reflection symmetry about the diagonal $p_{1\parallel}=p_{2\parallel}$ is retained.

In addition to that, Fig.~\ref{fig:schematic1} indicates a substantial overlap for (i) $M_l$ and $M_d$, (ii) $M_r$ and $M_u$, (iii) $M_u$ and $M_l$, and (iv) $M_r$ and $M_d$.  Therefore, their interference is expected to be substantial and will be calculated in pairwise coherent sums. The sums $M_{rl}=M_l+ M_r$ and $M_{ud}=M_u+ M_d$ of events separated by half a cycle will also overlap and interfere, but this interference is expected to be less relevant as the probability density is small outside the shaded regions \cite{Shaaran2010,Maxwell2015}.

\begin{figure}
    \centering
    \includegraphics[width=\columnwidth]{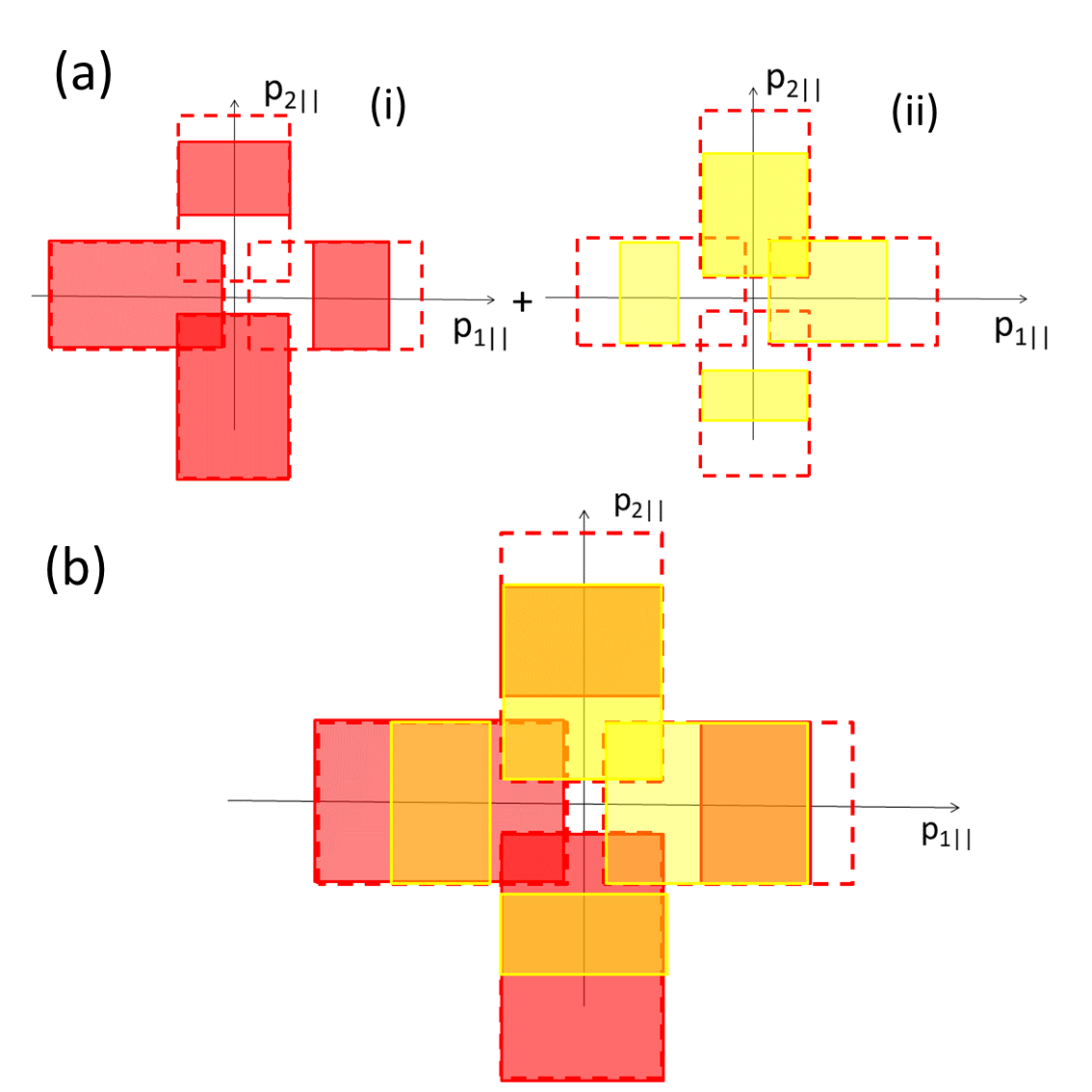}
    \caption{Schematic representation of the momentum-space regions associated with different events within the pulse [panel (a)], and those resulting from their coherent superposition [panel (b)]. The dashed rectangles indicate the momentum constraints that would hold for a monochromatic driving field, while the shaded areas show their counterparts for few-cycle pulses. We assumed that each subpanel in (a) gives the momentum regions occupied by events within a single cycle, and the events in (a)(i) and (a)(ii) are summed coherently.  
    Overlapping shaded regions indicate that quantum interference is potentially significant. We consider two events separated by a half cycle in each graph in panel (a), which eventually interfere.   }
    \label{fig:schematic2}
\end{figure}

Furthermore, the cycles within a pulse will not be identical and there may be several comparable rescattering and ionization events, which will potentially interfere. Thus, in principle, one must take their contributions to the electron-momentum distributions into consideration. A schematic representation is provided in Fig.~\ref{fig:schematic2}, which shows the momentum constraints determined by four events within a hypothetical pulse. Panel (a) illustrates the events whose contributions must be added up, and whose momentum constraints are shown as the red and yellow shaded areas. We assume that the event of each sub-panel  [(a)(i) and (a)(ii)] took place over a single cycle. The events in separate sub-panels are displaced by at least a cycle. The shaded regions around the same $p_{n\parallel}$, $n=1,2$ half axis (positive or negative) correspond to events displaced by an integer number of cycles (different colors), or none (same color), while those around opposite $p_{n\parallel}$, $n=1,2$ half-axis and highlighted by the same color are displaced by half a cycle.  

The momentum ranges occupied by the resulting events, shown in panel (b), illustrate several overlapping regions, for which quantum interference will occur. This interference may be pronounced, subtle, or negligible, depending on whether the events are comparable or whether some of them prevail. Outside the shaded regions, there will also be quantum interference, but due to the transition amplitudes being suppressed, it is expected to be negligible. One may control the dominant events by changing the pulse parameters.

\section{Generalized intra-channel interference conditions}
\label{sec:interfcondition}

Below we generalize the interference conditions derived in \cite{Maxwell2015} to an arbitrary driving field, with vector potential $\mathbf{A}(\tau)$ with $\tau=t,t'$. We make no assumption about its component frequencies, shape, or polarization, so the expressions below are entirely general. Due to their level of complexity, we first provide the key assumptions behind the derivations, together with a road map and a graphic representation to facilitate their understanding. 
\subsection{Diagrammatic representation}
Within a single cycle, quantum interference will stem from $|M_{ld}|^2$, $|M_{ru}|^2$, $|M_{ul}|^2$, and $|M_{rd}|^2$, where $M_{\mu \nu}=M_{\mu}+M_{\nu}$. If there is a shift of an integer number $n$ of cycles $T=2\pi/\omega$ in a specific transition amplitude, we use the notation $M_{\mu, nT}=M_{\mu}(t+nT)$. Each term in these coherent sums will be approximated by $e^{iS_{\mu}}$, where $\mu=l,d,u,r$, and the phase differences are given by $\alpha_{\mu \nu}=S_{\mu}-S_{\nu}$. This is justified by the underlying assumption that the prefactors vary much more slowly than the action, which is a key element of the approaches employed in this work. 
\begin{figure}[h]
     \centering
    \includegraphics[width=\columnwidth]{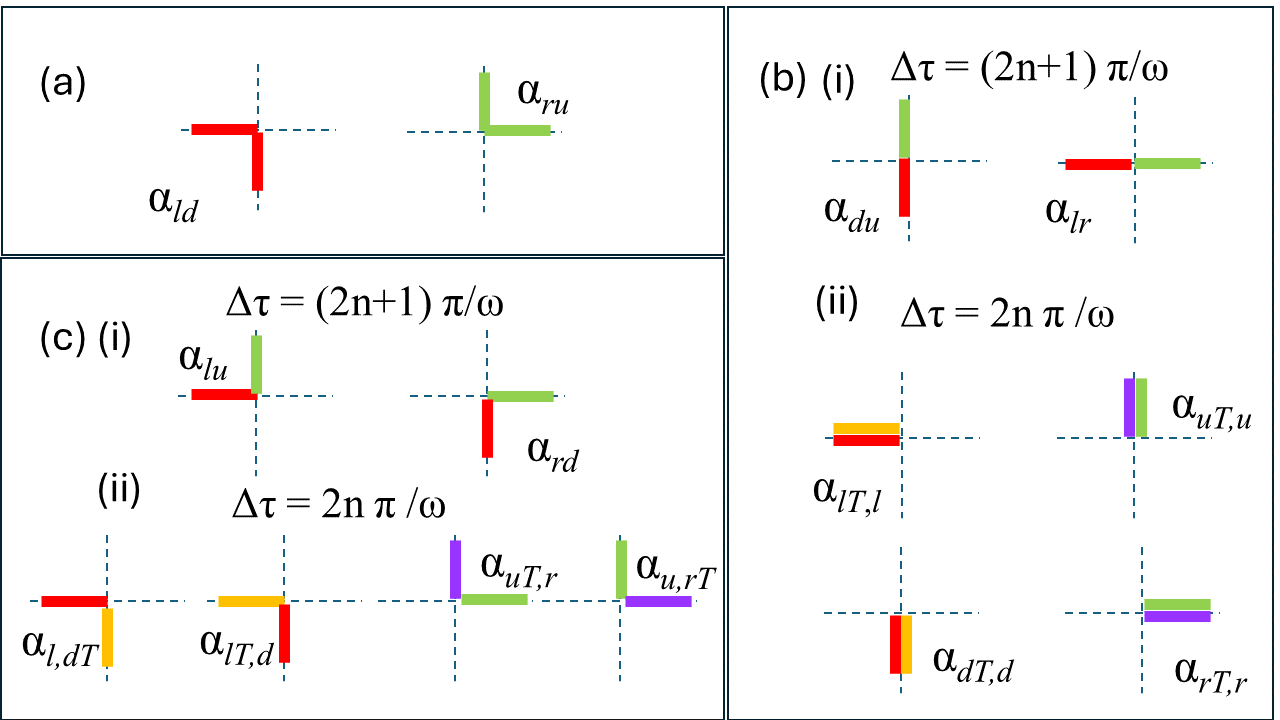}
    \caption{Diagrammatic representation of different types of pairwise single-channel interference that may occur for NSDI RESI. The momentum ranges occupied by a specific process are indicated by the thick solid lines, while the thin dashed lines are a simplified representation of the parallel momentum axes $p_{n\parallel}$, $n=1,2$. Panel (a) shows the processes associated with the electron momenta being swapped, but taking the same event. Panels (b) consider the interference of events shifted by an odd [(b)(i)] and even [(b)(ii)] number of half cycles, but without momentum exchange. Panels (c) represent interfering processes for which there are temporal shifts and momentum exchange,  considering odd and even numbers of half cycles [panels (c)(i) and (c)(ii), respectively]. In the diagrams, different colors indicate that the events are shifted in time, and the same color indicates the same event. The phase shifts $\alpha_{\mu \nu}$ are written in the figure. For simplicity, for events displaced by more than a cycle, we have only stated phase shifts associated with a single period, and, in (b)(ii), we omitted the events giving $\alpha_{\mu,\nu T}$ for which the second argument is delayed by a cycle. However, these events and those separated by longer times can be inferred using this diagram.
 }
    \label{fig:interfdiagram}
\end{figure}

In Fig.~\ref{fig:interfdiagram}, we present a graphic representation of the three qualitatively different types of intra-channel phase shifts that may occur, with the interfering transition amplitudes being indicated by thick solid lines. One should bear in mind that the phase differences $\alpha_{\mu \nu}$ illustrated in the figure are a sample of what may occur and the maximal temporal difference $\Delta \tau$ considered is a full cycle of the field.  However, this notation can be extrapolated to encompass larger time differences.  The diagrams are a simplification of the main momentum regions occupied by specific transition amplitudes. They allow a quick assessment of what is going on. Different colors indicate temporal shifts, while the same color indicates simultaneous occurrence.  

Fig.~\ref{fig:interfdiagram}(a) shows the phase differences purely associated with the electrons being exchanged. The phase shift  $\alpha_{ld}$ ($\alpha_{ru}$) is associated with the pairwise interference of $M_l$ and $M_d$ ($M_r$ and $M_u$). These coherent sums are expected to populate mainly the first and third quadrants of the $p_{1\parallel}p_{2\parallel}$ plane, and each pair represents the same event, but in one of the contributions $\mathbf{p}_1$ and $\mathbf{p}_2$ are interchanged. This is indicated by using the same colors for a pair.

Fig.~\ref{fig:interfdiagram}(b) shows the phase differences uniquely given by temporal shifts, without exchanging the electron momenta. In Fig.~\ref{fig:interfdiagram}(b)(i), we plot a diagrammatic representation of the pairwise interference of events displaced by a half cycle. These events are located in opposite momentum half axes. The phase difference $\alpha_{lr}$ is associated with the amplitudes  $M_l$ and $M_r$, located along the $p_{1\parallel}$ axis, and the phase shift $\alpha_{du}$ stems from the difference of  $M_d$ and $M_u$. This diagram can be generalized to an odd number of half-cycles. 
The diagram in Fig.~\ref{fig:interfdiagram}(b)(ii) represents the interference of events displayed by a full cycle, which we indicate by including the subscript $T$ in $\alpha$. For example, $\alpha_{lT,l}$ means that the interfering actions are $S_{lT}=S_l(\mathbf{p}_1,\mathbf{p}_2,t+T,t'+T,t''+T)$ and $S_l(\mathbf{p}_1,\mathbf{p}_2,t,t',t'')$, and the same holds for the other scenarios illustrated in that figure. In this case, the different events occupy the same parallel momentum half-axis, and therefore their overlap is significant. A similar pattern is encountered if the temporal shift is an arbitrary integer number of cycles. 

Finally, the diagrams in Fig.~\ref{fig:interfdiagram}(c) illustrate interfering processes for which there are temporal shifts \textit{and} electron exchange. Fig.~\ref{fig:interfdiagram}(c)(i) shows the interference of the amplitudes  $M_l$ and $M_u$ ($M_r$ and $M_d$), which correspond to processes displaced by half a cycle in which the two electron momenta have been swapped. Finally, the schematic representations in  Fig.~\ref{fig:interfdiagram}(c)(ii) depict interfering processes displaced by a full cycle, and whose momenta were exchanged. In summary, there are a few common patterns in Fig.~\ref{fig:interfdiagram}: L-shaped structures indicate that the electron momenta were exchanged in one of the interfering terms;  straight lines indicate processes displaced by either a full number of cycles (short lines), or an odd number of half cycles (long lines);  the same (different) colors indicate the absence (presence) of temporal shifts. We will apply this notation to concrete examples in subsequent sections, once we have outlined a way to select the relevant events within a pulse. They are also useful to understand the equations provided next. 

\subsection{Analytic expressions}
\label{sec:interfanalytic}

In the following, we derive analytic expressions for a wide range of phase differences that occur in the RESI amplitudes, and, as much as possible, we discuss what patterns are expected from their interference.

\subsubsection{Intra-cycle phase differences}

Let us commence by exploring the phase differences that occur within a single field cycle, and are illustrated by the diagram in Fig.~\ref{fig:interfdiagram}. The phase difference $\alpha_{ld}$ is purely associated with exchanging $\mathbf{p}_1$ and $\mathbf{p}_2$ and reads
\begin{equation}
  \alpha_{ld}=S_l-S_d=  \alpha^{(\mathrm{exch})}_{\mathbf{p}_1,\mathbf{p}_2}+ \alpha_{\mathbf{p}_1,\mathbf{p}_2}(t,t'),
\label{eq:phaseleftdown}
\end{equation}
where
\begin{equation}
\alpha^{(\mathrm{exch})}_{\mathbf{p}_1,\mathbf{p}_2}=\frac{1}{2}  \left(\mathbf{p}_{2 }^2 -\mathbf{p}_{1}^2\right)(t-t'),
\label{eq:alphaexch}
\end{equation}
\begin{equation}
\alpha_{\mathbf{p}_1,\mathbf{p}_2}(t,t')=  (\mathbf{p}_2-\mathbf{p}_1) \cdot \left[\mathbf{F}_{A}(t')-\mathbf{F}_{A}(t)\right]
  \label{eq:alphaexch2}
\end{equation}
and 
\begin{equation}
    \mathbf{F}_A(t)=\int^t \mathbf{A}(\tau)d\tau.
\end{equation}
is the temporal integral of the vector potential. 

Constructive interference between $M_l$ and $M_d$ will happen for $\alpha_{ld}=2n\pi$, where $n$ is an integer, and, for simplicity, we will investigate $n=0$. The phase shift  $\alpha^{(\mathrm{exch})}_{\mathbf{p}_1,\mathbf{p}_2}$ is field-independent, and requiring it to vanish gives equations of hyperbolae in the $p_{1\parallel}p_{2\parallel}$ plane, whose asymptotes are $p_{1\parallel}=\pm p_{2\parallel}$ and whose eccentricity is $\sqrt{2}$. The shift $ \alpha_{\mathbf{p}_1,\mathbf{p}_2}(t,t')$ is field dependent and states that $\mathbf{p}_2-\mathbf{p}_1$ must be orthogonal to $\mathbf{F}_A(t')-\mathbf{F}_A(t)$. For linearly polarized light, it reduces to $p_{1\parallel}=p_{2\parallel}$. The conditions upon the momenta are identical to those obtained in \cite{Maxwell2015} for a monochromatic field, and determine that bright fringes must be present in the diagonal of the parallel momentum plane. Their independence of the field profile makes these central fringes robust against focal averaging, as found in our previous publication \cite{Maxwell2016}. Nonetheless, the asymptotes and axes of the hyperbolae may shift as they depend on the transverse momentum integration. 

The other phase difference solely related to electron momentum exchange is 
\begin{equation}
  \alpha_{ru}=S_r-S_u=  -\alpha^{(\mathrm{exch})}_{\mathbf{p}_1,\mathbf{p}_2}+ \alpha_{\mathbf{p}_1,\mathbf{p}_2}(t+\frac{\pi}{\omega},t'+\frac{\pi}{\omega}),
\label{eq:phaserightup}
\end{equation}
which, if made to vanish, also leads to hyperbolae and a bright fringe at the main diagonal for linearly polarized fields. One should note, however, that the temporal shifts in the second term of Eq.~\eqref{eq:phaserightup} imply that $|\alpha_{ru}| \neq |\alpha_{ld}|$ unless $\mathbf{F}_A(t)=-\mathbf{F}_A(t + \pi/\omega)$. This condition, known as the half-cycle symmetry, renders $\alpha_{ru} = - \alpha_{ld}$. 
If this condition is broken, for instance for a driving field consisting of a wave and its second harmonic, or for the few-cycle pulses discussed in Secs.~\ref{sec:pulseshape} and \ref{sec:PMDpulses}, the hyperbolae symmetric about the main diagonal will be asymmetric about $(p_{1\parallel},p_{2\parallel}) \leftrightarrow (-p_{1\parallel},-p_{2\parallel})$, that is, in the first and third quadrants of the $p_{1\parallel}p_{2\parallel}$ plane.

Next, let us calculate the phase difference $\alpha_{lr}$ ($\alpha_{du}$)  between the right and the left (upward and downward) peaks [see diagrams in Fig.~\ref{fig:interfdiagram}(b)(i)].
The action $S_r$ ($S_u$) gives events shifted from those corresponding to $S_l$($S_d$) by half a cycle, but no momentum exchange between the first and second electron takes place. This means that $S_l(\mathbf{p}_1,\mathbf{p}_2,t,t',t'')=S_r(\mathbf{p}_1,\mathbf{p}_2,t+\pi/\omega,t'+\pi/\omega,t''+\pi/\omega)$ and that $S_r(\mathbf{p}_1,\mathbf{p}_2,t,t',t'')=S_l(\mathbf{p}_1,\mathbf{p}_2,t+\pi/\omega,t'+\pi/\omega,t''+\pi/\omega)$. For $\alpha_{lr}$, this gives 
\begin{align}
     \alpha_{lr}=S_l-S_r&=-\frac{1}{2(t'-t'')}\alpha^{(A^2)}_{\pi/\omega}(t',t'')- \frac{1}{2}\alpha^{(\mathrm{pond})}_{\pi/\omega}(t,t'')\nonumber \\&-\alpha^{(\mathrm{ene})}_{\pi/\omega}-\alpha^{(\mathbf{p}_1,\mathbf{p}_2)}_{\pi/\omega}(t,t'), 
     \label{eq:alphalr}
\end{align}
where the purely temporal phase shifts are defined as 
\begin{align}
\alpha^{(A^2)}_{\Delta\tau}(t',t'')&=\left[\mathbf{F}_A(t'+\Delta\tau)-\mathbf{F}_A(t''+\Delta\tau)\right]^2\nonumber \\ &-\left[\mathbf{F}_A(t')-\mathbf{F}_A(t'')\right]^2,
    \label{eq:alphaA2}
\end{align}
\begin{align}
    \alpha^{(\mathrm{pond})}_{\Delta \tau}(t,t'')&=F_{A^2}(t+\Delta\tau)+F_{A^2}(t''+\Delta\tau)\nonumber\\ &-F_{A^2}(t)-F_{A^2}(t''),
     \label{eq:alphapond}
\end{align}
and
\begin{equation}
 \alpha^{(\mathrm{ene})}_{\Delta\tau}=    \frac{\Delta\tau}{ 2 }  \left(2 \text{E}_{2 g}+2 \text{E}_{1 g}+\mathbf{p}_{1 }^2+\mathbf{p}_{2 }^2\right), 
 \label{eq:alphaene}
\end{equation}
with 
\begin{equation}
    F_{A^2}(t)=\int^t \mathbf{A}^2(\tau)d\tau.
\end{equation}
One should note that Eqs.~\eqref{eq:alphaA2}, \eqref{eq:alphapond} have temporal arguments, which indicate that the values of  $F_{A^2}$ and $F_{A}$ may vary within a pulse and this must be taken into consideration, while \eqref{eq:alphaene} depends only on the time difference $\Delta\tau$. The phase difference in Eq.~\eqref{eq:alphapond} is expected to give ponderomotive terms, which will add to that in Eq.~\eqref{eq:alphaene}. The latter is linear in time and contains the bound-state and kinetic energies. 

The phase difference $\alpha^{(\mathbf{p}_1,\mathbf{p}_2)}_{\pi/\omega}(t,t')$ is a particular case of 
  \begin{align}
\alpha^{(\mathbf{p}_1,\mathbf{p}_2)}_{\Delta \tau}(t,t')&=\mathbf{p}_1\cdot\left[\mathbf{F}_A(t'+\Delta \tau) - \mathbf{F}_A(t') \right] \nonumber \\&+ \mathbf{p}_2\cdot \left[\mathbf{F}_A(t+\Delta \tau) - \mathbf{F}_A(t) \right],
   \label{eq:alphap1p2tau}  
  \end{align}
which also incorporates the electron momenta. Similarly, 
\begin{align}
     \alpha_{du}=S_d-S_u&=-\frac{1}{2(t'-t'')}\alpha^{(A^2)}_{\pi/\omega}(t',t'')- \frac{1}{2}\alpha^{(\mathrm{pond})}_{\pi/\omega}(t,t'')\nonumber \\&-\alpha^{(\mathrm{ene})}_{\pi/\omega}-\alpha^{(\mathbf{p}_2,\mathbf{p}_1)}_{\pi/\omega}(t,t'), 
     \label{eq:alphadu}
\end{align}
where $\alpha^{(\mathbf{p}_2,\mathbf{p}_1)}_{\pi/\omega}$ indicates that the electron momenta have been swapped in Eq.~\eqref{eq:alphap1p2tau}, with $\Delta \tau=\pi/\omega$. For a monochromatic wave, we have found that the effect of these terms is minimal \cite{Shaaran2010,Maxwell2015}.

Computing the phase shifts $\alpha_{lu}$ and $\alpha_{rd}$ means that, in addition to exchanging $\mathbf{p}_1$ and $\mathbf{p}_2$, we must consider that one of the interfering events will be displaced by half a cycle [see Fig.~\ref{fig:interfdiagram}(c)]. Specifically,  we must note that $S_u(\mathbf{p}_1,\mathbf{p}_2,t,t',t'')=S_l(\mathbf{p}_2,\mathbf{p}_1,t+\pi/\omega,t'+\pi/\omega,t''+\pi/\omega)$ and that $S_r(\mathbf{p}_1,\mathbf{p}_2,t,t',t'')=S_d(\mathbf{p}_2,\mathbf{p}_1,t+\pi/\omega,t'+\pi/\omega,t''+\pi/\omega)$. These phase differences will mainly lead to fringes in the second and fourth quadrant of the $p_{1\parallel}p_{2\parallel}$ plane.  

Explicitly, this gives
\begin{align}
    \alpha_{lu}=S_l-S_u&=  -\alpha^{(\mathrm{exch})}_{\mathbf{p}_1,\mathbf{p}_2} -\frac{1}{2(t'-t'')}\alpha^{(A^2)}_{\pi/\omega}(t',t'')\nonumber \\& - \frac{1}{2}\alpha^{(\mathrm{pond})}_{\pi/\omega}(t,t'')-\alpha^{(\mathrm{ene})}_{\pi/\omega}-\alpha^{(\mathbf{p}_1 \leftrightarrow \mathbf{p}_2)}_{\pi/\omega} (t,t'),
    \label{eq:phaseupleft}
\end{align}
where the first four phase differences are given by Eqs.~\eqref{eq:alphaexch}, \eqref{eq:alphaA2}, \eqref{eq:alphapond} and \eqref{eq:alphaene}. 
The term
\begin{align}
 \alpha^{(\mathbf{p}_1 \leftrightarrow \mathbf{p}_2)}_{\Delta \tau}&(t,t')=\mathbf{p}_2\cdot \mathbf{F}_A(t'+\Delta \tau) + \mathbf{p}_1\cdot \mathbf{F}_A(t+\Delta \tau) \nonumber \\&-\mathbf{p}_2\cdot \mathbf{F}_A(t)-\mathbf{p}_1\cdot \mathbf{F}_A(t'),
   \label{eq:alphaexctau}
\end{align}
contains a momentum dependence and the double arrow in the superscript indicates that the momentum in the second row of Eq.~\eqref{eq:alphaexctau} are exchanged with regard to the temporal arguments  $t$ and $t'$ of $\mathbf{F}_A$. We have verified that $\alpha_{lu}=-\alpha_{rd}$ regardless of the field shape. This guarantees symmetry of these contributions in the second and fourth quadrant of the $p_{1\parallel}p_{2\parallel}$ plane. 

Eq.~\eqref{eq:phaseupleft} shows that, in addition to $\alpha^{(\mathrm{exch})}_{\mathbf{p}_1,\mathbf{p}_2}$, which is also present for $\alpha_{ld}$ and leads to hyperbolic interference fringes, the phase shift  $\alpha_{lu}$ contains several terms which come from the temporal shifts. The effects of these equations are more difficult to visualize. For linearly polarized fields, the scalar products in Eq.~\eqref{eq:alphaexctau} will reduce to the products of $p_{2\parallel}$ and $p_{1\parallel}$ with field-dependent terms. Imposing that $\alpha^{(\mathbf{p}_1 \leftrightarrow \mathbf{p}_2)}_{\Delta \tau}(t,t')=0$ will lead to conditions that depend on the field symmetry, although a first inspection may give the impression that there should be fringes located at the axes $p_{n\parallel}=0$. 

For instance, one may verify that, for fields satisfying the half cycle symmetry, the condition $\mathbf{F}_A(t)=-\mathbf{F}_A(t + \pi/\omega)$ leads to $ \alpha^{(A^2)}_{\pi/\omega}(t',t'')=0$ and 
\begin{equation}
   \alpha^{(\mathbf{p}_1 \leftrightarrow \mathbf{p}_2)}_{\pi/\omega}(t,t')=-(\mathbf{p}_2+\mathbf{p}_1) \cdot \left[\mathbf{F}_A(t')+\mathbf{F}_A(t)\right].
     \label{eq:alphaulc2}
\end{equation}
For linearly polarized fields satisfying the half-cycle symmetry, the scalar product stated above becomes
$(p_{2\parallel}+p_{1\parallel})\left[F_A(t')+F_A(t)\right]$, and the requirement that $ \alpha^{(\mathbf{p}_1 \leftrightarrow \mathbf{p}_2)}_{\pi/\omega}$ vanishes for all times would imply that $p_{2\parallel}=-p_{1\parallel}$.

If, further to that, the field is taken to be monochromatic and linearly polarized, that is $A(t)=2 \sqrt{U_p}\cos \omega t$, one can show that
\begin{align}
       \alpha^{(\mathrm{mono})}_{lu}&=-\frac{\pi}{2 \omega} \left[4U_p+2E_{1g}+2E_{2g}+\mathbf{p}_{1 }^2+\mathbf{p}_{2 }^2\right]\nonumber \\
      &-\frac{1}{2}  \left(\mathbf{p}_{2 }^2 -\mathbf{p}_{1}^2\right)(t-t') \nonumber \\&+\frac{\sqrt{U_p}}{\omega} (p_{2\parallel}+p_{1\parallel})  \left[\sin \omega t'+\sin \omega t\right],
      \label{eq:phaseleftupmono}
\end{align}
where the extra term in $U_p$ comes from $\alpha^{(\mathrm{pond})}_{\pi/\omega}(t,t'')= 4U_p \pi / \omega$ in this case. 

Due to several field- and energy-dependent terms in $\alpha_{lu}$, it is not straightforward to derive simple analytic expressions for constructive interference between $M_u$ and $M_l$. There is, however, a term in $\mathbf{p}^2_{2}-\mathbf{p}^2_{1}$, which, if required to vanish, leads to hyperbolic fringes. Interestingly, because, for a few-cycle pulse,  the half-cycle symmetry is broken, we cannot use the argument in \cite{Maxwell2015}, in which imposing that the last term in \eqref{eq:phaseleftupmono} vanishes leads to a fringe along the anti-diagonal $p_{1\parallel}=-p_{2\parallel}$. 

\subsubsection{Inter-cycle phase differences}

Furthermore, in case we are considering more than one cycle of the field, we must also assess what happens for contributions displaced by at least a full cycle ($\Delta \tau \geq 2 \pi/\omega$), without and with momentum exchange [see Figs.~\ref{fig:interfdiagram}(b)(ii) and (c)(ii), respectively]. The contributions from events separated by a full number of cycles overlap considerably in momentum space, which potentially makes their interference substantial. This is particularly important in the present scenario, as the pulse envelope renders the cycles not equivalent. 

Shifting the action associated with the left peak in a full cycle gives $S_{l,T}(\mathbf{p}_1,\mathbf{p}_2,t,t',t'')=S_l(\mathbf{p}_1,\mathbf{p}_2,t+2\pi/\omega,t'+2\pi/\omega,t''+2\pi/\omega)$, where we have defined $T=2\pi/\omega$. 
Similarly, for the downward peak one obtains $S_{d,T}(\mathbf{p}_1,\mathbf{p}_2,t,t',t'')=S_d(\mathbf{p}_1,\mathbf{p}_2,t+2\pi/\omega,t'+2\pi/\omega,t''+2\pi/\omega)$. The phase difference  $\alpha_{lT,l}$ then reads
\begin{align}
\alpha_{lT,l}=S_{l,T}-S_{l}&=\frac{1}{2(t'-t'')}\alpha^{(A^2)}_{2\pi/\omega}(t',t'')\nonumber \\
&+\frac{1}{2}\alpha^{(\mathrm{pond})}_{2\pi/\omega}(t,t'')+\alpha^{(\mathrm{ene})}_{2\pi/\omega}+\alpha^{(\mathbf{p}_1,\mathbf{p}_2)}_{2\pi/\omega}(t,t'), 
\label{eq:phaseperiodl}
\end{align}
where the phase differences have been defined in Eqs.~\eqref{eq:alphaA2},\eqref{eq:alphapond}, \eqref{eq:alphaene} and \eqref{eq:alphap1p2tau}. 
 The phase difference $\alpha_{dT,d}=S_{d,T}-S_{d}$ resembles Eq.~\eqref{eq:phaseperiodl}, but with the momenta swapped in the last building block, i.e., $\alpha^{(\mathbf{p}_2,\mathbf{p}_1)}_{2\pi/\omega}(t,t')$.  
The phase differences $\alpha_{rT,r}=S_{r,T}-S_{r}$ and  $\alpha_{uT,u}=S_{u,T}-S_{u}$ are calculated similarly, but the arguments of the functions $F_A$ and $F_{A^2}$ must be shifted in half a cycle with regard to those used in Eq.~\eqref{eq:phaseperiodl}. This implies that  one must take $\alpha^{(A^2)}_{2\pi/\omega}(t'+\pi/\omega,t''+\pi/\omega)$, $\alpha^{(\mathrm{pond})}_{2\pi/\omega}(t+\pi/\omega,t''+\pi/\omega)$ and $\alpha^{(\mathbf{p}_i,\mathbf{p}_j)}_{2\pi/\omega}(t+\pi/\omega,t'+\pi/\omega) $ in the above equation. 

For fields of period  $T=2\pi/\omega$,  $\alpha^{(A^2)}_{2\pi/\omega}$ and $\alpha^{(\mathbf{p}_i,\mathbf{p}_j)}_{2\pi/\omega}$, with $i \neq j$, will vanish, while $ \alpha^{(\mathrm{pond})}_{2\pi/\omega}$ will give ponderomotive shifts which will add to $\alpha^{(\mathrm{ene})}_{2\pi/\omega}$. This will happen regardless of whether the half-cycle symmetry is broken, but will not apply to the pulse studied in this work. An example of this specific scenario is a monochromatic field, for which 
\begin{equation}
\alpha^{\mathrm{mono}}_{lT,l}= \alpha^{\mathrm{mono}}_{dT,d} =\frac{\pi}{\omega} \left[4U_p+2E_{1g}+2E_{2g}+\mathbf{p}_{1 }^2+\mathbf{p}_{2 }^2\right]
\end{equation}
is the equation of a six-dimensional hypersphere in momentum space.  
Re-writing the condition for interference maxima $\alpha^{\mathrm{mono}}_{lT,l}=2n\pi$ in terms of the momentum components parallel to the driving-field polarization yields
\begin{equation}
p_{1\parallel}^2+p_{2\parallel}^2 =2n\omega-(4U_p+2E_{1g}+2E_{2g}+\mathbf{p}^2_{1\perp}+\mathbf{p}^2_{2\perp}), 
\end{equation}
whose projection in the $p_{1\parallel}p_{2\parallel}$ plane gives circular fringes if the integer $n$ is sufficiently large so that the right-hand side is positive. 
Interference fringes taking into account electron exchange and events displaced by a full cycle can be obtained by considering, for instance, $\alpha_{lT,d}=S_{l,T}-S_d=\alpha_{lT,l}+\alpha_{ld}$, or  $\alpha_{uT,l}=S_{u,T}-S_l=\alpha_{uT,u}+\alpha_{ul}$. The diagrams in Fig.~\ref{fig:interfdiagram} are useful as a guidance to construct phase shifts associated with events displaced by even or odd numbers of half cycles. Alternatively, if all ionization and rescattering times are computed directly, they may be used to construct these shifts. 

One should note that the time shifts employed in this section are an approximation for the times calculated for different events in few-cycle laser pulses. However, we verified that this works reasonably well for the dominant events (not shown).  Understanding these events and their dominance will be the main topic of the next section. 
\section{Investigating the pulse}
\label{sec:pulseshape}
In the results that follow, we employ a linearly polarized few-cycle pulse $\mathbf{E}(t)=-d\mathbf{A}(t)/dt$, whose  vector potential is determined by
\begin{equation}
\mathbf{A}(t)= A_0\sin^2\left(\frac{\omega t}{2 N}\right)\sin{(\omega t + \phi)}\hat{e}_z,
    \label{eq:Apulse}
\end{equation}
where $A_0$ is the vector-potential amplitude, $N$ is the number of cycles, $\omega$ is the field frequency, $\phi$ the carrier-envelope phase.  Throughout this work, we have made the approximation  $A_0 = 2\sqrt{U_p}$. This is exact for a monochromatic linearly polarized field, but not for a few cycle pulse. By varying the parameters in \eqref{eq:Apulse}, we can influence the main rescattering and ionization events.

\begin{figure}
    \centering
\includegraphics[width=\columnwidth]{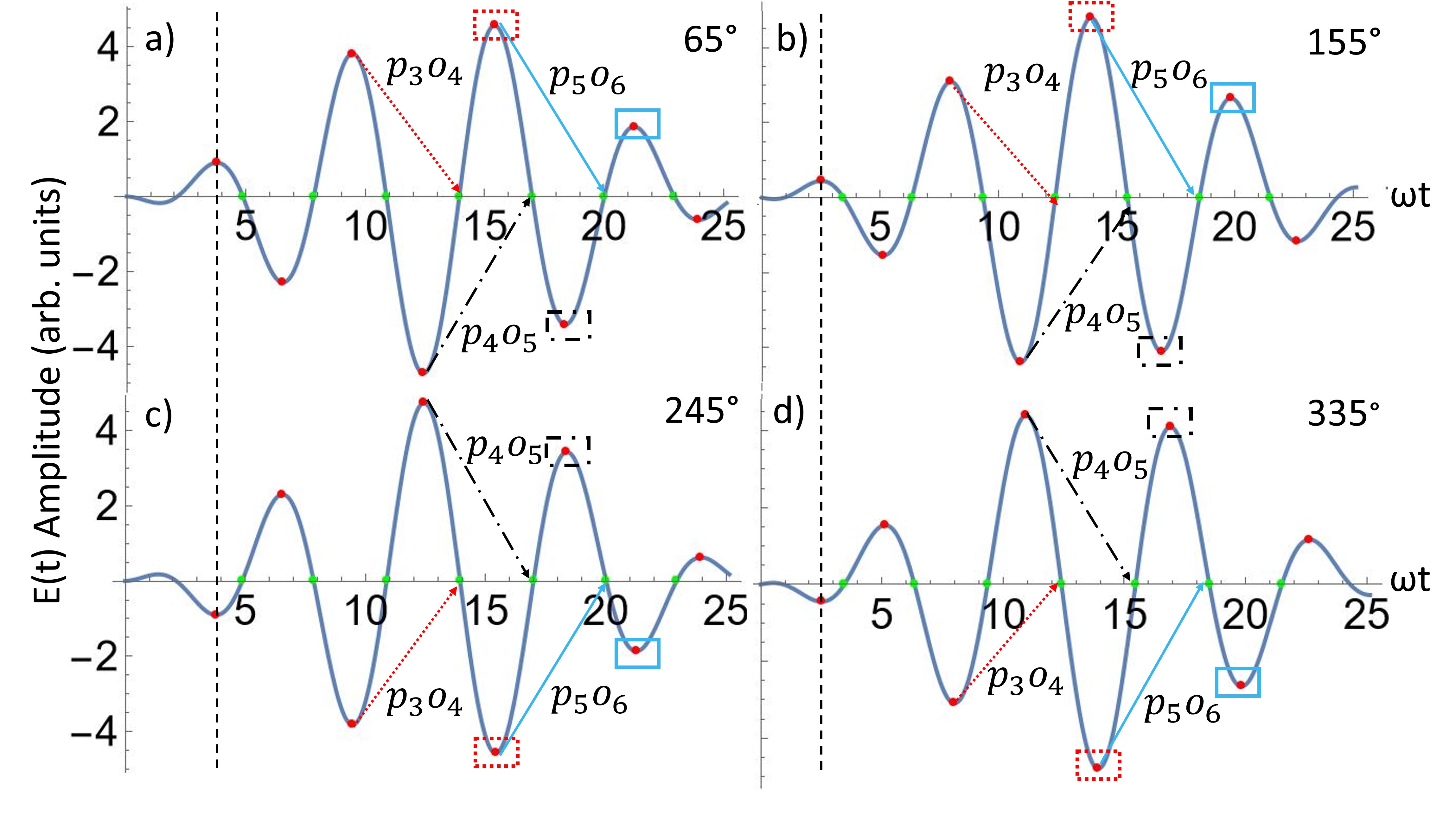}
    \caption{Few-cycle pulse given by (\ref{eq:Apulse}), with peak intensity $I=1.5 \times 10^{14}\mathrm{W/cm}^2$, wavelength $\lambda=800$ nm ($\omega=0.057$ a.u.), $N=4.3$ and carrier-envelope phases $\phi_1=65^{\circ}, 155^{\circ}, 245^{\circ}, 335^{\circ}$ [panels (a), (b), (c) and (d), respectively].  The ionization and rescattering times at the peak and zero crossings of the field are indicated by the red and green dots. The three main events $p_io_j$ towards the center of the pulse are labeled with their corresponding pair and orbit numbers. The ionization and return times associated with the pairs of orbits $p_3$, $p_4$, $p_5$ of the first electron are indicated by arrows, and the most relevant ionization times for the orbits $o_j$ of the second electron, with $(j=4,5,6)$, are signposted by rectangles. The initial numbers chosen for the indices $i,j$ refer to the extremum of the field for which the counting starts.  Matching styles and colors have been used for different events $\varepsilon_k=p_io_j$. }
    \label{fig:pulseshape}
\end{figure}

In Fig.~\ref{fig:pulseshape}, we plot the specific $\mathrm{sin}^2$ pulse employed in this work.  Its length ($N=4.3$) and carrier-envelope phases (CEPs) have been chosen in order to facilitate a comparison with our previous publication \cite{Faria2012}, although we have taken a lower peak intensity to ensure that RESI is the dominant mechanism. This means that the maximal kinetic energy of the returning first electron is lower than the second ionization potential. Overall we observe that $E(t,\phi)=-E(t,\phi+\pi)$. This symmetry will influence the electron momentum distributions so that $\mathcal{P}(p_{1\parallel},p_{2\parallel},\phi)=\mathcal{P}(-p_{1\parallel},-p_{2\parallel},\phi+\pi)$. Thus, without loss of generality, we may consider only two of the CEPs employed in \cite{Faria2012}.  Here, we follow the same convention as in our previous publication \cite{Faria2012}  and define the carrier-envelope phases as $\phi=\phi_1-\phi_0$, where $\phi_0=60^{\circ}$ is an
offset value. When discussing the CEP in the following figures, we refer to $\phi_1$ without the offset phase. 

The classical counterparts $\mathrm{Re}[t'']$,  $\mathrm{Re}[t']$ and $\mathrm{Re}[t]$ of the ionization and rescattering times computed with Eqs.~\eqref{eq:sp1}-\eqref{eq:sp4} are indicated in the figure for both electrons, where the orbits are classified in increasing numerical order starting from the beginning of the pulse. This classification has been used in \cite{Faria2012}. The orbits associated with the first electron occur in pairs \cite{Shaaran2012,Faria2012} and, for that reason, are indicated by $p_i$. They start after a maximum of the field and end near a zero crossing roughly three quarters of a cycle later (see arrows in the figure). Each pair is composed of a short and a long orbit, whose  classical ionization and times read $(\mathrm{Re}[t'_s],\mathrm{Re}[t''_s])$ and $((\mathrm{Re}[t'_l],\mathrm{Re}[t''_l])$, respectively.  The classical times shown in the figure have been estimated using the tangent construction \cite{Faria1999}.

The most relevant orbits associated with the second electron start around the field peak closest to the rescattering times, and are denoted as $o_j$. An event encompassing the two electrons is classified as $\varepsilon_k=p_io_j$, where $i,j,k$ are integers. For example, $\varepsilon_1=p_3o_4$ designates the first event in the pulse we incorporate, in which the first electron returns following the pair or orbits $p_3$ and the second electron is released along the orbit $o_4$ (see Fig.~\ref{fig:pulseshape} for clarity).  Note that, in this work, we only consider the ionization event $o_j$ for the second electron immediately following the rescattering of the first electron, as, due to bound-state depletion, the probability of later events is smaller and therefore less important.

\subsection{Finding dominant orbits}
\label{sec:dominance}

A central question when dealing with pulses is how to find the events whose contributions to the single-channel probability density $\mathcal{P}^{(\mathcal{C})}(p_{1\parallel},p_{2\parallel})$ are dominant. This provides a roadmap of what can be neglected or incorporated without compromising the physics. In previous publications \cite{Shaaran2012,Faria2012}, we have performed qualitative studies of this issue. Here, we aim to quantify the dominance of an event within a single excitation channel by proposing a dominance parameter.
An orbit's contribution will prevail in $\mathcal{P}^{(\mathcal{C})}(p_{1\parallel},p_{2\parallel})$ if three conditions are satisfied.

First, the ionization probability for the first electron must be high \cite{Faria2004,Faria2012}. For a specific orbit, the ionization probability is proportional to $\exp [-2\mathrm{Im}S^{(\mathcal{C})}(\mathbf{p}_{1},\mathbf{p}_{2},\mathbf{k},t,t^{\prime },t^{\prime\prime })]$. In order to consider the first electron individually, one must note that the action will be proportional to the ionization times. Thus,  $\exp[-2\mathrm{Im}[t'']]$ is a good indicator of whether ionization is significant for the first electron along a specific orbit [see also the saddle-point equation (\ref{eq:sp1})]. A small imaginary part for $t''$ means a high ionization probability. For simplicity, one may consider $ |1/\mathrm{Im}[t'']|$.

Second, the classically allowed region (CAR) for the (returning) first electron must be large\footnote{The classically allowed region can be identified by nearly vanishing $\mathrm{Im}[t']$. At the boundary of this region, there is a sudden increase in $|\mathrm{Im}[t']|$ and $\mathrm{Re}[t']$ nearly coalesce. For details see \cite{Faria2012}. }. This means that the first electron has returned with enough kinetic energy for the process to have a classical counterpart over a large momentum range. Therefore, according to Eq.~\eqref{eq:sp3}, there is a substantial energy transfer to the core and the excitation of the second electron is very likely. We take a rough estimate of the width of the region in momentum space, where the difference between imaginary parts of the rescattering times $t'$ for the long and short orbits is minimal. The width is found by taking the difference between the lowest and highest classically allowed momenta $p^{(\mathrm{min})}_{1\parallel}$ and $p^{(\mathrm{max})}_{1\parallel}$ for $p_{1\perp}=0$. A more precise value of these momenta can be found by solving the saddle-point equation (\ref{eq:sp3}) for $p_{1\perp}=0$.  However, we found that this added precision did not affect the dominance parameter value significantly so it is sufficient to take an estimate.
 On the other hand, it may happen that the complex times obtained from solving the saddle-point equations have no classical counterpart and this region collapses. In that case, we take the most probable momentum. This momentum is chosen so that $|\mathrm{Im}[t^{\prime}]|$ has a minimum. Physically, this implies that rescattering is most probable at that time, although it has no classical equivalent \cite{Faria2004,Faria2012}. Because we are taking the momentum for which rescattering is most probable, this is likely an overestimation of the actual probability for this process.

Third, there must be a high ionization probability for the second electron. This is inferred similarly to the first condition, with the difference that, instead, we take $ |1/\mathrm{Im}[t]|$ and the saddle-point equation (\ref{eq:sp4}). One should also prioritize the first ionization event after recollision, as those in subsequent half-cycles will be suppressed by bound-state depletion.

Putting all these criteria together, we can define the dominance parameter as a product of the partial dominance parameters for the rescattered and direct electrons
\begin{equation}
\mathcal{D}(p_i,o_j) = \mathcal{D}^{(1)}(p_i) \mathcal{D}^{(2)}(o_j),
\end{equation}
where the indices $p_i,o_j$ refer to specific events (classified as pairs and orbits, respectively - see Fig \ref{fig:pulseshape}) and
\begin{equation}
\mathcal{D}^{(1)}(p_i) = \frac{|\mathrm{Im}[\bar{t}_{p_i}'']|^{-1}| |p^{(\mathrm{max})}_{1\parallel}-p^{(\mathrm{min})}_{1\parallel}| }{ \mathcal{D}^{(\mathrm{e_{1}})}_{\mathrm{max}}}
\end{equation}
\begin{equation}
\mathcal{D}^{(2)}(o_j) = \frac{|\mathrm{Im}[t_{o_j}]^{(min)}|^{-1}}{ \mathcal{D}^{(\mathrm{e_{2}})}_{\mathrm{max}}}
\end{equation}

The denominator $\mathcal{D}^{(e_n)}_{(\mathrm{max})}$, $n=1,2$, is the maximum dominance parameter for the corresponding electron, for the channel in question, which we have used as a normalization. 
In the partial dominance parameter $\mathcal{D}^{(1)}(p_i)$ for the first electron, as the orbits occur in pairs, we have considered the average of $\mathrm{Im}[t_{\varepsilon_1}'']$ within a pair $p_i$ of orbits at vanishing parallel and perpendicular momenta. Vanishing perpendicular momenta are used to obtain the largest possible classically allowed region as non-zero values simply shift the ionization potential \cite{Shaaran2010}. For the second electron, we have taken the minimum of the imaginary part of the tunneling time as this will lead to the greatest probability.

\begin{figure*}
    \centering
\includegraphics[width=\textwidth]{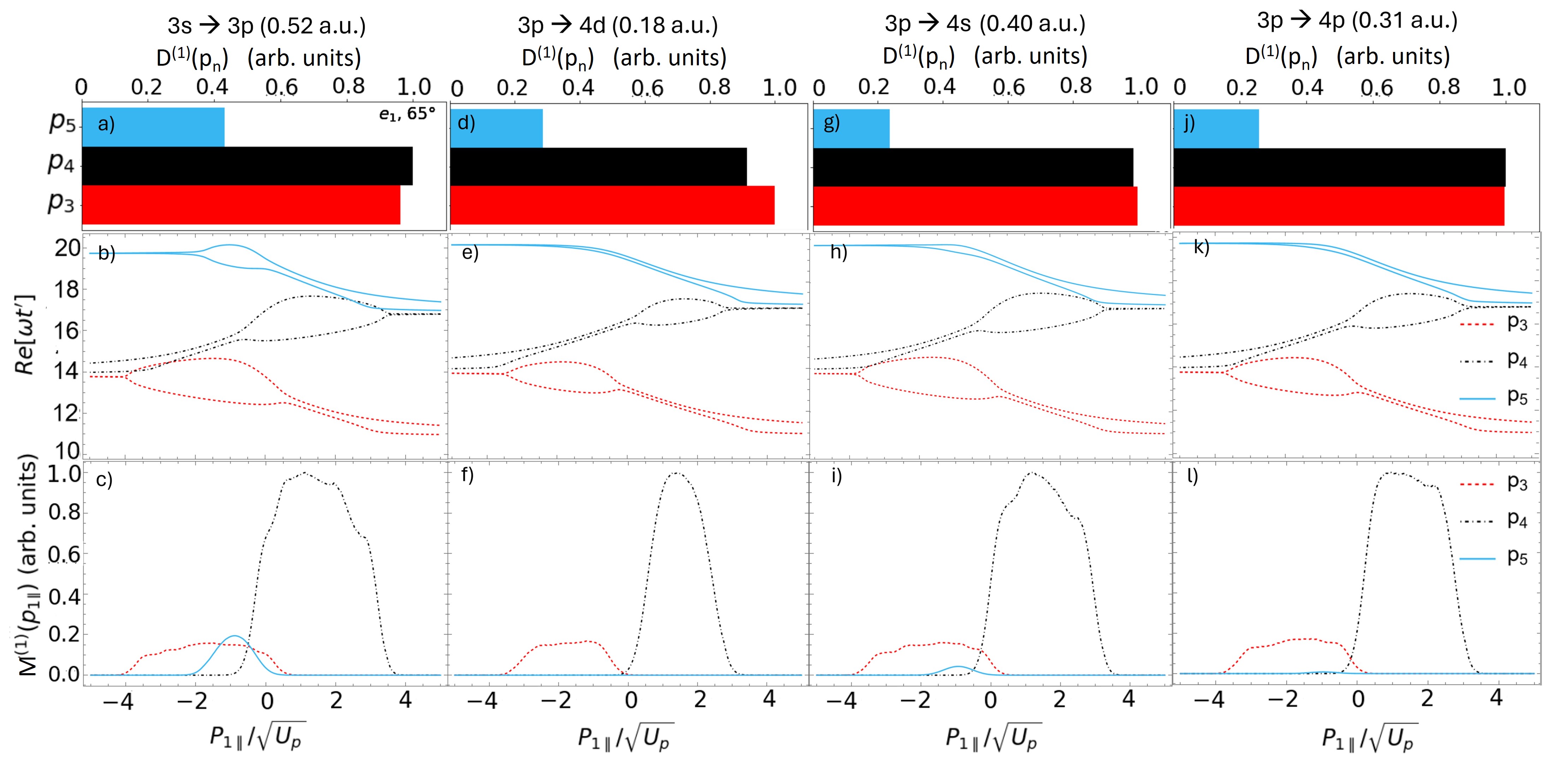}
    \caption{Estimates of the partial dominance parameter $\mathcal{D}^{(1)}(p_j)$ (top row), the real part of the rescattering time calculated for $p_{1\perp}=0$(middle row) and the partial momentum distribution [Eq.~\eqref{eq:partialdistr}] calculated for the first electron as a function of the parallel momentum without prefactors (bottom row) for three events in the center of a pulse with CEP $\phi=65$, for the $3p\rightarrow4p$ (panels (a)-(c)), $3p\rightarrow4d$ (panels (d)-(f)), $3p\rightarrow4s$ (panels (g)-(i)) and $3s\rightarrow3p$ (panels (j)-(l)) transitions. The values of the dominance parameter and the momentum map have been normalized with respect to the most dominant event for each channel.}
    \label{fig:dpelectron1}
\end{figure*}

We will now discuss the dominance parameter for each electron separately.  Fig.~\ref{fig:dpelectron1} shows the dominance parameter for the first electron and four of the six excitation channels, together with the real parts of the ionization times and the partial momentum distributions for the first electron (first, second, and third row, respectively). The $3p\rightarrow5s$ 
and $3p\rightarrow3d$ 
transitions have been omitted in Fig.~\ref{fig:dpelectron1} because they strongly resemble (are almost identical to) the $3p\rightarrow4d$ 
and $3p\rightarrow4s$ 
transitions, given their very similar energy gaps.
This renders the analysis redundant unless prefactors are incorporated. Indeed, we have verified that the partial distributions and dominance of pairs are almost identical. 

For all channels, the least dominant contribution comes from $p_5$ arising from the rescattering of the first electron towards the edge of the pulse. This can be seen from the real parts of the rescattering times, displayed in the middle row of Fig.~\ref{fig:dpelectron1}. 
 The behavior of these times provides a good illustration of whether there is or not a classically allowed region. If a classically allowed region is present, $\mathrm{Re}[t']$ will surround the momentum value for which rescattering is most probable, and nearly coalesce at two values of $p_{1\parallel}$ which mark the minimal and maximal momentum for which rescattering has a classical counterpart. If there is no classical counterpart, 
$\mathrm{Re}[t']$ associated with a particular pair of orbits will be practically the same \cite{Faria2003}.
For $3s\rightarrow3p$, we see a classically allowed region for $p_5$ owing to the relatively small energy gap $\Delta E^{(\mathcal{C})} $ [Fig.~\ref{fig:dpelectron1}(b)]. 

For the $3p\rightarrow4d$, $3p\rightarrow4s$ and  $3p\rightarrow4p$ transitions the energy gap between the ground and excited state of the second electron is larger, and the classically allowed region collapses. This is reflected in the value of the dominance parameter [Figs.~\ref{fig:dpelectron1}(d), (g), (j)], which is lower than when a classically allowed region exists.  For Fig.~\ref{fig:dpelectron1}(g), pair 5 has the lowest dominance parameter. This happens because the classically allowed region is substantial for the other pairs [see Fig.~\ref{fig:dpelectron1}(h)], which brings it down due to the normalization employed. Still, one should note that Fig.~\ref{fig:dpelectron1}(i) exhibits a faint blue peak associated with $p_5$, while the probability density associated with this event is vanishingly small in Fig.~\ref{fig:dpelectron1}(f).  As the energy gap between the ground and excited states in Table \ref{tab:channels} increases, the residual kinetic energy for the first electron, and, consequently, the CAR will decrease. This will bring the dominance parameter associated with pair 3 slightly up. 
We do not take pair 5 into account in the final momentum maps if the CAR collapses. 

The contributions from $p_3$ and $p_4$ appear to be strongly competing for all channels studied - this is also indicated by comparable widths of the classically allowed regions for all channels. This can be intuitively understood by looking at Fig.~\ref{fig:pulseshape}(a), as these are the two events closest to the peak of the pulse. The ratio of the dominance parameter values for $p_3$ and $p_4$ is directly proportional to the difference in width of the classically allowed regions for these events.

However, the partial momentum distributions $M^{(1)}(p_{1\parallel})$ for the three events, displayed in the bottom row of Fig.~\ref{fig:dpelectron1},  show that, in reality, the contributions from $p_4$ prevail in all channels. This agrees with Fig.~\ref{fig:pulseshape}(a), which shows that the local extremum associated with $p_{4}$ leads to the largest absolute value of the electric field $E(t)$ and intuitively translates into a large ionization probability for the first electron. However, $\mathrm{Im}[t'']$ is very similar for both $p_3$ and $p_4$, and this leads to competing dominance parameters. On the other hand, the partial distributions illustrate the collapse of the classically allowed region for $p_5$ accurately. This region is present in  Fig.~\ref{fig:dpelectron1}(b), which leads to a small, but visible peak in Fig.~\ref{fig:dpelectron1}(c) associated with $p_5$. In Fig.~\ref{fig:dpelectron1}(i) the peak is suppressed as the CAR collapses [see Fig.~\ref{fig:dpelectron1}(h)]. Furthermore, the peaks of the distributions are roughly centered at the most probable momenta, in agreement with \cite{Faria2012}.

The results discussed above indicate that the dominance parameter is an oversimplification of the actual contributions. However, it provides a good indication of which events to exclude in the final momentum distributions. 
When the dominance parameter is less than $0.4$, we can neglect the contribution of that event. However, when the value is greater than $0.9$, it is difficult to ascertain which pair is most dominant.

\begin{figure}
    \centering
\includegraphics[width=\columnwidth]{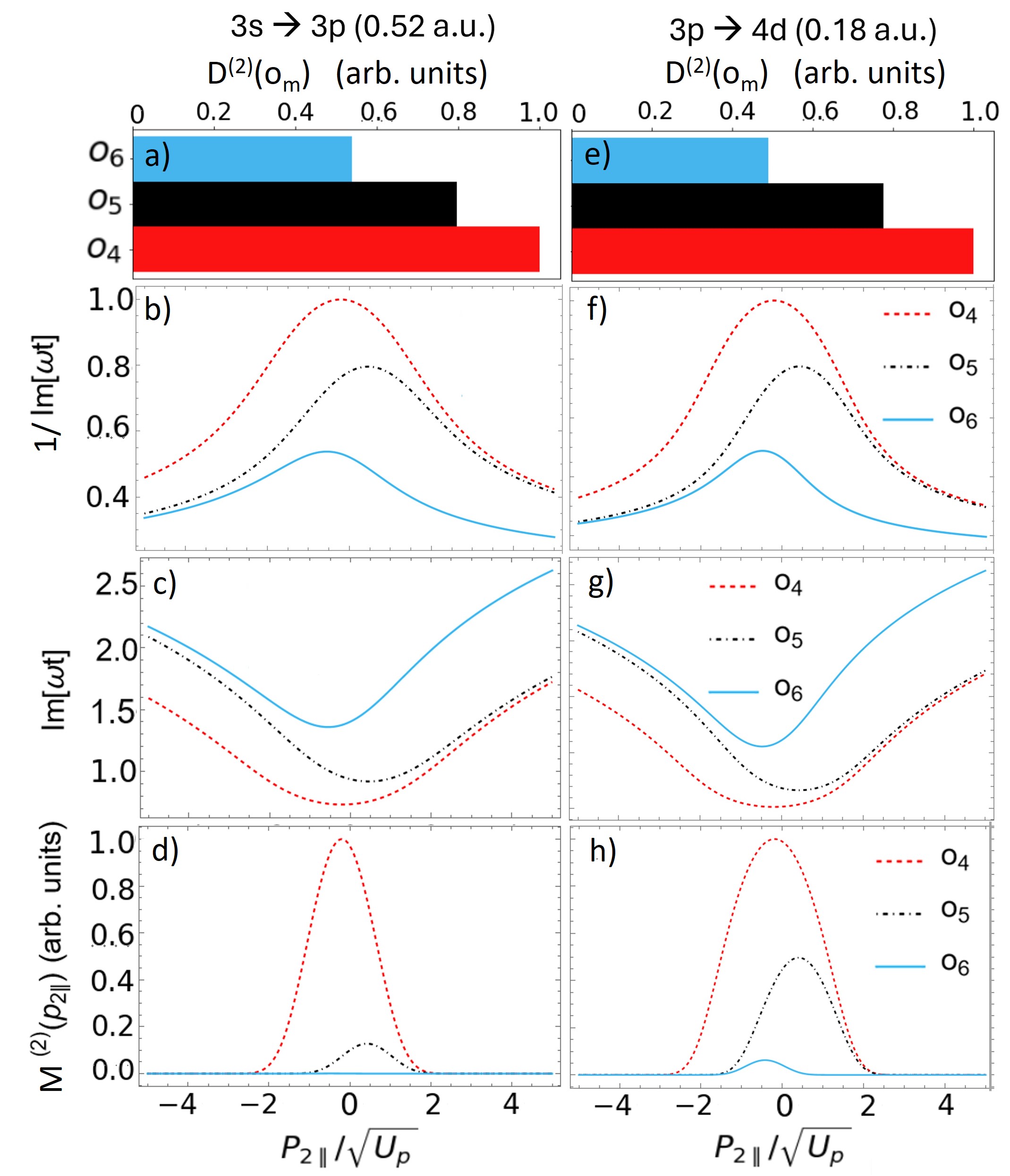}
    \caption{Estimates of the dominance parameter $\mathcal{D}^{(2)}(o_j)$ for the second electron using the minimum imaginary ionization time $t$ (top row), the dominance parameter as a function of the parallel momentum of the second electron (second row),
    the imaginary part of the ionization time calculated for $p_{2\perp}=0$ (third row) and the partial momentum distribution as a function of the parallel momentum without prefactors (bottom row) for three events in the center of a pulse with CEP $\phi=65$, for the $3p\rightarrow3p$ (panels (a)-(d)) and $3p\rightarrow4p$ (panels (e)-(h)) transitions. The values of the dominance parameters and the momentum map have been normalized to the most dominant event for each channel.}
    \label{fig:dpelectron2}
\end{figure}

Next, we will consider the dominance parameter for the second electron. Estimates of this parameter are shown in Fig.~\ref{fig:dpelectron2} for the $3s\rightarrow3p$ [panel (a)] and $3p\rightarrow4d$ [panel (e)] transitions. In both cases, we see that $o_6$ is the least dominant orbit, followed by $o_5$ then $o_4$. Physically, this agrees with Fig.~\ref{fig:pulseshape}(a) which shows that $o_6$ lies towards the edge of the pulse whilst $o_4$ is at the center. It is noteworthy that the ratio between $o_4$ and $o_5$ values is larger than for the corresponding events for the first electron, with values for $o_5$ around 0.8. This implies that these events are not competing, and we can state that $o_4$ has a higher tunneling probability than $o_5$. This was not the case for $p_3$ and $p_4$. Because we are taking the minimum imaginary time $t$ to calculate the dominance parameter, we find that the plots look almost identical for both excitation pathways, despite the difference in their binding energies. 

Nonetheless, upon closer inspection of how $1/\mathrm{Im}[t]$ normalized concerning the maximum value varies with $p_{2\parallel}$ for each of these orbits [Fig.~\ref{fig:dpelectron2}(b)], we can see that, for $o_4$, it is centered and almost completely symmetric about the origin $p_{2\parallel}=0$. This is expected as $o_4$ is located near the center of the pulse. However, for $o_5$ and $o_6$, the maxima deviate from the origin, explained by the asymmetry of the pulse. More importantly, we note that the gradient is not the same on either side of the maxima. In the negative (positive) momentum region, the normalized values of $1/\mathrm{Im}[t]$ approach each other for $o_5$ and $o_6$ ($o_4$ and $o_5$), with a large deviation in the positive (negative) region. This implies that the dominance of orbits is momentum dependent and a single value, as in Figs.~\ref{fig:dpelectron2}(a) and (e), is not enough to capture the entire physics. Therefore the dominance parameter must be used as an indicator of which events to exclude from the momentum map rather than a way to find the location of the brightest primary maxima.  

The asymmetry effect is much stronger for the $3p\rightarrow4d$ pathway [Fig.~\ref{fig:dpelectron2}(f)] than for $3s\rightarrow3p$ [Fig.~\ref{fig:dpelectron2}(b)], because the second electron tunnels from a much more loosely bound state, leading to a steeper gradient. Thus, the imaginary parts of the tunneling times will follow each other more closely once they move away from the origin [see Fig.~\ref{fig:dpelectron2}(g) in comparison to (c)]. This broadens the dominant partial momentum distribution $M^{(2)}(p_{2\parallel})$ from $o_4$, and enhances the remaining contributions, as expected from a narrower potential barrier. Interestingly, practically identical dominance parameters lead to different distributions. Regardless, the estimates arising from the dominance parameter give a clear hierarchy of which events contribute the most.

\begin{figure*}
    \centering
\includegraphics[width=\textwidth]{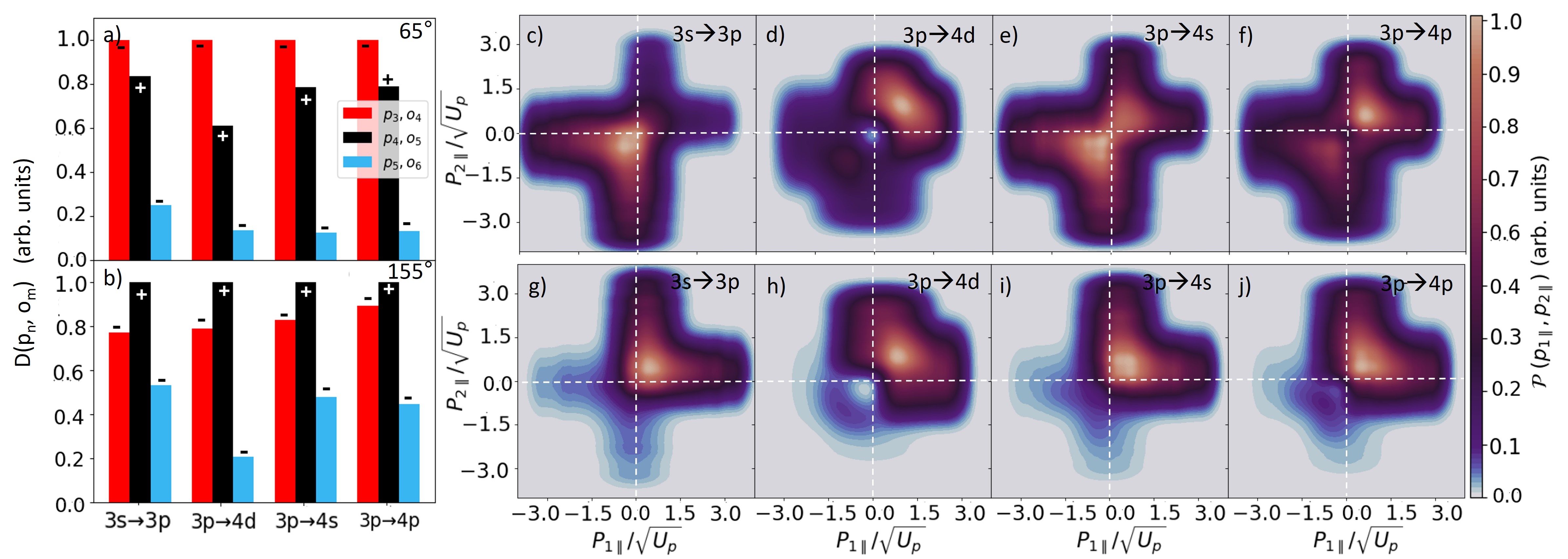}
    \caption{Estimates of the total dominance parameter $\mathcal{D}(\varepsilon_1, \varepsilon_2)$  for three events in the center of a pulse with CEP $\phi_1=65^{\circ}$ [panel (a)] and $\phi_1=155^{\circ}$ [panel (b)] for the $3p\rightarrow4p$, $3p\rightarrow4d$, $3p\rightarrow4s$ and $3s\rightarrow3p$ transitions. For each pair $p_3$, $p_4$ and $p_5$, only the ionization event immediately after has been taken. The incoherent-incoherent momentum maps without prefactors have been calculated for each channel and are displayed in panels (c)-(f) for $\phi_1=65^{\circ}$ and panels (g)-(j) for $\phi_1=155^{\circ}$. The axes are indicated with white dashed lines. The values of the dominance parameters and the momentum map have been normalized with respect to the most dominant event for each channel.}
    \label{fig:dptotal}
\end{figure*}

In Fig.~\ref{fig:dptotal}, we inspect the combined estimates for both electrons (note that we consider degenerate energies $E_{2g}$ for $3s$, $3p$ here). Figs.~\ref{fig:dptotal}(a) and (b) show the total dominance parameter for the channels discussed previously, for CEP $\phi=65^{\circ}$ and $\phi=155^{\circ}$, respectively. For $\phi=65^{\circ}$, according to Fig.~\ref{fig:dptotal}(a) the $p_3 o_4$ event is prevalent, followed by $p_4 o_5$ and $p_5 o_6$ for all channels. This happens because, for $p_{3} o_{4}$, the second electron is freed near a very prominent field maximum [see red rectangle in Fig.~\ref{fig:pulseshape}(a)] bringing up the dominance of this event significantly. For $p_{5} o_{6}$, the ionization time for the second electron is too close to the end of the pulse for it to be as dominant as the other two events, decreasing this value. The dominance of this event is larger for $3s\rightarrow3p$, as the collapse of the CAR suppresses this event for the other excitation pathways. The second electron brings down the value for the $p_4 o_5$ event for all channels too. However, there are also differences due to the excited states' binding energies. The dominance of $p_4 o_5$ for the $3p\rightarrow4d$ pathway is smaller than that of the other channels, due to a smaller CAR for this event.

When the CEP is increased to $\phi_1=155^{\circ}$ [Fig.~\ref{fig:dptotal}(b)], the dominance parameter indicates that, for most excitation channels, $p_{4} o_{5}$ prevails, followed by $p_{3} o_{4}$. Furthermore, the values obtained for the least dominant event $p_5 o_6$ are higher than for $\phi_1=65^{\circ}$. Physically, this may be understood by inspecting Fig.~\ref{fig:pulseshape}(b), which shows that  $p_{4} o_{5}$ has moved towards the center of the pulse, with an increase in the field amplitude for the second electron's ionization time (see black rectangle therein). Although $p_3$ has moved further to the left edge of the pulse, $o_4$ is still towards the center, bringing up $\mathcal{D}(p_3,o_4)$. 
Meanwhile, $p_5 o_6$ has a larger field amplitude for both electrons' ionization times. For this phase, we also have a classically allowed region for the $p_5 o_6$ event for all channels except $3p\rightarrow4d$ (which is the most loosely bound), explaining the increase in the relative size of the $p_5 o_6$ dominance parameter. 

To assess the validity of the dominance parameter estimates, we look at the correlated two-electron momentum distributions without any prefactors [Figs.~\ref{fig:dptotal}(c)-(f) and (g)-(j)]. This will give a good idea of the key variables used to construct this parameter without introducing further biases, leading to distributions whose shapes are roughly described by the constraints in Fig.~\ref{fig:schematic1}. For $\phi_1=65^{\circ}$, an overall feature is that the agreement is better for more deeply bound excited states. Concretely, the distribution for the $3s\rightarrow3p$ pathway, shown in Fig.~\ref{fig:dptotal}(c), agrees with the prediction of the dominance parameter. The distribution is brightest in the lower left quadrant, associated with contributions from $p_3 o_4$ and $p_5 o_6$. The strong tail in the positive momentum region comes from the $p_4 o_5$ event. 
Similarly, the $3p\rightarrow4s$ transition populates both negative and positive momentum regions [see the probability density displayed in Fig.~\ref{fig:dptotal}(e)]. Since the CAR collapses for this region, the loss of relevance of $p_5$ and $o_5$ becoming more prominent leads to distributions almost equally occupying the four quadrants of $p_{1 \parallel}, p_{2 \parallel}$ plane, although they are still slightly brighter in the lower left quadrant. 

For the $3p\rightarrow4p$ transition [Fig.~\ref{fig:dptotal}(f)], we see a similar distribution. However, now the brightest intensity has shifted to the positive momentum region despite $p_3 o_4$ predicted to be the most dominant. This shift is even more pronounced for $3p\rightarrow3d$ distribution, depicted in Fig.~\ref{fig:dptotal}(d), for which the intensity is more strongly located in the top right quadrant, despite the strong tail in the negative momentum region. These shifts can be accounted for by the gradient arising from the $o_5$ event not being taken into account by the dominance parameter.  
As discussed previously this effect is most strongly observed for smaller binding energies. Further evidence of a strongly asymmetric gradient is the fact that the maxima of the move further and further away from the axes $p_{n\parallel}=0$ as the binding energy decreases. 
From these distributions, a main limitation of the dominance parameter becomes evident: a single value is not sufficient to explain the dominance of events in a few-cycle pulse, given that the variables used to construct it are strongly dependent on the momentum region. For instance, the dominance parameter $\mathcal{D}^{(2)}(o_n)$ for the second electron considers only the minimum $\mathrm{Im}[t]$, which, for the SFA, is around $p_{2\parallel}=0$, but this quantity varies drastically away from this value.  Furthermore, the threshold for when orbits become `competing' or large enough to contribute is fairly arbitrary for the combined dominance parameter. When the value is less than 0.2, we can safely omit the pair from the distribution. If the value is greater than 0.6, the events are competing but due to how sensitive the parameter is with respect to the estimates of the width of the CAR, the parameter is prone to over or under-estimation. 

For $\phi_1=155^{\circ}$ [bottom panels in Fig.~\ref{fig:dptotal}], the top-right quadrant of the momentum plane is occupied for all channels arising from the dominance of the $p_4 o_5$ event. In all cases, we see a tail in the negative momentum region arising from the $p_3 o_4$ event. As the binding energy increases, this tail becomes longer and more localized along the axes, so that the entire distribution becomes more cross-shaped. However, because, for a few-cycle pulse,  the gradient of $\mathrm{Im}[t]$ is not symmetric around its minima, as it would be for a monochromatic field, there will never be a perfect symmetry around the axes. With decreasing binding energy, the gradients of $\mathrm{Im}[t]$ as functions of $p_{2\parallel}$ become steeper and we see the maxima of the distribution deviate away from the axes. It should be noted, however, that the parameter does not consider, at least explicitly, the brightness of the distributions. 

\subsection{Prefactor effects and mapping}
\label{sec:prefmapping}

 The shape of the momentum distributions in Fig.~\ref{fig:dptotal}(c)-(j) is altered by the ionization ($V_{\mathbf{p}_{2}e}$) and excitation ($V_{\mathbf{p}_{1}e,kg}$) prefactors, whose expressions are given in the appendix.  For full derivations see \cite{Shaaran2010,Maxwell2015}. The prefactors introduce additional momentum biases which will modify the length, width and maxima of the correlated distributions. They also bring additional phase shifts to those discussed in Sec.~\ref{sec:interfcondition}.

 The six channels investigated in this study (Table \ref{tab:channels}) involve the excitation of the second electron to states of very different spatial geometry. This geometry is reflected in the prefactor shapes via the radial and angular nodes.  However, these nodes do not directly map onto the two-electron momentum distributions. In this section we perform a more detailed analysis on the mapping of the $V_{\mathbf{p}_{2}e}$ and $V_{\mathbf{p}_{1}e,kg}$ prefactors into the $p_{1\parallel}p_{2\parallel}$ plane to understand how the maxima and nodes translate to the momentum distributions explicitly.

\begin{figure*}
    \centering
\includegraphics[width=\textwidth]{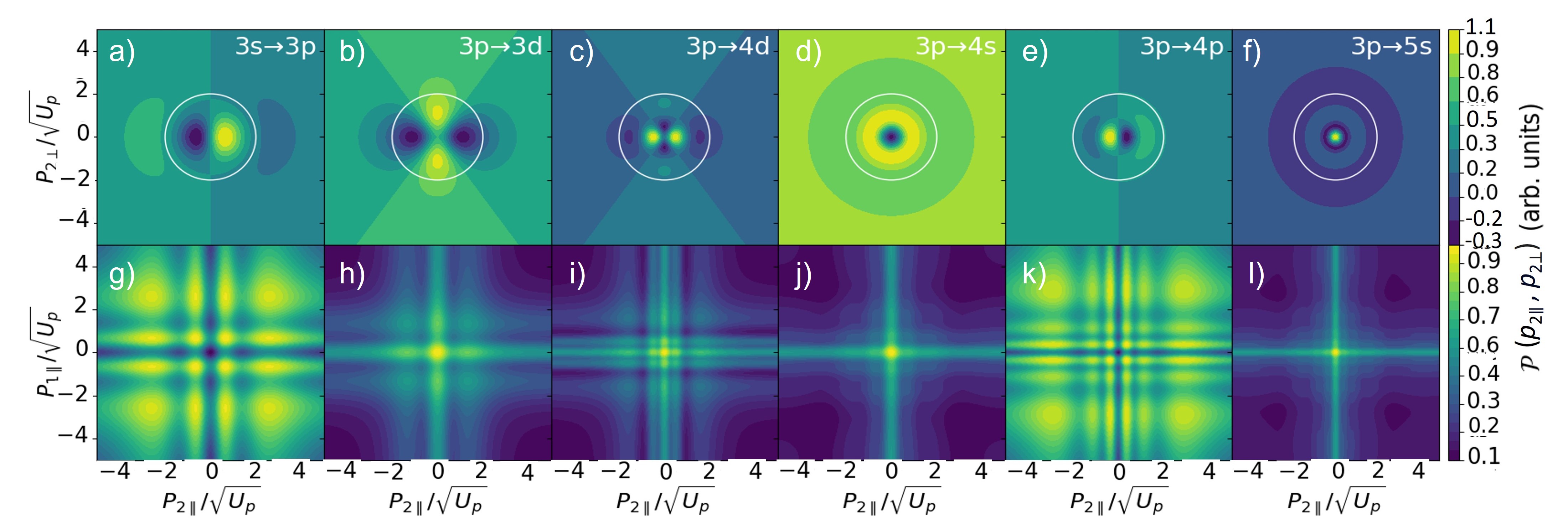}
    \caption{The imaginary part of the ionization prefactor $V_{\mathbf{p}_2e}$ for the second electron, as a function of the momentum components $p_{2\parallel},p_{2\perp}$ (upper panels), and corresponding mapping into the $p_{1\parallel},p_{2\parallel}$ plane. From left to right, the panels refer to channel 1 (panels (a) and (g)), channel 2 (panels (b) and (h)), channel 3 (panels (c) and (i)), channel 4 (panels (d) and (j)), channel 5 (panels (e) and (k)), channel 6 (panels (f) and (l)). The white circle shows the direct ATI cut-off.}
    \label{fig:pf2}
\end{figure*}

 First, we will consider the ionization prefactor $V_{\mathbf{p}_{2}e}$, displayed in Fig.~\ref{fig:pf2}(a)-(f) for the six channels. This prefactor is the most important in determining the shapes of the electron momentum distributions, and has specific nodes. There are $l_e$ angular nodes associated with the spherical harmonics in the wave function $\psi_{n_{e}l_{e}m_{e}}(\mathbf{r}_2)=R_{n_el_e}(r_2)Y_{l_e}^{m_e} (\theta_2, \phi_2)$ of the second electron, where the subscripts $e$ indicate that it belongs to an excited state. The figure shows that, for $s$ excited states [Figs.~\ref{fig:pf2}(d) and (f)], there are no angular nodes for the prefactor, as expected. For $p$ states [Figs.~\ref{fig:pf2}(a) and (e)], there is a single angular node and a strong suppression along the $p_{2\perp}$ axis, which leads to a phase shift with regard to $p_{2\parallel}\rightarrow -p_{2\parallel}$. Finally, for $d$ states [Figs.~\ref{fig:pf2}(b) and (c)], there are two nodes that cross each other at the origin approximately at angles $ \pm 55^{\circ}$. Clear positive and negative regions can be seen from the imaginary part of the prefactor, corresponding to phase shifts of $\pi$. 
 
 Further to angular nodes, there is 
an $n_{e}-l_{e}-1$ number of radial nodes which manifest as circles in the $p_{2\parallel} p_{2\perp}$ plane and a phase jump of $\pi$ if a radial node is crossed. The relevant radial nodes for our problem must be located at energies lower than that of the direct ATI cutoff, indicated by the white circle at $p^2_{2\parallel}+\mathbf{p}^2_{2\perp}=4 U_p$.
Changes in the shapes of the momentum distributions will be caused by nodes within the direct ATI cutoff. The greater the number of nodes, the more phase shifts are present.

The mapping to the $p_{1\parallel}p_{2\parallel}$ plane is performed by integrating  $|V_{p_{2e}}|^2$ over the plane spanned by the momenta perpendicular to the laser-field polarization.  We account for particle exchange leading to four-fold symmetry in the plots. Explicitly, we consider
\begin{equation}
   | V_{p_{2\parallel e}}|^2\hspace*{-0.1cm}=\hspace*{-0.15cm}\iint \hspace*{-0.1cm} d p_{1\perp} d p_{2\perp}\left|p_{1\perp}\right|\left|p_{2 \perp}\right|\left(\left|V_{\mathbf{p}_{2e}}\left(\mathbf{p}_1\right)\right|^2+\left|V_{\mathbf{p}_{2e}}\left(\mathbf{p}_2\right)\right|^2\right).
    \label{eq:prefactormappingintegral}
\end{equation}
A similar mapping can be done for the excitation prefactor, resulting in $| V_{p_{1\parallel e, kg}}|^2$ with the difference that the integrand in \eqref{eq:prefactormappingintegral} is replaced by $V_{\mathbf{p}_{1}e,\mathbf{k}g} (\mathbf{p}_1)$  and its symmetrized version taking $\mathbf{p}_1 \leftrightarrow \mathbf{p}_2$.

The outcomes of these integrals are displayed in the bottom row of Fig.~\ref{fig:pf2}. From the mappings of Figs.~\ref{fig:pf2}(g)-(i), we see that, in agreement with \cite{Maxwell2015}, $s$ states lead to cross-shaped distributions in momentum space, with maxima at the origin and along the axes [Figs.~\ref{fig:pf2}(j) and (l)]. However, their widths will depend on the number and energy positions of the radial nodes.  Figs.~\ref{fig:pf2}(g) and (k) show that, for $p$ states, the strong suppression along $p_{2\parallel}=0$ survives the transverse momentum integration. For these states there are pronounced maxima along all diagonals $p_{1\parallel}=\pm p_{2\parallel}$, between $\pm 2 \sqrt{U_p}$ and $4 \sqrt{U_p}$. In the region of interest (i.e. within the ATI cutoff), there are maxima at around $\sqrt{U_p}$ and $\sqrt{U_p}/2$ for these states respectively.
For $d$ states, the x-shaped nodes intersecting at the origin are mapped to maxima along the axes with the brightest point at the origin similar to the $s$ states. However, unlike the $s$ states, there are additional secondary k maxima parallel to the axes as well. For every orbital angular momentum, any additional radial node within the ATI cut-off leads to narrower distributions. This effect is particularly prominent for highly excited states with small binding energies. For example, Fig.~\ref{fig:pf2}(k) corresponding to a $5s$ transition, shows a narrower cross compared to Fig.~\ref{fig:pf2}(j) with a $4s$ transition.

\begin{figure*}
    \centering
\includegraphics[width=\textwidth]{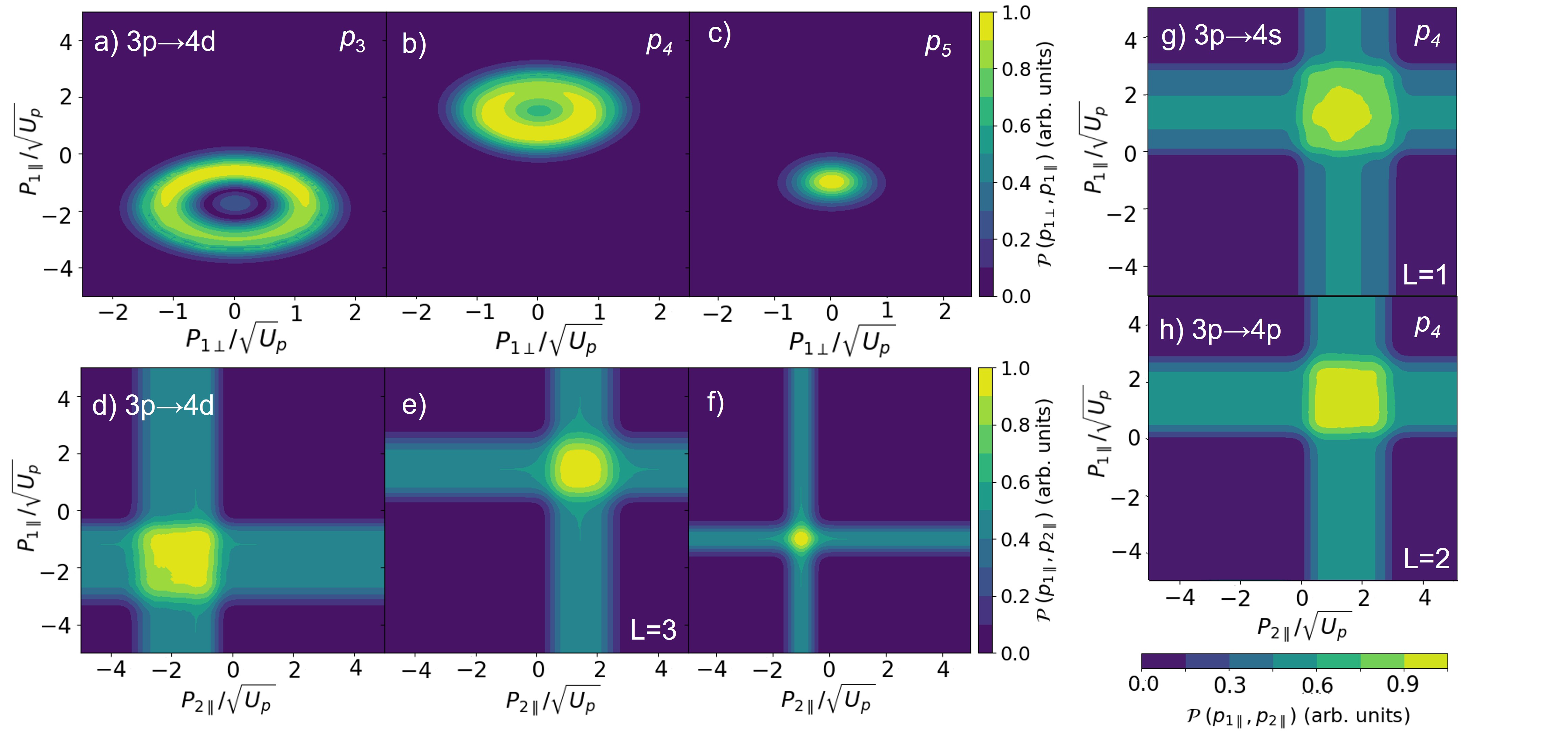}
    \caption{Absolute values of the excitation contribution ($V_{ec}$) accounting for the excitation prefactors $V_{\mathbf{p}_1ekg}$ for the first electron (panels (a)-(c)) and its corresponding mapping into the $p_{1\parallel} p_{2\parallel}$ plane (panels (d)-(h)) for three different pairs $p_3, p_4, p_5$ respectively. The ionization and rescattering times associated with the short orbit were used to compute these plots. Panels (a)-(f) are associated with channel 3, 
    whilst (g) and (h) show mappings for channels 4 and 5 
    with $p_4$ as this is most dominant for $\phi_1=65^{\circ}$. The mapping axes have been transposed relative to the other figures to make a clearer connection with the location of the events in the pulse (upper or lower half). The orbital angular momenta L indicated in panels (e), (g), and (h) are obtained from the sums over $l_g$ and $l_e$ given in Eq.~\eqref{eq:angmom} in the first Appendix.}
\label{fig:excitationcontribution}
\end{figure*}    

Next, we consider the excitation prefactor. From the explicit expression of $V_{\mathbf{p}_{1}ekg}$ given in the appendix, we can see that the shape of this prefactor is not trivial given that $\mathbf{q }= \mathbf{p}_1 - \mathbf{k}$ where $\mathbf{k}$ has an implicit dependence on the times $t'$ and $t''$ [see Eq.~\eqref{eq:sp2}]. This makes this prefactor orbit dependent, giving very different shapes for the long and short orbits. These pairs of orbits are incorporated collectively in a uniform saddle-point approximation, as they nearly coalesce at the boundaries of the classically allowed region \cite{Faria2002}. The nodes of this prefactor are mostly washed out, even before integration over transverse momentum is performed \cite{Shaaran2010,Maxwell2015}.  However, $V_{\mathbf{p}_{1}ekg}$ moves the centers of the electron momentum distributions, and care must be taken if the driving field is a pulse. The contribution of this prefactor to the overall momentum distribution can be better understood by looking at how it changes the transition amplitude, when incorporated in the uniform approximation. This contribution (referred to as `excitation contribution' ($V_{ec}$) henceforth) is obtained by subtracting the amplitude calculated using the coefficient with the prefactor from the amplitude calculated using the coefficient without the prefactor.  The mapping is done as in Eq.~\eqref{eq:prefactormappingintegral}. 

 In $V_{ce}$, any radial or angular nodes from the excitation prefactor are washed out with insignificant differences between different excitation pathways. Instead, the prefactor serves to cause a slight shift of the transition amplitude up or down, depending on the location of the event in the pulse. Fig.~\ref{fig:excitationcontribution}(a)-(c) shows $V_{ec}$ for three different pairs, for the $3p\rightarrow4d$ pathway.
The locations of these rings are pulse-dependent, with $p_3$ and $p_5$ located in the region with negative $p_{2\parallel}$ and $p_4$ in the positive $p_{2\parallel}$ region. This is implied by Fig.~\ref{fig:pulseshape}(a) for this phase. The size of the rings is also dependent on the width of the CAR for each pair and agrees with Fig.~\ref{fig:dpelectron1}(e). The mappings for each pair in Figs.~\ref{fig:excitationcontribution}(d)-(f) are therefore also pulse-dependent, with sizes proportional to the width of the CAR. This can be explicitly seen in Figs.~\ref{fig:excitationcontribution}(g) and (h), which show the mapping of the excitation contribution with $p_4$, the most dominant pair, for two more excitation channels, whose energy gaps are smaller than that considered in  Fig.~\ref{fig:excitationcontribution}(e). The shapes are almost identical, but occupy broader regions in momentum space.   This indicates the excitation prefactor has little impact on the shapes of the final momentum distributions, other than through narrowing them - the amount it narrows is proportional to the binding energy.  

\section{Photoelectron momentum distributions}
\label{sec:PMDpulses}
 
In this section, we will focus on the correlated electron momentum distributions for each of the channels in Table \ref{tab:channels} and the few-cycle pulse given by Eq.~(\ref{eq:Apulse}), bringing together the dominant events, the different types of intra-channel interference and the phase shifts and momentum biases associated with prefactors.  We will start by discussing the main features [Sec.~\ref{sec:overall}], and then delve deeper into quantum interference [Secs.~\ref{sec:PMDinterfexch}, \ref{sec:PMDinterfev}
and \ref{sec:PMDinterfexchev}]. Guided by the results from Sec.~\ref{sec:dominance}, we take the three most dominant (combined) events for both CEP $65^{\circ}$ and CEPs $155^{\circ}$. Although the event $p_5 o_6$ does not have a CAR for most channels computed with CEP $65^{\circ}$, it has been included in this investigation for completeness and contributes very little to the total momentum distribution.

\subsection{Prefactors, dominance and quantum interference}
\label{sec:overall}

\begin{figure*}
    \centering
   \includegraphics[width=\textwidth]{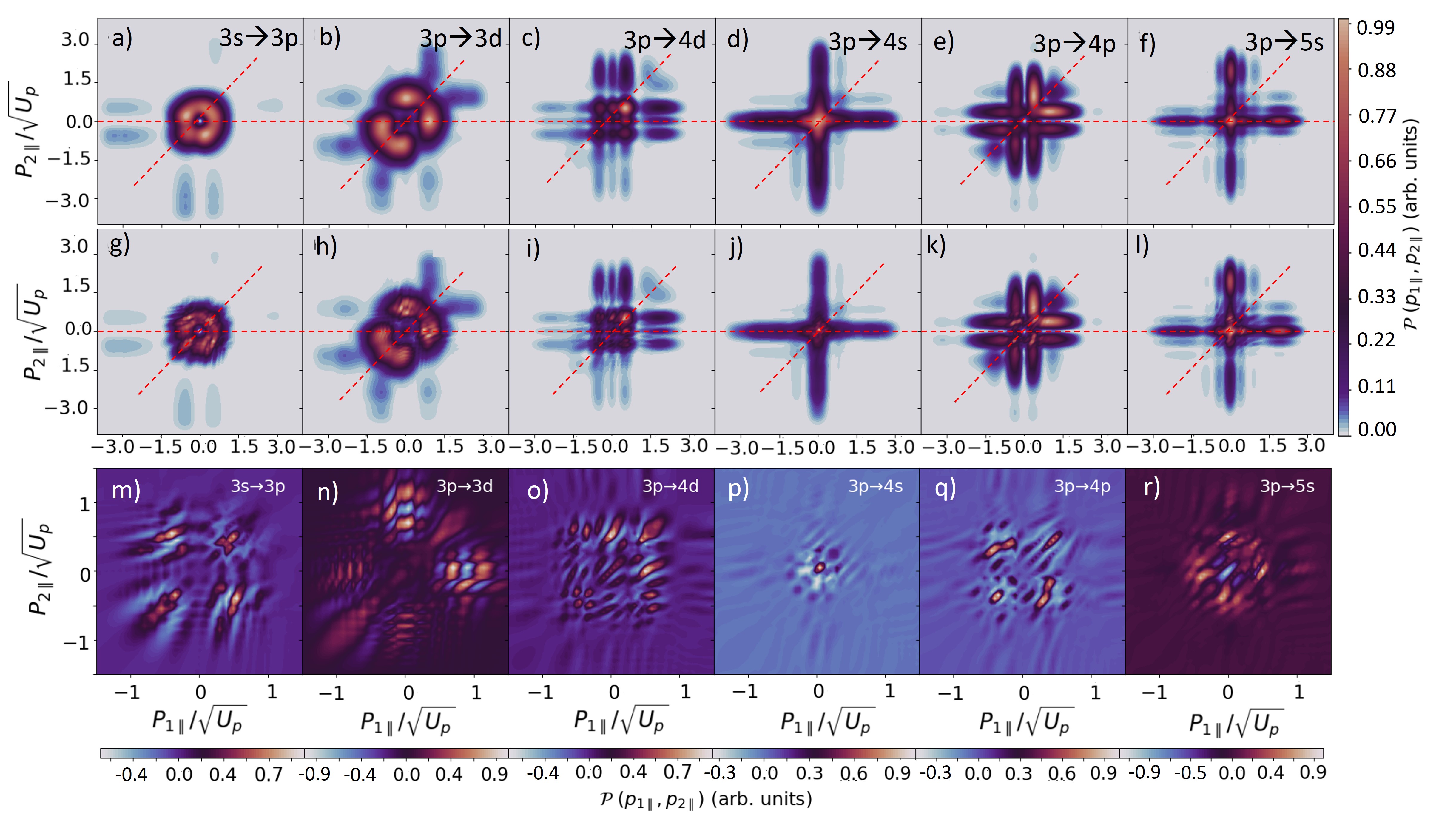}
    \caption{Correlated two-electron momentum distributions as functions of the momentum components $p_{1\parallel}$,$p_{2\parallel}$ parallel to the driving field polarization, calculated for a single excitation channel using a $\sin^2$ pulse of 4.3 cycles, intensity $I=1.5 \times 10^{14}\mathrm{W/cm}^2$, angular frequency $\omega=0.057$ a.u. and CEP $\phi_1=65^{\circ}$. The first and second rows correspond to fully incoherent and coherent sums of amplitudes as given in Eqs~\eqref{eq:1ii}, and \eqref{eq:1coherent}, respectively. The third row shows the difference between the maps in the first two rows, showing the interference arising from summing events and particle exchange terms coherently. All six RESI channels described in Table \ref{tab:channels} are shown from panels (a)-(f), (g)-(l) and (m)-(r). The signal in each panel has been normalized with regard to its maximum. The diagonals $p_{1\parallel} = p_{2\parallel}$ and the $p_{1\parallel}$ axes are indicated with red dashed lines in each panel of the upper and middle rows.}
    \label{fig:cep65singlechannel}
\end{figure*}

\begin{figure*}
    \centering
   \includegraphics[width=\textwidth]{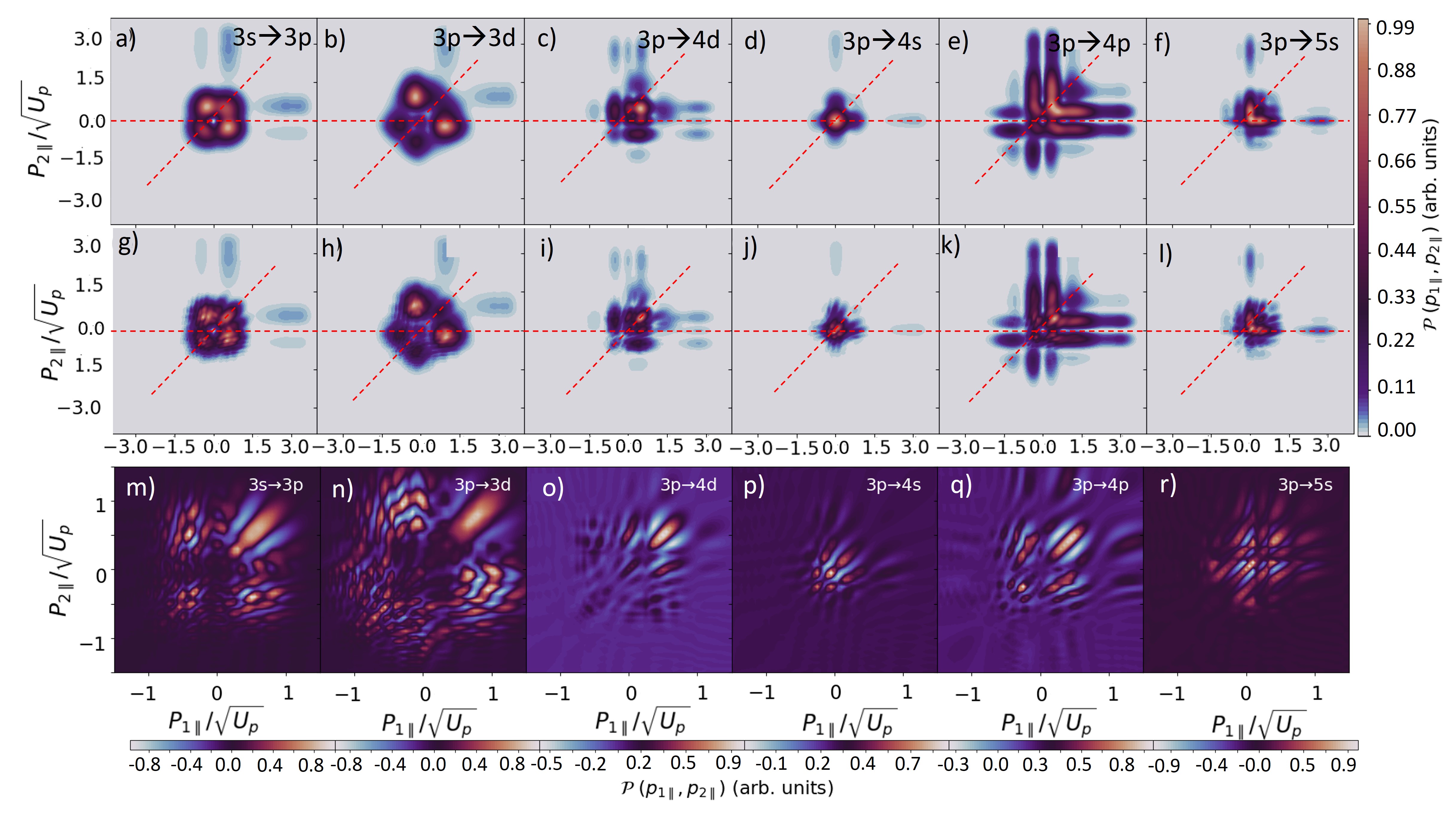}
    \caption{Correlated two-electron momentum distributions as functions of the momentum components $p_{1\parallel}$,$p_{2\parallel}$ parallel to the driving field polarization, calculated for a single excitation channel for the same pulse parameters employed in \ref{fig:cep65singlechannel} and $\phi_1=155^{\circ}$. The first and second rows correspond to fully incoherent and coherent sums of amplitudes as given in Eqs~\eqref{eq:1ii}, and \eqref{eq:1coherent}, respectively. The third row shows the difference between the first two rows, showing the interference arising from summing events and particle exchange terms coherently. All six RESI channels described in Table \ref{tab:channels} are shown from panels (a)-(f) and (g)-(l). The signal in each panel has been normalized with regard to its maximum. The diagonals $p_{1\parallel} = p_{2\parallel}$ and the $p_{1\parallel}$ axes are indicated with red dashed lines in each panel of the upper and middle rows. }
    \label{fig:cep155singlechannel}
\end{figure*}

Figs.~\ref{fig:cep65singlechannel} and \ref{fig:cep155singlechannel}  exhibit the fully incoherent [Eq.~\eqref{eq:1ii}] and coherent [Eq.~\eqref{eq:1coherent}] single-channel sums for CEPs $65^{\circ}$ and $155^{\circ}$, respectively. 
The incoherent sums, plotted in panels (a)-(f) of both figures, exhibit the symmetry about the diagonal $p_{1\parallel}=p_{2\parallel}$, although the four-fold symmetry associated with monochromatic fields is absent. Furthermore, they display the expected shapes, depending on the second electron's excited states. Distributions computed for $s$ excited states are localized at the axes $p_{n\parallel}=0$ [see Figs.~\ref{fig:cep65singlechannel} and \ref{fig:cep155singlechannel}, panels (d) and (f)], distributions involving $p$ excited states exhibit a strong suppression at the axes and maxima around the diagonals $p_{1\parallel}=\pm p_{2\parallel}$ [see Figs.~\ref{fig:cep65singlechannel} and \ref{fig:cep155singlechannel}, panels (a) and (e)], and distributions for which the second electron leaves from a $d$ state, shown in Figs.~\ref{fig:cep65singlechannel} and \ref{fig:cep155singlechannel}, panels (b) and (c), exhibit maxima around the axes and the diagonals. 

Moreover, the distributions including the excitation and ionization prefactors are much narrower than those in Sec.~\ref{sec:dominance}, where only the momentum space constraints were taken. This is due to the $n_{e}-l_{e}-1$  radial nodes and $l_{e}$ angular nodes in $V_{\mathbf{p}_2,e}$, which will be mapped to vertical or horizontal nodes in $|V_{p_{2\parallel},e}|^2$ (see Fig.~\ref{fig:pf2}). 
Many nodes in the momentum regions of interest will split the momentum distributions, or translate into narrower electron momentum distributions. For instance, according to Fig.~\ref{fig:pf2}, narrow distributions are expected for channels 3,  4, 5 and 6. This is very visible in Figs.~\ref{fig:cep65singlechannel}(c) and (e), for which the distributions split into several maxima, and in Figs.~\ref{fig:cep65singlechannel}(d) and (f), for which the distributions are very narrow. Due to a higher number of nodes, the distribution in  Fig.~\ref{fig:cep65singlechannel}(f) has more substructure and is narrower than that in Fig.~\ref{fig:cep65singlechannel}(d).

The dominance of specific events within a pulse potentially dictates the length of the distributions, which is associated with the CAR, and the brightest peaks. 
For example, for the $3s\rightarrow3p$ transition and CEP $65^{\circ}$ [Fig.~\ref{fig:cep65singlechannel}(b)], the combination of a classically allowed region for $p_{5}$ [Fig.~\ref{fig:dpelectron1}(b)] and a high ionization probability for the $o_4$ event of the second electron [Fig.~ \ref{fig:dpelectron2}(a)] causes the primary maxima of the correlated two-electron distribution to be stronger in negative momentum regions. In contrast, the distribution resulting from the  $3p\rightarrow4p$ transition  [Figs.~\ref{fig:cep65singlechannel}(d)] shows much stronger maxima in the top right quadrant (the positive momentum region) because the CAR for the pair $p_5$ has collapsed and the $p_{4}o_{5}$ event dominates. The $p_{3} o_{4}$ transition, by itself, is not strong enough to cause the spread of the maxima in the lower half quadrant, as understood by the PMD in Fig.~\ref{fig:dpelectron1}(l). Competing events, together with the prefactor, may give the impression that a specific distribution is fourfold symmetric. This is the case, for instance, in Fig.~\ref{fig:cep65singlechannel}(d), for the $3p \rightarrow 4s$ transition. 

Varying the CEP changes the spread of the signal quite drastically for all channels, and can change the quadrant in which the maxima is localized. This is directly linked to the change in dominance of events as discussed previously. The overall locations of the maxima and nodes in the distributions remain the same as with $\phi_1=65^{\circ}$. However, the shapes for almost all the distributions in Fig.~\ref{fig:cep155singlechannel}, most notably panels (d) and (f) (the $s$ states) become much shorter; we no longer see the long extended distributions along the axes. Instead, they are concentrated mostly at the origin with some secondary maxima along the positive axes. Similarly, in Figs.~\ref{fig:cep155singlechannel}(c) and (e), we have less of a tail in the lower-left quadrant, and the majority of the distribution, though narrower, is in the positive momentum region. This is reflective of the $p_4 o_5$ event being much stronger for this phase, compared to their $\phi_1=65^{\circ}$ counterparts where the $p_3 o_4$ and $p_4 o_5$ events are competing.

Panels (g)-(l) of Figs.~\ref{fig:cep65singlechannel} and \ref{fig:cep155singlechannel} show coherent sums over symmetrization and events for $\phi_1=65, 155^{\circ}$, given by Eq.~\eqref{eq:1coherent}. This means that, for the dominant events of the pulse considered here, the interference types outlined in Fig.~\ref{fig:interfdiagram} and the phase types specified in Sec.~\ref{sec:interfcondition} are present. Throughout, one can see superimposed interference fringes near the diagonals and the origin. To visualize the fringes more clearly, the difference between the fully coherent and fully incoherent sums is plotted in the third row of these figures. 

Overall, there exists an intricate tapestry of patterns, which, however, share some common features.  In all panels, we can see a bright fringe along the diagonal, surrounded by hyperbolas, and are reflection-symmetric about the diagonal $p_{1\parallel}=p_{2\parallel}$.
These features are particularly clear in the third quadrant of Figs.~\ref{fig:cep65singlechannel} (m), (n), and the first quadrants of Figs.~\ref{fig:cep155singlechannel}(m), (n), (o), and (q). Furthermore, because the half-cycle symmetry is broken, these features are not symmetric about $(p_{1\parallel},p_{2\parallel}) \rightarrow (-p_{1\parallel},-p_{2\parallel}) $ and there is no bright interference fringe along the anti-diagonal $p_{1\parallel}=-p_{2\parallel}$. These are key differences from the single-channel interference observed for monochromatic fields \cite{Maxwell2015,Maxwell2016}.  Other features are phase shifts associated with the prefactor, which can be seen very clearly for excitations to $p$ states [see panels (m) and (q) in Figs.~\ref{fig:cep65singlechannel} and \ref{fig:cep155singlechannel}], and convoluted patterns in the second and fourth quadrant.  

Next, we will have a closer look at the different types of interference present in Figs.~\ref{fig:cep65singlechannel} and \ref{fig:cep155singlechannel}.  According to Sec.~\ref{sec:dominance}, there are at most three relevant events within the pulse, $p_3o_4$, $p_4o_5$ and $p_5o_6$, each of which is roughly shifted by half a cycle from the previous one. Pairwise, they may interfere in fifteen main ways, which are listed in Table.~\ref{tab:phasetab}. This table can be used as a roadmap to understand the subsequent discussions. The left column of Table.~\ref{tab:phasetab} gives the phase differences discussed in Sec.~\ref{sec:interfcondition}. The second column illustrates schematically how these phases can be applied to the specific pulse studied in this work, following the convention used in Fig.~\ref{fig:interfdiagram}. The colors and labelling of events are as in Sec.~\ref{sec:dominance}. The third column details the specific sources of pairwise interference, including electron exchange, and the temporal shift, $\Delta \tau$ which quantifies the relative (approximate) position of interfering events in a pulse. Another analytical device employed to temporally order the events within the pulse is the temporal locations $ \tau_{\mu}$. They are not directly employed in the numerical computations, but serve as a measure of where the times $t, t', t''$ are located for a specific event. For instance, for  $p_3o_4$  we take $\tau_{\mu}$ where $\mu = l,d$ to be zero given that this is the first dominant event in the pulse to be considered. The later event $p_4o_5$ is separated by half a cycle from $p_3o_4$, so that  $t+\pi/\omega, t'+\pi/\omega, t''+\pi/\omega$, where $\tau_{\mu}=\pi/\omega$ and $\mu=r,u$. Similarly for $p_5o_6$, $\tau_{\mu}=2\pi/\omega$ where $\mu=lT,dT$.

There exist three main types of interference.  Phase differences numbered from 1 to 3 are associated with individual events within the pulse, but for which the electron momenta have been exchanged. Because we are only considering one event, the temporal shift $\Delta \tau=0$. 
Next, there are phase differences associated solely with temporal shifts, numbered 4 to 9. For pairs of events that are separated by half a cycle (i.e. $p_3o_4$+$p_4o_5$ and $p_4o_5$+$p_5o_6$), the relative temporal shift associated is, therefore, $\Delta \tau=\pi/\omega$. For the inter-cycle events, the relative temporal shift is $\Delta \tau=2\pi/\omega$. Finally, the last five phase differences (10 to 15) account for a combination of electron exchange and temporal shifts. 

\begin{table*}[t]
\centering
\begin{tabular}{cccc}
\hline \hline
Phases & Schematic Representation & Type & S-E \\
\hline
\parbox[c]{2cm}{\begin{enumerate}
    \item $\alpha_{l, d}$
    \item $\alpha_{r, u}$
    \item $\alpha_{lT, dT}$
\end{enumerate}} & 
        \begin{tabular}{@{}c@{}}
            \includegraphics[height=2cm]{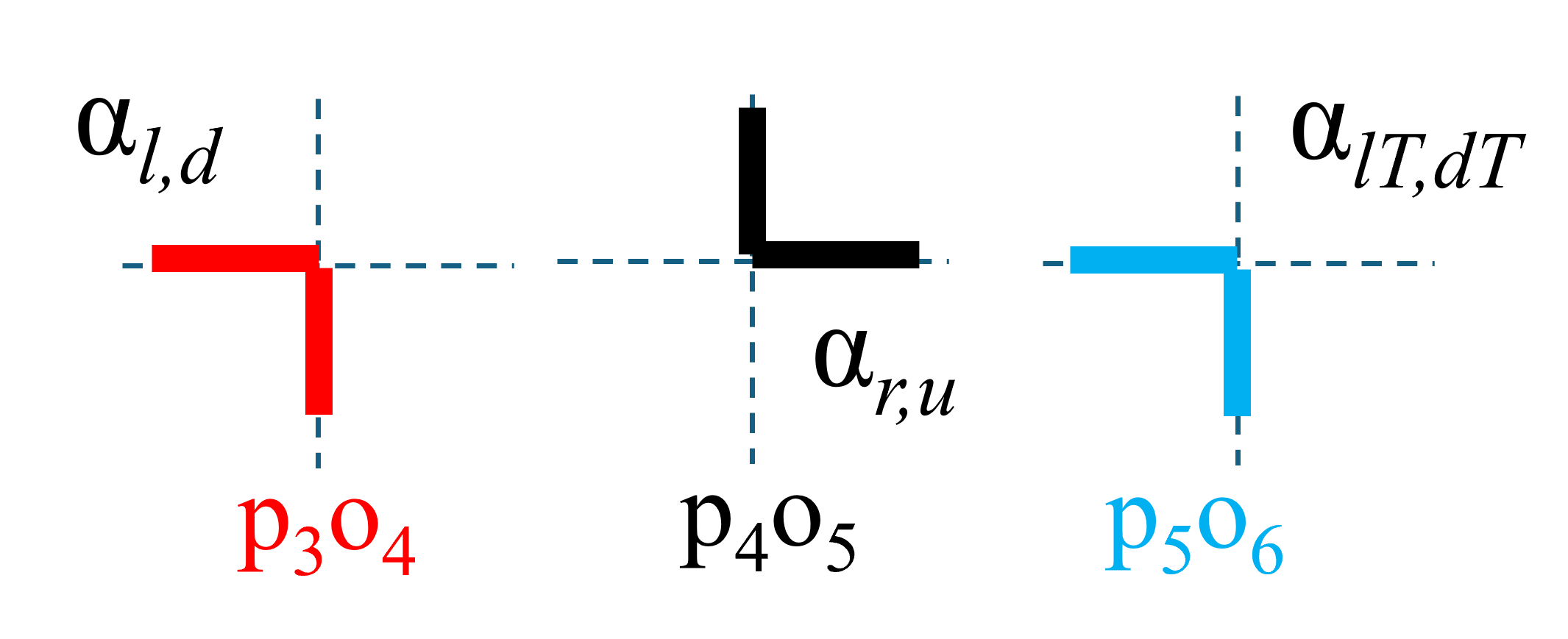} \\
        \end{tabular} & 
\begin{tabular}{@{}c@{}} \\
    Intracycle Exchange \\ \\
    No temporal shift $\Delta\tau = 0$ \\
    \parbox[]{4.5cm}{\begin{enumerate}
    \item $\tau_{l}=0$, $\tau_{d}=0$
    \item $\tau_{r}=\pi/\omega$, $\tau_{u}=\pi/\omega$
    \item $\tau_{lT}=2\pi/\omega$, $\tau_{dT}=2\pi/\omega$
\end{enumerate}} \\
\end{tabular} & 
\begin{tabular}{@{}c@{}}
    ci, cc
\end{tabular} \\
\hline
\parbox[c]{2cm}{\begin{enumerate}\addtocounter{enumi}{3}
\item $\alpha_{l, r}$
\item $\alpha_{d, u}$ \\ 
\vspace{0.25cm}
\item $\alpha_{lT, r}$
\item $\alpha_{dT, u}$ \\
\vspace{0.25cm}
\item $\alpha_{lT, l}$
\item $\alpha_{dT, d}$
\end{enumerate}} & 
        \begin{tabular}{@{}c@{}}
            \includegraphics[height=3.5cm]{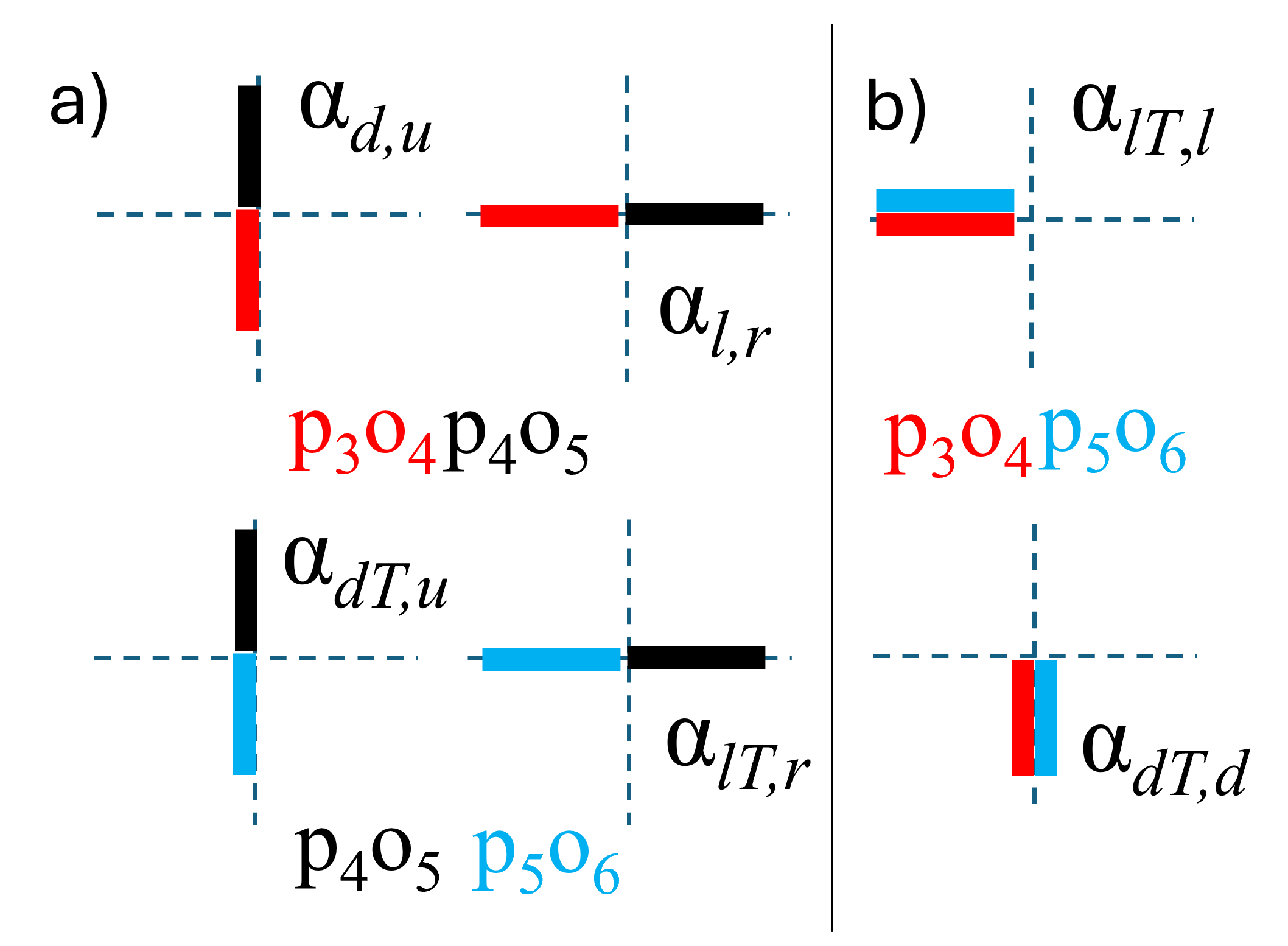} \\
        \end{tabular} & 
\begin{tabular}{@{}c@{}}
\\ No Exchange \\ \\
a) Intracycle shift $\Delta\tau = \pi/\omega$ \\
\parbox[]{4.5cm}{\begin{enumerate}\addtocounter{enumi}{3}
\item $\tau_{l}=0$, $\tau_{r}=\pi/\omega$
\item $\tau_{d}=0$, $\tau_{u}=\pi/\omega$ \vspace{0.25cm}
\item $\tau_{lT}=2\pi/\omega,\tau_{r}=\pi/\omega$ 
\item $\tau_{dT}=2\pi/\omega,\tau_{u}=\pi/\omega$ \\
\end{enumerate}} 
\\
b) Intercycle shift $\Delta\tau = 2\pi/\omega$ \\
\parbox[]{4.5cm}{
\begin{enumerate}\addtocounter{enumi}{7}
\item $\tau_{lT}=2\pi/\omega$, $\tau_{l}=0$ 
\item $\tau_{dT}=2\pi/\omega$, $\tau_{d}=0$ \\
\end{enumerate}} \\
\end{tabular} & 
\begin{tabular}{@{}c@{}}
ic, cc
\end{tabular} \\
\hline
\parbox[c]{2cm}{\begin{enumerate}\addtocounter{enumi}{9}
\item $\alpha_{l, u}$
\item $\alpha_{r, d}$ \\ \vspace{0.25cm}
\item $\alpha_{lT, u}$
\item $\alpha_{dT, r}$ \\ \vspace{0.25cm}
\item $\alpha_{lT, d}$
\item $\alpha_{dT, l}$
\end{enumerate}} & 
        \begin{tabular}{@{}c@{}}
            \includegraphics[height=3.5cm]{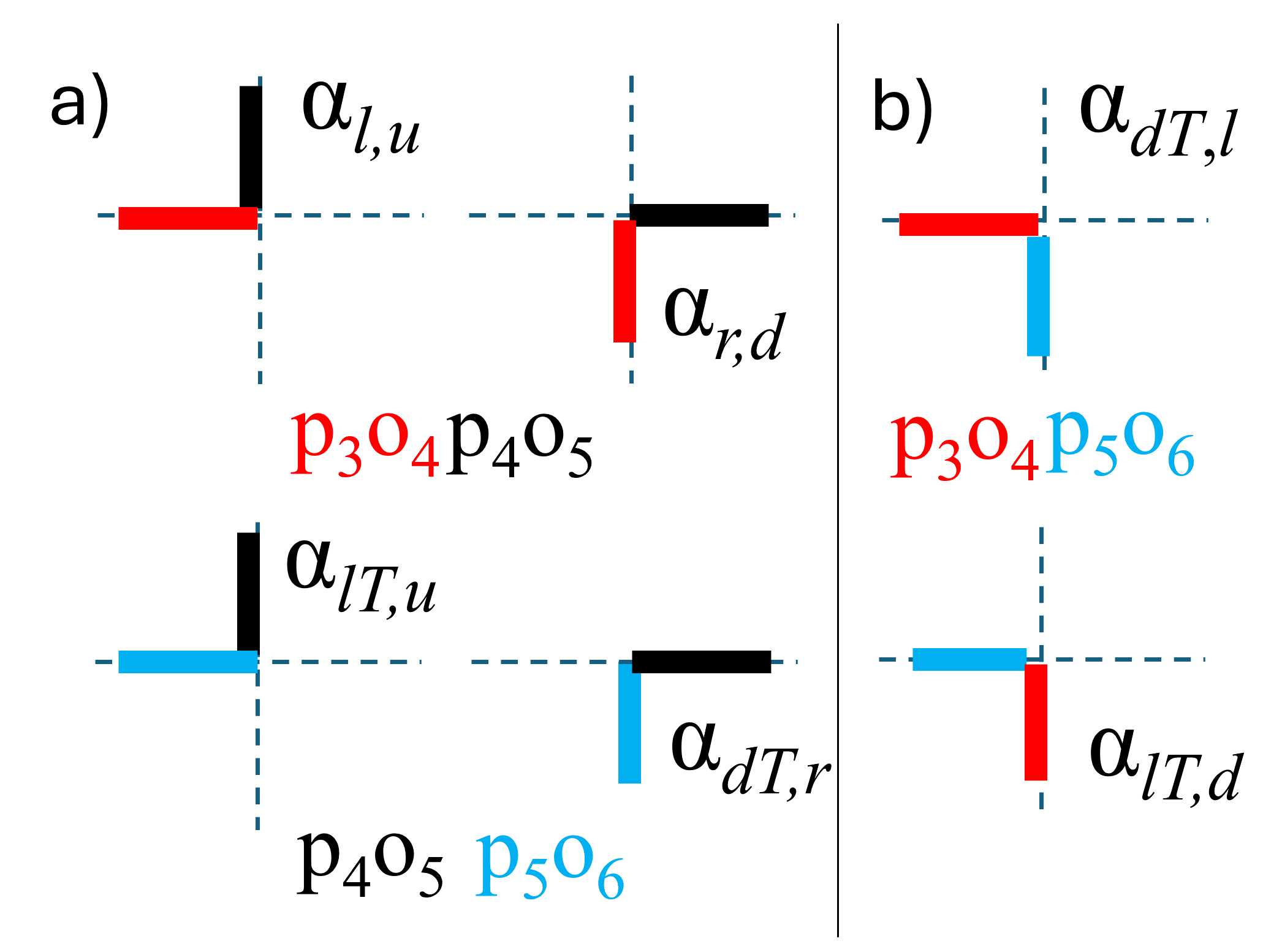} \\
        \end{tabular} & 
        \begin{tabular}{@{}c@{}} \\
a) Intracycle shift $\Delta\tau = \pi/\omega$ \\
\parbox[]{4.5cm}{\begin{enumerate}\addtocounter{enumi}{9}
\item $\tau_{l}=0$, $\tau_{u}=\pi/\omega$
\item $\tau_{d}=0$, $\tau_{r}=\pi/\omega$ \\ \vspace{0.25cm}
\item $\tau_{lT}=2\pi/\omega,\tau_{u}=\pi/\omega$ 
\item $\tau_{dT}=2\pi/\omega,\tau_{r}=\pi/\omega$ \\
\end{enumerate}} \\
b) Intercycle shift $\Delta\tau = 2\pi/\omega$ \\
\parbox[]{4.5cm}{\begin{enumerate}\addtocounter{enumi}{13}
\item $\tau_{lT}=2\pi/\omega$, $\tau_{d}=0$ 
\item $\tau_{dT}=2\pi/\omega$, $\tau_{l}=0$ \\
\end{enumerate}} \\
\end{tabular} & 
\begin{tabular}{@{}c@{}}
cc
\end{tabular} \\       
\hline \hline
\end{tabular}
\caption{All the 15 phase differences present for the few-cycle pulse, associated with just exchange (row 1), temporal shifts (row 2) and a combination of both exchange and shifts (row 3). From left to right, the first column provides the phases for the interfering processes involving the three most relevant events in the pulse given by Eq.~\eqref{eq:Apulse}, $p_3o_4$, $p_4o_5$ and $p_5o_6$. The second column shows a schematic representation of these phases. The colors used in the diagrams and the labeling of the events are the same as in the section discussing the dominant orbits, i.e.,  $p_3o_4$ is represented in red, $p_4o_5$ in black and $p_5o_6$ in blue.  When more than one event is interfering, the two events are placed between the corresponding diagrams. The third column indicates the type of interference i.e. whether it transpires due to particle exchange, temporal shifts of either $\pi/\omega$ or $2\pi/\omega$ or a combination of both. The specific shifts, $\tau_{\mu}$ of $t,t',t''$ for each phase $\mu$ are stated - the shifts associated $p_3 o_4$ ($\tau_l, \tau_d$) are taken to be 0.
The final column indicates when the corresponding type of interference can be observed i.e. with a fully coherent sum of symmetrization or events, or combination of incoherent and coherent symmetrization and events.}
\label{tab:phasetab}
\end{table*}

\subsection{Exchange Interference}
\label{sec:PMDinterfexch}

\begin{figure}[]
    \centering
    \includegraphics[width=\columnwidth]{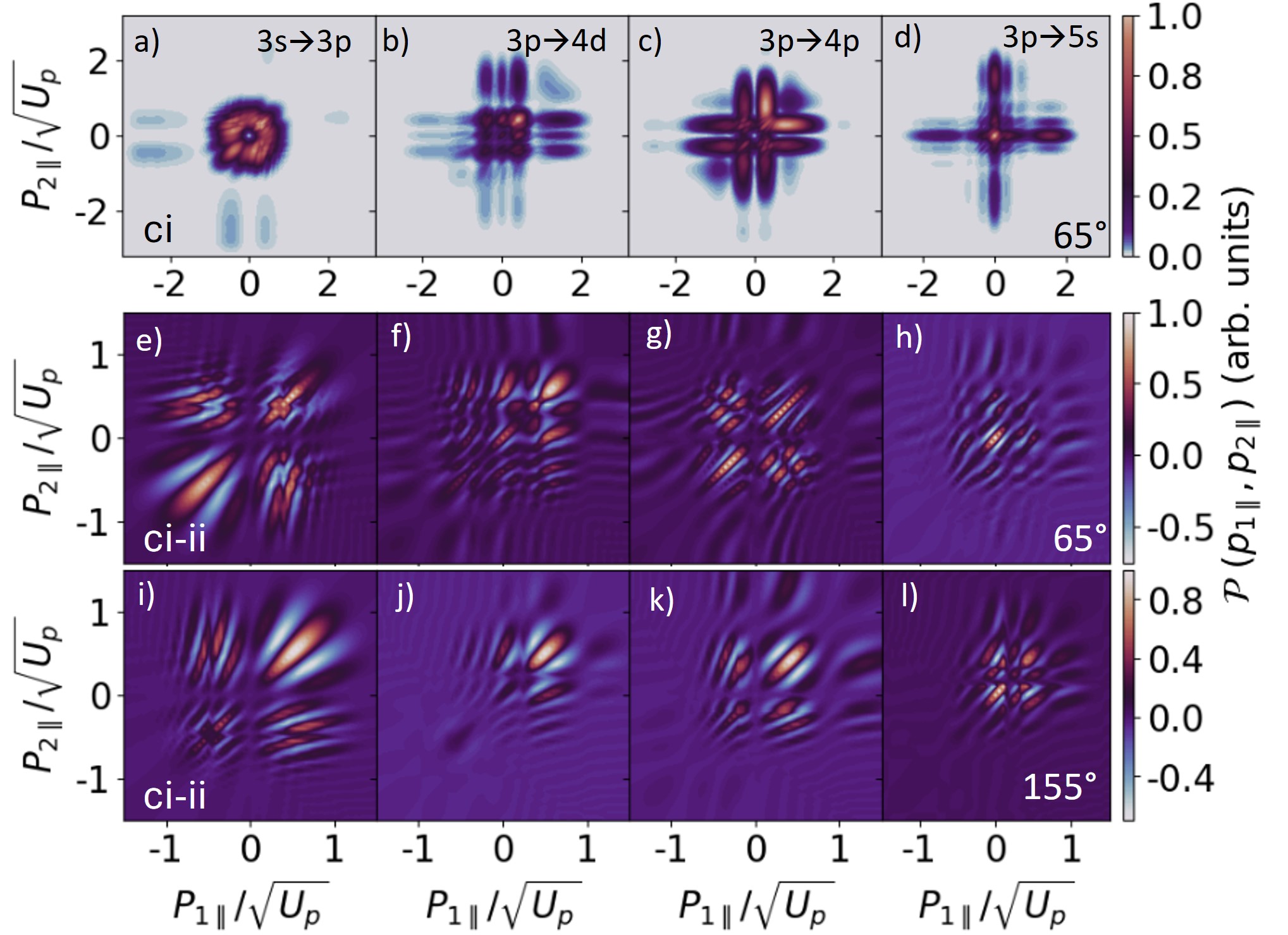}
    \caption{Correlated two-electron momentum distributions as functions of the momentum components $p_{1\parallel}$,$p_{2\parallel}$ parallel to the driving field polarization, calculated for a single excitation channel for the same pulse parameters employed in \ref{fig:cep65singlechannel}. The first row corresponds to amplitudes where events are summed incoherently, and symmetrization is done coherently as given in Eq~\eqref{eq:1ci}, calculated for $\phi_1=65^{\circ}$, and illustrates that the interference fringes are different from those in \ref{fig:cep65singlechannel}. The second and third rows show the difference between the first row and the corresponding fully incoherent sums [see panels (a)-(f) in Figs.\ref{fig:cep65singlechannel}, \ref{fig:cep155singlechannel}], with CEP $65^{\circ}$ and $155^{\circ}$. These rows show the interference arising from summing particle exchange terms coherently and events incoherently. All six RESI channels described in Table \ref{tab:channels} are shown from panels (a)-(f) and (g)-(l). The signal in each panel has been normalized with regard to its maximum. 
    \label{fig:exchangeonlyoverall}}
\end{figure}

 Fig.~\ref{fig:exchangeonlyoverall} shows similar coherent sums and difference plots as in Figs.~\ref{fig:cep65singlechannel} and \ref{fig:cep155singlechannel}, but summing only the exchange processes coherently. This isolates the interference effects arising from the   $\alpha_{\mathbf{p}_1,\mathbf{p}_2}^{(\mathrm{exch})}$ and $\alpha_{\mathbf{p}_1, \mathbf{p}_2}(t, t')$ 
 phase differences in Eqs.~\eqref{eq:alphaexch}, ~\eqref{eq:alphaexch2}. 
 In all plots, we observe superimposed hyperbolas and a bright fringe along the diagonal. Upon inspection of Eq.~\eqref{eq:alphaexch}, field-independent hyperbolae arise when $\alpha_{\mathbf{p}_1,\mathbf{p}_2}^{(\mathrm{exch})} = 2n\pi$
whilst the bright fringe (the ``spine"), which increases in width along the diagonal can be attributed to the field-dependent 
$\alpha_{\mathbf{p}_1, \mathbf{p}_2}(t, t')$ phase. 
  As expected for a pulse with no half cycle symmetry, the structures are asymmetric with regard to $(p_{1\parallel},p_{2\parallel}) \leftrightarrow (-p_{1\parallel},-p_{2\parallel})$ and there is no bright fringe along the anti-diagonal. Interestingly, the asymptotes and focal points of the hyberbolae do not lie at the diagonals, but at the axes. This feature was also present for monochromatic driving fields \cite{Maxwell2015}, and it is likely caused by the transverse momentum integration. 
 It is also noteworthy that many of the intricate patterns in the second and fourth quadrants of the $p_{1\parallel}p_{2\parallel}$ plane are absent, which is evidence that they stem from other types of quantum interference. 
 We note that for CEP $65^{\circ}$ [Figs.~\ref{fig:exchangeonlyoverall}(e)-(h)] the fringes are well distributed in the four quadrants and the patterns are more intricate suggesting competing events within the pulse, whilst the distributions and the interference structures for CEP $155^{\circ}$ [Fig.~\ref{fig:exchangeonlyoverall}(i)-(l)] are primarily located in the third quadrant of the parallel momentum plane, and their cleanness suggests a dominant event.
 
 To understand the effects in Fig.~\ref{fig:exchangeonlyoverall}(d)-(i) in greater detail, the interferences associated with $\alpha_{l,d}$, $\alpha_{r,u}$ and $\alpha_{lT,dT}$ are separated. Therein, we choose the $3s \rightarrow 3p$ channel as Fig.~\ref{fig:exchangeonlyoverall}(e) shows obvious hyperbolic structures and central nodes at the axes, and the $3p \rightarrow 4p$ excitation channel for which there are also visible hyperbolas and extra nodes stemming from the higher principal quantum number $n_e=4$ [Fig.~\ref{fig:exchangeonlyoverall}(g)].
\begin{figure}[]
     \centering
    \includegraphics[width=\columnwidth]{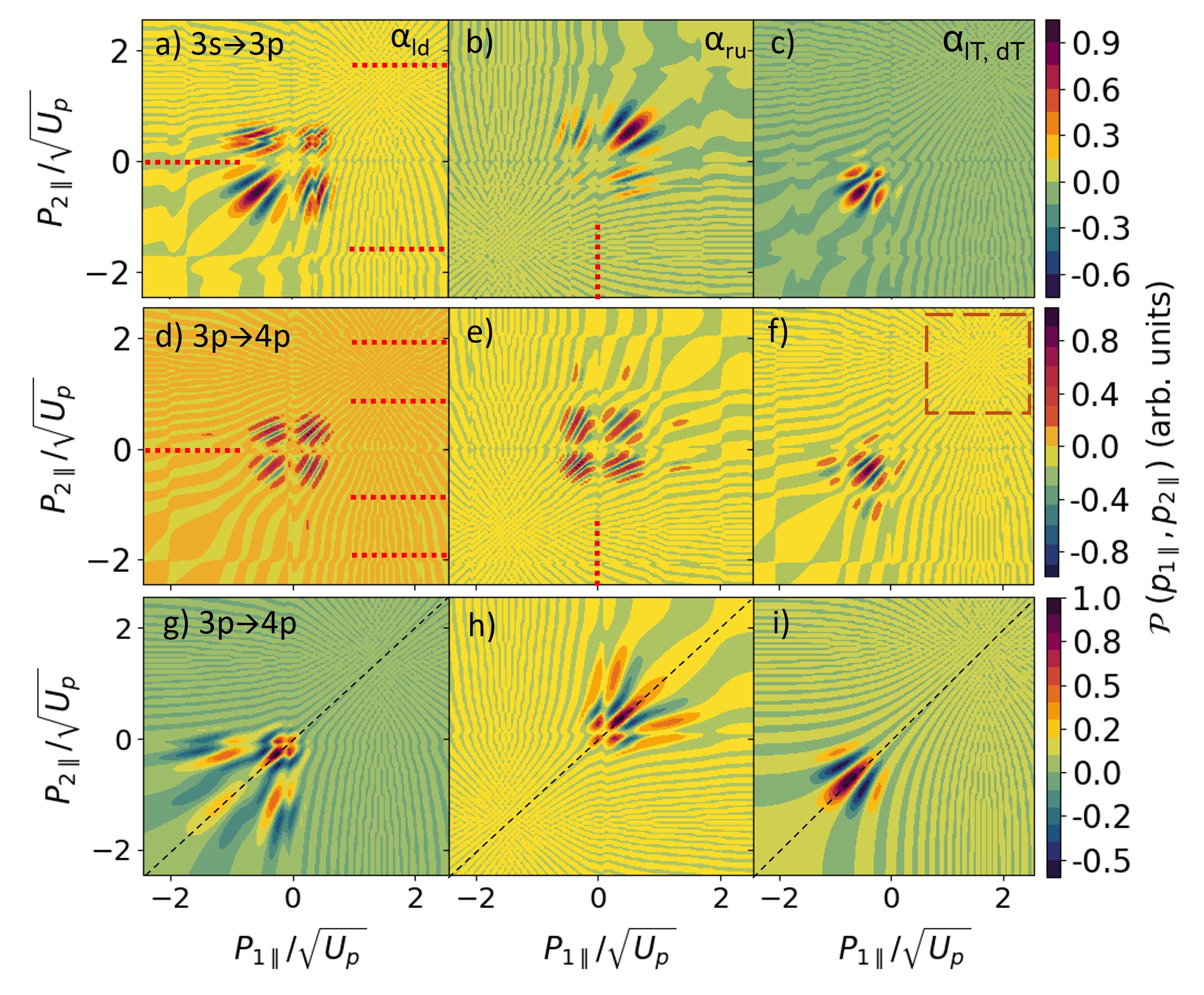}
    \caption{Interference patterns associated with the purely exchange phase differences for the three most dominant events within a pulse for the $3s \rightarrow 3p$ transition incorporating both the excitation and ionization prefactors (top row) and the $3p \rightarrow 4p$ transition, with both prefactors (middle row) and with neither prefactor (bottom row). The same pulse parameters are employed as in Fig.~\ref{fig:cep65singlechannel}, and the carrier-envelope phase is $\phi_1=65^{\circ}$. The signal in each panel has been normalized with regard to its maximum. The diagonals $p_{1\parallel} = p_{2\parallel}$ are indicated with black dashed lines in the bottom row. The locations of the nodes and phase shifts arising from the prefactors are marked by red dotted lines in (a), (b), (d), (e). The box in panel (f) shows the location of finely spaced hyperbole.  
    \label{fig:exchangebreakdown}}
\end{figure}
Figs.~\ref{fig:exchangebreakdown}(a)-(c) and  Figs.~\ref{fig:exchangebreakdown}(d)-(f) show $\alpha_{l,d}, \alpha_{r,u}$ and $\alpha_{lT,dT}$ for the two chosen excitation pathways incorporating both the  $V_{\mathbf{p}_1,ekg}$ and  $V_{\mathbf{p}_2,e}$ prefactors. Both prefactors have been omitted in Figs.~\ref{fig:exchangebreakdown}(g)-(i) for the excitation channel $3p \rightarrow 4p$ to highlight the phase jumps and nodes associated with the ionization prefactor, arising from locations of the radial nodes and the mapping of the angular nodes of $V_{\mathbf{p}_2,e}$ [Fig.~\ref{fig:pf2}(g)] respectively. These are indicated by the red dotted lines in Figs.~\ref{fig:exchangebreakdown}(a), (b), (d) and (e). The effects of the prefactor are present and the same for all three interference types discussed in this work. 

 For $3s \rightarrow 3p$, one can see that the fringes stemming from the $p_3o_4$ event, are the brightest and occupy the largest momentum region, located primarily in the third quadrant of the $p_{1\parallel}p_{2\parallel}$ plane, but also spilling in the other quadrants [see the left column in Fig.~\ref{fig:exchangebreakdown}]. This is consistent with the findings in Fig.~\ref{fig:dptotal}, which indicate that this event is dominant. Figs.~\ref{fig:exchangebreakdown}(b) and (c) shed light on the other events. The fringes arising from the second dominant event, i.e., $p_4o_5$ lead to well-defined hyperbolas in the first quadrant of the parallel momentum plane. These occupy a slightly smaller momentum region but are also quite bright, in agreement with the findings that both events are comparable. Lastly, $p_5o_6$ is much less dominant than the other two events, which is reflected in the much smaller momentum region in which the fringes caused by $\alpha_{lT,dT}$ are significant. For the excitation pathway $3p \rightarrow 4p$ [Figs.~\ref{fig:exchangebreakdown}(d)-(f)], the RESI distributions are more symmetric. This is due to the events $p_3o_4$ and $p_4o_5$ competing and the collapse of the CAR for $p_5o_6$  due to a higher energy gap between the ground and excited states [see Fig.~\ref{fig:dptotal}(a) and (f)],  together with the influence of the prefactors, which have additional radial nodes.

We retain the diagonal symmetry of the interference patterns for all three phases, indicated by the black dashed lines in Fig.~\ref{fig:exchangebreakdown}(g)-(i). Additionally, within the direct ATI cut-off, the hyperbolae associated with alpha exchange can only be seen for $n=0$ - these are most clear in quadrants opposite to where the events are dominant - an example is indicated by the box in Fig.~\ref{fig:exchangebreakdown}(f). This is because $\alpha_{\mathbf{p}_1,\mathbf{p}_2}(t, t')$ plays the smallest role in these regions, and thus the bright spine no longer masks or skews the hyperbolae caused by $\alpha_{\mathbf{p}_1,\mathbf{p}_2}^{(\mathrm{exch})}$. 

\begin{figure}[]
     \centering
\includegraphics[width=\columnwidth]{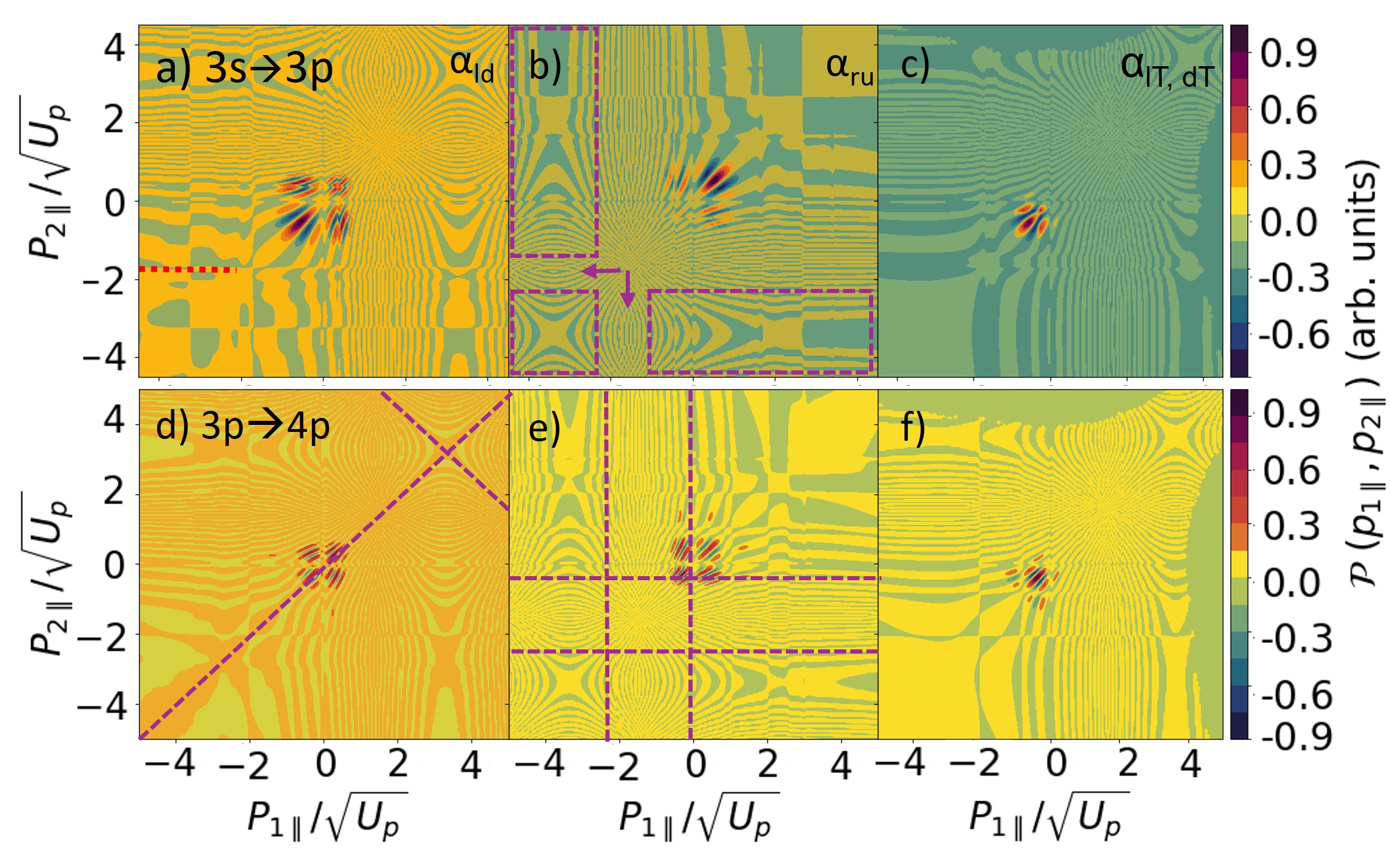}
    \caption{Interference patterns associated with the purely exchange phase differences for the three most dominant events within a pulse for the $3s \rightarrow 3p$ transition  (top row) and the $3p \rightarrow 4p$ transition (bottom row), for a larger momentum region than in Fig.~\ref{fig:eventannotations}. The same pulse parameters are employed as in Fig.~\ref{fig:cep65singlechannel}, and the carrier-envelope phase is  $\phi_1=65^{\circ}$. The signal in each panel has been normalized with regard to its maximum.  The locations of phase shifts arising from the prefactors are marked by red dotted lines in (a). Purple dashed boxes in (b) show the different hyperbole arising from $\alpha_{\mathbf{p}_1,\mathbf{p}_2}^{(\mathrm{exch})} = 2n\pi$ with $n>0$. The arrows indicate where these hyperbolae meet leading to convex shapes in the interference pattern. The asymptotes of one such hyperbola are marked in (d) with purple dashed lines along the diagonal and parallel to the anti-diagonal. Panel (e) marks a cross-shape with purple lines, between which we see the finest fringes.    
    \label{fig:exchangezoomedout}}
\end{figure}

Further information about the hyperbolae can be gleaned from the interference patterns at larger momentum values. In Fig.~\ref{fig:exchangezoomedout}, we see additional hyperbolae with shifted centers, presumably arising from greater values of $n$ indicated by the purple boxes in panel (b). 
These hyperbolae have asymptotes at or parallel to the diagonal and antidiagonal [see Fig.~\ref{fig:exchangezoomedout}(d)] with decreasing fringe spacing. At the locations where two hyperbolae meet, we see a convex shape [see arrows in Fig.~\ref{fig:exchangezoomedout}(b)]. The narrowest fringe spacings form a rough cross [marked in Fig.~\ref{fig:exchangezoomedout}(e)], the shape of which is not unlike the mapping of the excitation prefactor in Fig.~\ref{fig:excitationcontribution} with the center of the cross located in the momentum quadrant where the associated event is dominant. We also note that at locations of phase shifts, the fringes become slightly jagged [see red dotted line in Fig.~\ref{fig:exchangezoomedout}(a)].

Finally, this analysis reveals several additional key observations regarding the combination of the three events (and thus the three exchange phases) shown in Figs.~\ref{fig:exchangeonlyoverall}(d)-(k). 

First, faint hyperbolae observed in the second and fourth quadrants result from the incoherent combination of events and are associated with $\alpha_{\mathbf{p}_1,\mathbf{p}_2}^{(\mathrm{exch})}$. These shapes are therefore not actually interference effects but a consequence of combining the interference effects from $\alpha_{l,d}, \alpha_{r,u}$ and $\alpha_{lT,dT}$ incoherently, since their fringe direction in the second and fourth quadrants varies depending on the events' locations in the pulse. The spine, along with alternating intense positive and negative fringes in the diagonal (both of which arise from $\alpha_{\mathbf{p}_1, \mathbf{p}_2}(t, t')$) contribute to the first and third quadrants, with the most dominant events leading to the brightest and biggest contributions. 

Second, in the quadrants where both $\alpha_{\mathbf{p}_1,\mathbf{p}_2}^{(\mathrm{exch})}$ and $\alpha_{\mathbf{p}_1, \mathbf{p}_2}(t, t')$ are present, the effect of $\alpha_{\mathbf{p}_1, \mathbf{p}_2}(t, t')$ tends to be stronger causing the bright fringes to obfuscate the hyperbolae. However, $\alpha_{\mathbf{p}_1,\mathbf{p}_2}^{(\mathrm{exch})}$ appears to bend these parallel fringes creating large bright hyperbole-like structures on either side of the spine. Where $\alpha_{\mathbf{p}_1, \mathbf{p}_2}(t, t')$ is less strong (i.e. in the second and fourth quadrants) there are more clearly observable hyperbolae. 

\subsection{Event Interference}
\label{sec:PMDinterfev}

\begin{figure}[]
     \centering
\includegraphics[width=\columnwidth]{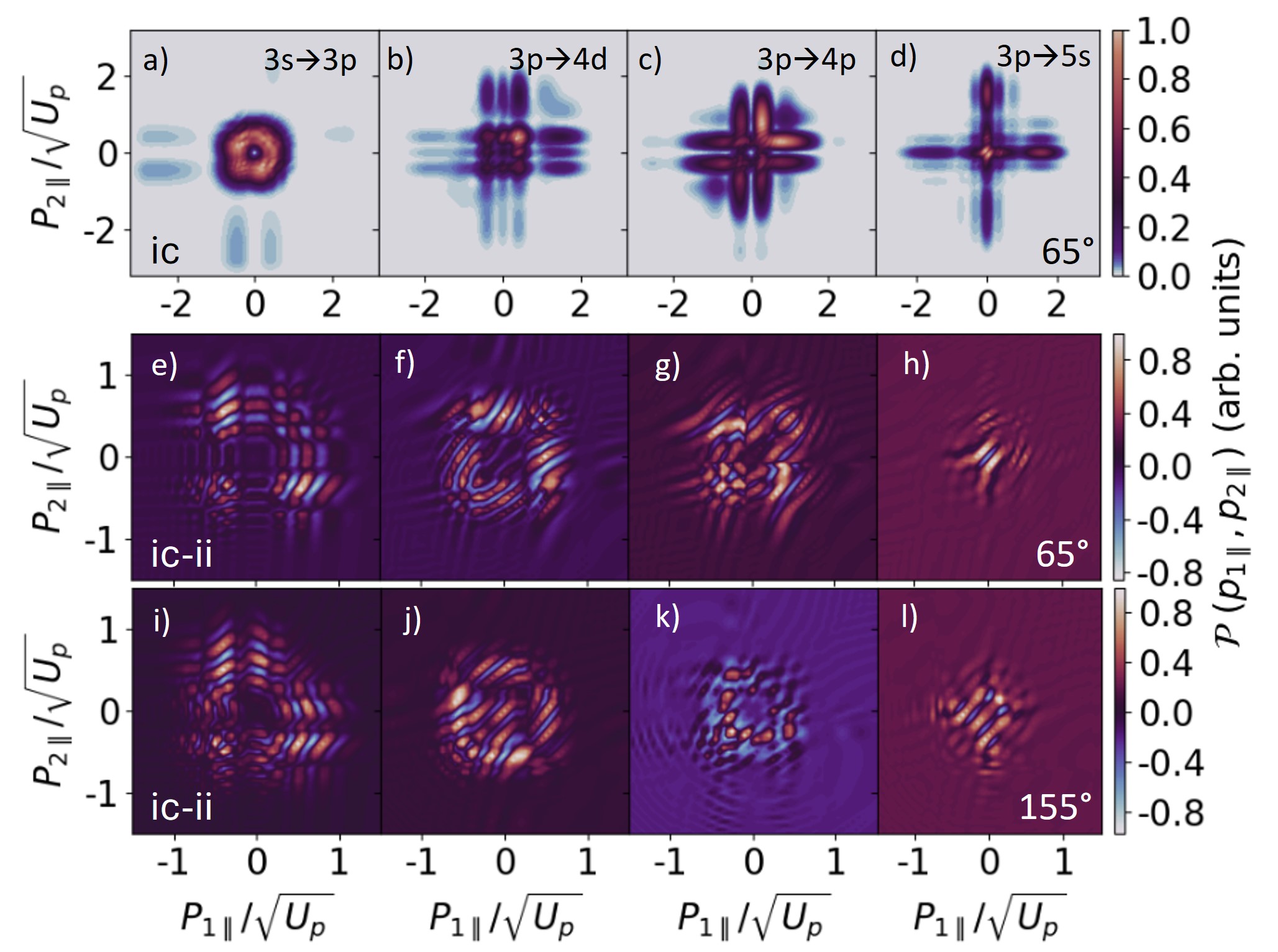}
    \caption{Correlated two-electron momentum distributions as functions of the momentum components $p_{1\parallel}$,$p_{2\parallel}$ parallel to the driving field polarization, calculated for a single excitation channel for the same pulse parameters employed in \ref{fig:cep65singlechannel}. The first row corresponds to amplitudes where events are summed coherently, and symmetrization is performed incoherently as given in Eq.~\eqref{eq:1ic}, and was calculated for $\phi_1=65^{\circ}$. The second and third rows show the difference between the first row and the corresponding fully incoherent sums [see panels (a)-(f) in Figs.\ref{fig:cep65singlechannel}, \ref{fig:cep155singlechannel}], with CEP $65^{\circ}$ and $155^{\circ}$. These rows show the interference arising from summing particle-exchange terms incoherently and events coherently. All six RESI channels described in Table \ref{tab:channels} are shown from panels (a)-(f) and (g)-(l). The signal in each panel has been normalized with regard to its maximum. 
    \label{fig:eventonlyoverall}}
\end{figure}

Next, we isolate the effects of the phase differences associated with the temporal shifts: $\alpha_{l, r}, \alpha_{lT, r}, \alpha_{lT, l}$ and their identical but transposed (symmetrized) counterparts: $\alpha_{d, u}, \alpha_{dT, u}, \alpha_{dT, d}$. Fig.~\ref{fig:eventonlyoverall} shows the momentum distribution with incoherent symmetrization and coherent events, as well as the differences between these distributions and the fully incoherent sum for both CEPs. This enables us to investigate interference effects emerging from four different phase differences: $\alpha^{(\mathrm{ene})}_{\Delta\tau}$, $\alpha^{(A^2)}_{\Delta\tau}(t',t'')$, $\alpha^{(\mathrm{pond})}_{\Delta \tau}(t,t'')$ and $\alpha^{(\mathbf{p}_1,\mathbf{p}_2)}_{\Delta \tau}(t,t')$. 
Figs.~\ref{fig:eventonlyoverall}(e)-(l) show distributions with alternating positive and negative fringes, that are brightest in the region(s) associated with the dominant event(s) for each excitation pathway, for both CEPs. 
There appears to be a darker ``box" in the center of the distributions - most prominent for channels with a higher number of angular nodes. All other prefactor effects, such as nodes and relative distribution size remain. 
A faint circular substructure is also somewhat visible, particularly in Fig.~\ref{fig:eventonlyoverall}(k). This is associated with $\alpha^{(\mathrm{ene})}_{\Delta\tau}$ [Eq.~\eqref{eq:alphaene}] which gives rise to circular fringes with a target-dependent radius. This is expected because Eq.~\eqref{eq:alphaene} is that of a hypersphere. The radius of the largest circular fringe is obtained at vanishing perpendicular momenta. These distributions occupy similar momentum regions to the fully coherent distributions in Fig.~\ref{fig:cep65singlechannel}(m)-(r) and Fig.~\ref{fig:cep155singlechannel}(m)-(r). 

\begin{figure}[]
     \centering
     \includegraphics[width=\columnwidth]{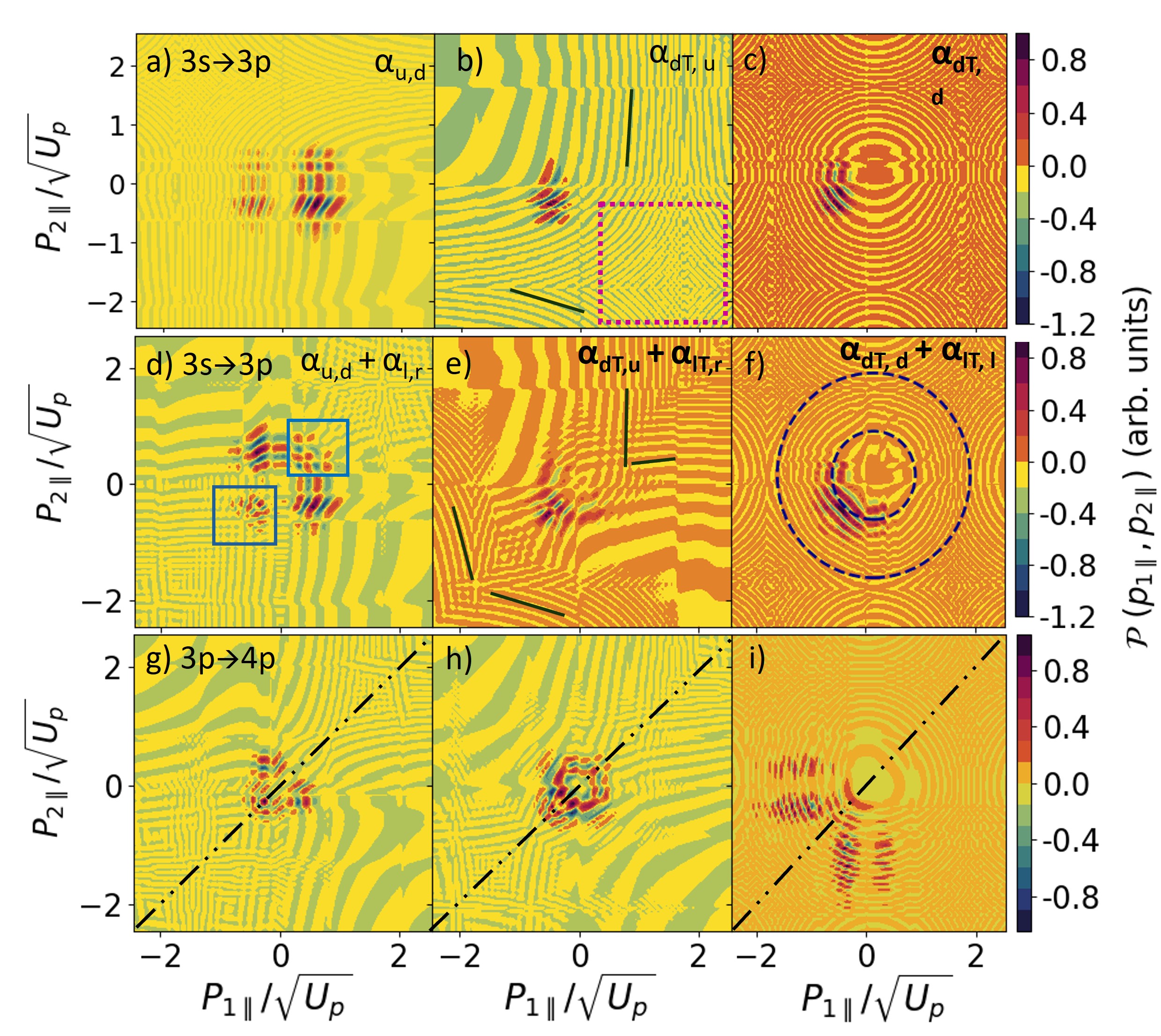}
    \caption{Interference patterns associated with the temporal-shift phase differences utilizing interfering pairs of the three most dominant events within a pulse for the $3s \rightarrow 3p$ transition (top, middle rows) and the $3p \rightarrow 4p$ transition (bottom row). The top row shows the unsymmetrized distributions, whilst the middle and bottom rows account for the transposed counterpart of the phase differences. The same pulse parameters are employed as in Fig.~\ref{fig:cep65singlechannel}, using a CEP of $65^{\circ}$ for the $3s \rightarrow 3p$ transition [top and middle row] and a CEP of $155^{\circ}$ for the $3p \rightarrow 4p$ transition [bottom row]. The signal in each panel has been normalized with regard to its maximum. The gradients of the fringes are marked in (b) and (e) with solid black lines. The red dotted box in (b) indicates the location of fine fringes coming from $\alpha^{(\mathbf{p}_1,\mathbf{p}_2)}_{\Delta \tau}(t,t')$, which are obfuscated by the larger 'wings' in the symmetrized distributions in (e) and (h).  The solid blue boxes in (d) indicate the non-interference chequerboard patterns coming from $\alpha^{(A^2)}_{\Delta\tau}(t',t'')$. Purple dashed circles in (f) indicate the largest circular fringes resulting from $\alpha^{(\mathrm{ene})}_{\Delta\tau}= 2\pi n$ where $n=3, 4$. The symmetry along the diagonal of the symmetrized distributions can also be seen by the dash-dot lines in (g)-(i).
    \label{fig:eventannotations}}
\end{figure}

As with the exchange-only case, we now isolate the effects of $\alpha_{l, r}, \alpha_{lT, r}, \alpha_{lT, l}$ with the $3s \rightarrow 3p$ with CEP $65^{\circ}$ [Fig.~\ref{fig:eventonlyoverall}(e)]. However, this time $3p \rightarrow 4p$ is taken with CEP $155^{\circ}$  [Fig.~\ref{fig:eventonlyoverall}(k)]. Figs.~\ref{fig:eventannotations}(a)-(c) show the separated phases,  $\alpha_{u, d}, \alpha_{dT, u}, \alpha_{dT, d}$ with the $3s \rightarrow 3p$ transition. The gradient of some of the fringes is indicated by the lines in Fig.~\ref{fig:eventannotations}(b). As expected for up-down interference, the majority of the lines in Figs.~\ref{fig:eventannotations}(a),(b) are roughly located in the up-down direction. In quadrants two and four (depending on the events in question), we see thinner fringes, coming from different directions and combining [Fig.~\ref{fig:eventannotations}(b)]. For their identical but transposed phases, we would expect these distributions to also be transposed. An incoherent sum of the temporal-shift phases with their transposed counterparts, $\alpha_{u, d} + \alpha_{l, r}, \alpha_{dT, u} + \alpha_{lT, r}, \alpha_{dT, d} + \alpha_{lT, l}$ is provided in the second row for $3s \rightarrow 3p$, and in the third row for $3p \rightarrow 4p$. In the first quadrants of Figs.~\ref{fig:eventannotations}(e), (h), we see widening v-shaped fringes which overlap to form complex fine criss-crossed patterns. These are actually not due to interference, but arise as a consequence of incoherent symmetrization - in panel (e), the gradients of the fringes associated with $\alpha_{dT, u}$ and $\alpha_{lT, r}$ individually are indicated with lines. The fine fringes in the red box in panel (b) are almost completely obfuscated by the thicker fringes in that quadrant coming from the symmetrized phase difference, forming a ``wing" shape along the anti-diagonal in panels (d), (e), (g) and (h). These patterns occur mostly for intra-cycle shifts as in the first two columns of this figure. 

With inter-cycle shifts i.e. $\alpha_{dT, d}$ and $\alpha_{lT, l}$, almost circular fringes are observed arising from $\alpha^{(\mathrm{ene})}_{\Delta\tau}$, the largest radius of which is positive only for $n>2$ for the target in question. The largest fringes are plotted with $n=3, 4$ in Fig.~\ref{fig:eventannotations}(f) and appear to be a fairly good match. The other rings may arise from varying values of transverse momenta. We note that the fringes are not centered at the exact origin, since the temporal shift of $2\pi/\omega$ from $p_3 o_4$ to $p_5 o_6$ is an approximation for the few-cycle pulse.  

\begin{figure}[]
     \centering
    \includegraphics[width=\columnwidth]{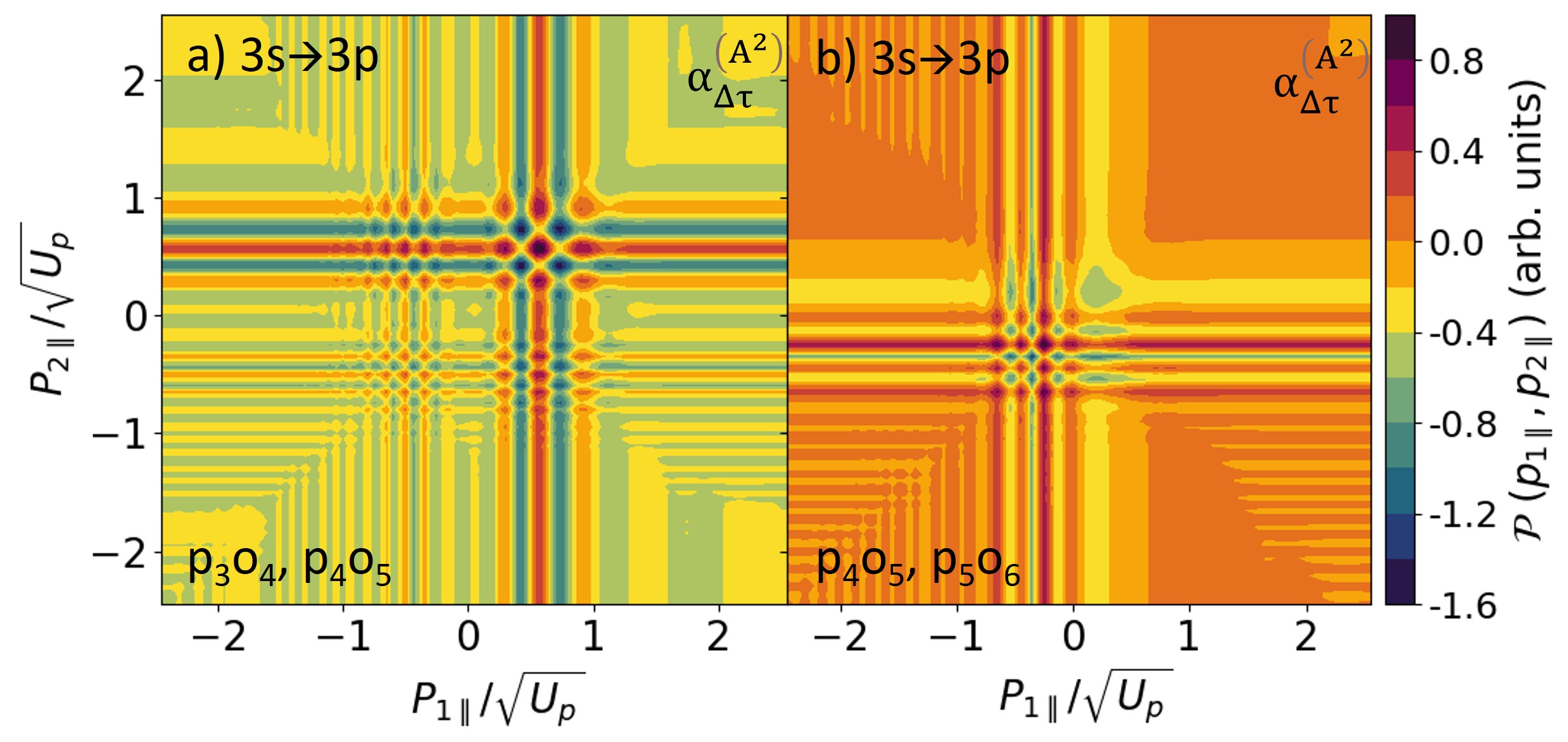}
    \caption{Interference patterns associated with a pure temporal-shift phase difference, $\alpha^{(A^2)}_{\Delta\tau}$ computed for pairwise combinations of the most dominant events that separated by half a cycle for the $3s \rightarrow 3p$ transition and CEP $65^{\circ}$. All other pulse parameters employed are the same as in Fig.~\ref{fig:cep65singlechannel}. 
    \label{fig:eventalphaA2}}
\end{figure}

We note that $\alpha^{(A^2)}_{\Delta\tau}(t',t'')$ depends only on the ionization and rescattering times of the first electron. Therefore, we can detangle the interference stemming from $\alpha^{(A^2)}_{\Delta\tau}(t',t'')$
 by calculating $\alpha_{l, r} ( + \alpha_{u, d})$ and $\alpha_{lT, r} ( + \alpha_{dT, u})$ for the first and second electron separately. The distribution with the first electron only contains effects from all four phase differences associated with temporal shifts including $\alpha^{(A^2)}_{\Delta\tau}(t',t'')$ whilst the distribution from the second electron excludes the $\alpha^{(A^2)}_{\Delta\tau}(t',t'')$ contribution. For more details, see Fig.~\ref{fig:eventelectronbreakdown} in the second Appendix\ref{appendix:interf}. 

The one-electron unsymmetrized distributions (i.e. associated with interference along either the $p_{1\parallel}$ or $p_{2\parallel}$ axes) consist of straight fringes parallel to the associated axis, varying in intensity. When incoherently symmetrized, these fringes form a cross shape. At the point where the fringes intersect, we see a chequerboard-type pattern. When $\alpha^{(A^2)}_{\Delta\tau}(t',t'')$
 is isolated, as in Fig.~\ref{fig:eventalphaA2}, this pattern becomes even more detailed and complex due to the subtraction of the two individual electrons' fringes. We can therefore deduce that this chequerboard, indicated by the blue boxes in Fig.~\ref{fig:eventannotations}d), for instance, is not actually an interference effect but a symptom of incoherent symmetrization. 

The wings of the distribution are likely to be brought about by 
$\alpha^{(\mathbf{p}_1,\mathbf{p}_2)}_{\Delta \tau}(t,t')$ (Eq.\eqref{eq:alphap1p2tau}), the only purely temporal-shift phase which is both momentum and field dependent. Thus, intuitively, when the momenta are large and have the same sign (as in quadrants one and three), one expects that this phase difference will be large and the fringes will be finer. In contrast, if both momenta are small (close to the origin), or if they are large but have opposing signs (second and fourth quadrant),  a small $\alpha^{(\mathbf{p}_1,\mathbf{p}_2)}_{\Delta \tau}(t,t')$ will lead to thicker fringes. This effect contributes to the appearance of the dark "box" visible in some distributions of Fig.~\ref{fig:eventonlyoverall} - there are simply fewer interference fringes at play around the origin.

\begin{figure}[]
     \centering
    \includegraphics[width=\columnwidth]{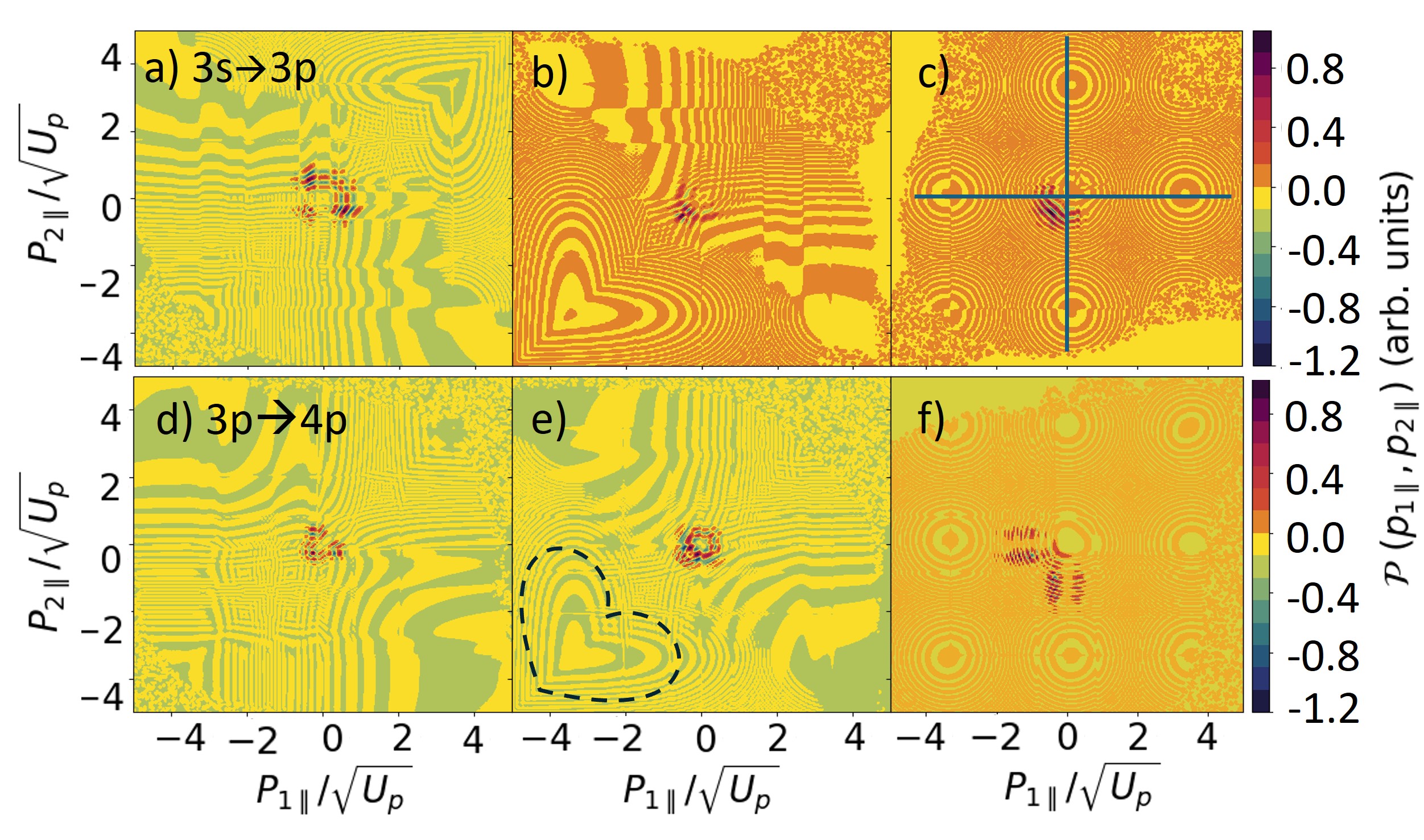}
    \caption{Interference patterns associated with the temporal-shift phase differences utilizing interfering pairs of the three most dominant events within a pulse for the $3s \rightarrow 3p$ transition (top row) and the $3p \rightarrow 4p$ transition (bottom row). A larger momentum region is taken than in Fig.~\ref{fig:eventannotations}. The same pulse parameters are employed as in Fig.~\ref{fig:cep65singlechannel}, using a CEP of $65^{\circ}$ for the $3s \rightarrow 3p$ transition [top row] and a CEP of $155^{\circ}$ for the $3p \rightarrow 4p$ transition [bottom row]. The signal in each panel has been normalized with regard to its maximum. The blue lines in (c) along the axes are drawn to emphasize the presence of the slightly off-axes circular fringes presumably coming from $\alpha^{(\mathrm{ene})}_{\Delta\tau}$. Panel (e) annotates the heart shape of the interference which is also present in (a) and (b).
    \label{fig:eventzoomedout}}
\end{figure}

Fig.~\ref{fig:eventzoomedout} shows the temporal-shift phases for both channels, in a larger momentum space. As before, the wings can be seen. In addition to the wings, the intracycle shifts also exhibit heart shapes in the first or third quadrants (depending on the dominance of events involved), with 'static' like patterns in the opposing quadrant. In panel (d) the heart disappears, presumably because for CEP $155^{\circ}$, $p_3o_4$ loses dominance. The origin of this shape is unclear. However, this behavior indicates that they may arise from a combination of the field-dependent temporal-shift phase differences. Chequerboards appear only for small momentum values indicating $\alpha^{(A^2)}_{\Delta\tau}(t',t'')$
 is only influential in the regions where the events are strongest, which makes sense as this phase difference is field-dependent only.
 
The inter-cycle shifts (Fig.~\ref{fig:eventzoomedout}(c), (f)) show multiple sets of circular fringes due to $\alpha^{(\mathrm{ene})}_{\Delta\tau}$, located along the axes, and faintly even in the quadrants. They are centered slightly off-axis due to the asymmetry of the pulse. Since $\alpha^{(\mathrm{ene})}_{\Delta\tau}$ [Eq.\eqref{eq:alphaene}] gives the equation of a hypersphere, it is not trivial to obtain the centers and sizes of these extra circular fringes. Additionally, due to the strong dependence of $\alpha^{(\mathrm{pond})}_{\Delta \tau}(t,t'')$ on the field shape, it is difficult to predict or detangle the effects of this phase difference explicitly. However, all remaining patterns stem from this term or from a combination of the other phase differences. 

\subsection{Exchange and Event Interference}
\label{sec:PMDinterfexchev}

\begin{figure}[]
     \centering
    \includegraphics[width=\columnwidth]{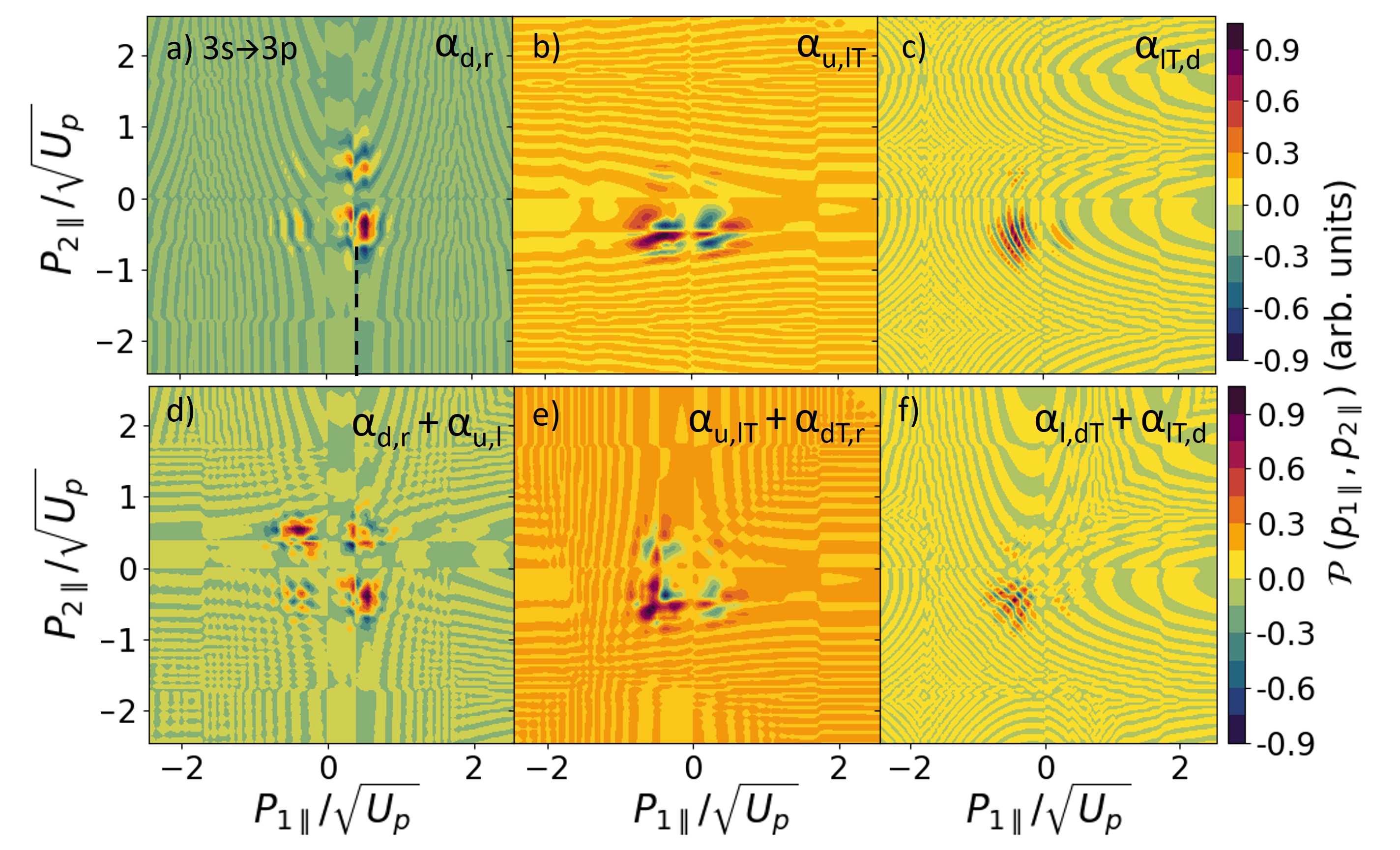}
    \caption{Interference patterns associated with the combined exchange and temporal-shift phase differences utilizing interfering pairs of the three most dominant events within a pulse for the $3s \rightarrow 3p$ transition and a carrier-envelope phase of $\phi_1=65^{\circ}$. The distributions shown are unsymmetrized (top row) and symmetrized (bottom row).  The same pulse parameters are employed as in Fig.~\ref{fig:cep65singlechannel}. The signal in each panel has been normalized with regard to its maximum. The dashed black lines in (a) indicate an example of a ``cut" in the distribution, which also translates to the symmetrized distribution in (d). 
    \label{fig:exchangeshiftoverall}}
\end{figure}

The remaining six phases associated with a combination of exchange and temporal shift terms, $\alpha_{l, u}, \alpha_{lT, u}, \alpha_{lT, d}$ and their identical transposed counterparts $\alpha_{r, d}, \alpha_{r, dT}, \alpha_{dT, l}$ are considered in Fig.~\ref{fig:exchangeshiftoverall}, which shows the unsymmetrized and incoherently symmetrized effects arising from these phases.
Due to the localization of the events within specific momentum quadrants, we expect this phase to contribute the least to the total coherent distributions. However, the effect is stronger than for the monochromatic wave, as the asymmetry of the pulse (and the effect of prefactors) causes the events to occupy regions near the origin and sometimes spill into other quadrants. This enhances the events' overlap. 

\begin{figure*}[]
    \centering
\includegraphics[width=\textwidth]{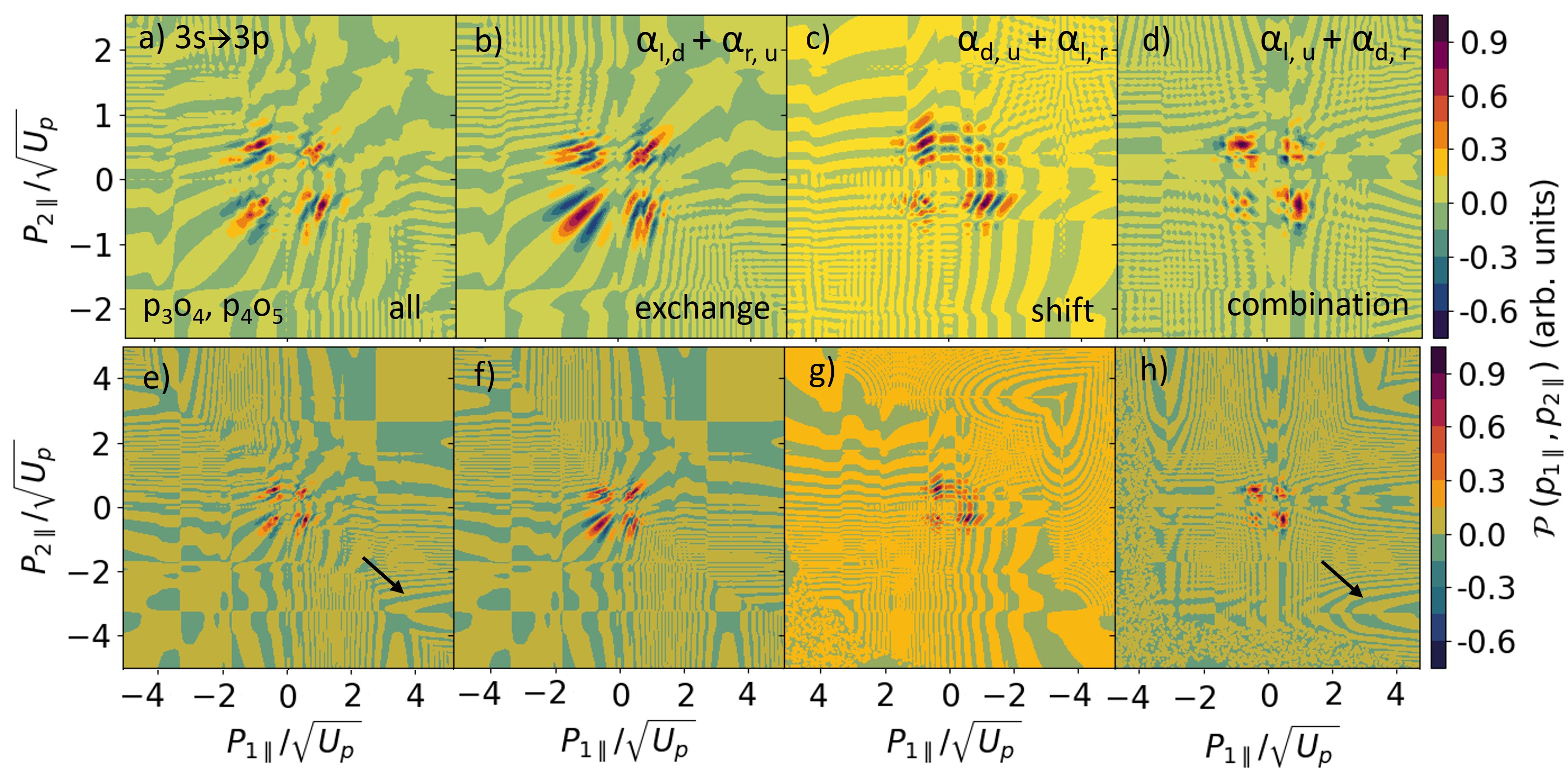}
    \caption{Interference patterns arising within the region of interest (top row) and in a larger momentum region (bottom row) considering a fully coherent sum of events and symmetrization [panels (a), (e)], and isolating interference effects due to exchange [panels (b), (f)],  temporal shifts [panels (c), (g)] and those arising from a combination of exchange and event interference [panels (d) and (h)] for the same pulse parameters employed in \ref{fig:cep65singlechannel}. The two most dominant events for the channel ($3s \rightarrow 3p$) and CEP ($65^{\circ}$) have been taken. The signal in each panel has been normalized with regard to its maximum. The contribution to the fully coherent interference pattern at larger momentum values, arising from a combination of exchange and shift, has been indicated by arrows in (e) and (h).
    \label{fig:interfcomparison}}
\end{figure*}

Fig.~\ref{fig:exchangeshiftoverall}(a)-(c) shows fringes parallel to the $p_{n\parallel}$ axes. This is expected since only $\alpha^{(\mathbf{p}_1 \leftrightarrow \mathbf{p}_2)}_{\Delta \tau}$ (Eq.~\eqref{eq:alphaexctau}) contributes now. The locations of the brightest spots and their widths are dependent, however, on the exact pulse field symmetry. The distributions also appear to be cut (an example is given by the dashed line in panel (a)).

The question then remains: which of these three types of interference contribute most to the total coherent-coherent distribution? A comparison of exchange, event (temporal shift) and combined exchange and shift interferences to the total coherent interference is shown in Fig.~\ref{fig:interfcomparison}. The top row shows the momentum region of interest and the bottom row displays a larger momentum region containing the secondary effects. 

The pure exchange phases are the most influential, leading to the spine and to small hyperbolic patterns in the total coherent distributions. These features will remain most resistant to focal averaging. The pure temporal shifts serve to make the spine a bit more jagged and blurred, whilst also shifting the locations of the brightest maxima within the fringes themselves. The non-interference effects coming from summing symmetrization incoherently further enhance the complexity of the fringes in the first and third quadrants. The most notable impact of temporal shifts emerges in the second and fourth quadrants, where the large wings overshadow the smaller, overlapping hyperbolic patterns from pure exchange phase differences. 

Predictably, the combined effect of exchange and temporal shifts yields the smallest influence, lightly reinforcing some effects from the other types of phase differences, such as overlapping fringes in quadrants two and four. Notably, even the discernible``cut" observed when isolating this phase dissipates within the total coherent plot. For larger momentum regions, the heart and static patterns disappear although the effect of exchange and shift combined in skewing the fringes becomes slightly more noticeable (indicated by the arrows in Fig.~\ref{fig:interfcomparison}(e) and (h)). We have verified that these results hold for other events, and also the other excitation channels. 

\section{Conclusions}
\label{sec:conclusions}
In conclusion, we investigate quantum interference in recollision excitation with subsequent ionization (RESI) for few-cycle pulses, focusing on two-electron interference patterns associated with a single excitation channel. This type of interference stems from quantum phase shifts which can be derived analytically, thus adding predictive power and transparency to the problem. This considerably extends our previous investigations for monochromatic fields \cite{Maxwell2015,Maxwell2016}, and reveals much more convoluted patterns. The added complexity results from three key differences regarding the driving field. First, the field cycles are not equivalent, which means that there will be different kinds of interfering events and many types of fringes. Second, the dominant events within the pulse will be strongly influenced by the carrier-envelope phase (CEP). Third, the symmetries associated with monochromatic driving fields will be broken. A monochromatic field remains invariant under a translation of half a cycle followed by a reflection about the time axis, also known as the half-cycle symmetry; a time reflection around its extrema, and a time reflection around its zero crossings \cite{Rook2022}. These three symmetries translate into fourfold symmetric RESI distributions for monochromatic fields, if all transition amplitudes are summed incoherently. For a few-cycle pulse, only symmetries associated with electron exchange will be retained. Furthermore, because the field gradients will be unequal around its maxima or zero crossings, the electron-momentum distributions may be skewed towards a specific momentum region. 

With that in mind, we relaxed some key assumptions in \cite{Maxwell2015}, including the field shape and polarization, and derived a myriad of phase shifts for arbitrary driving fields, which gave rise to generalized interference conditions. These phase shifts stem from exchanging the momenta of the two electrons, which is necessary due to their indistinguishability, from events occurring at different times within the pulse, and from electron exchange \textit{and} temporally shifted events. The analytic conditions obtained show a multitude of features, some of which are field independent, and some of which depend on its polarization, shape and symmetry.  We have also identified building blocks for those phase shifts. With pure exchange, these are $\alpha_{\mathbf{p}_1,\mathbf{p}_2}^{(\mathrm{exch})}$ and $\alpha_{\mathbf{p}_1, \mathbf{p}_2}(t, t')$, where the first term has a linear dependence and the latter depends on the field via the temporal arguments. With different events and purely temporal shifts, we have identified 
$\alpha^{(\mathrm{ene})}_{\Delta\tau}$, $\alpha^{(A^2)}_{\Delta\tau}(t',t'')$, $\alpha^{(\mathrm{pond})}_{\Delta \tau}(t,t'')$ and $\alpha^{(\mathbf{p}_1,\mathbf{p}_2)}_{\Delta \tau}(t,t')$ as the sources of interference. The term  $\alpha^{(\mathrm{ene})}_{\Delta\tau}$ is associated with the electronic bound-state and kinetic energies, $\alpha^{(A^2)}_{\Delta\tau}(t',t'')$ is a field-dependent term involving only the first electron, $\alpha^{(\mathrm{pond})}_{\Delta \tau}(t,t'')$ gives ponderomotive energy shifts and $\alpha^{(\mathbf{p}_1,\mathbf{p}_2)}_{\Delta \tau}(t,t')$ a term encompassing both the field and momentum dependence. Finally, with both exchange and temporally shifted events both types of phase shifts will be present, but, instead of  $\alpha^{(\mathbf{p}_1,\mathbf{p}_2)}_{\Delta \tau}(t,t')$, it is described by the $\alpha^{(\mathbf{p}_1 \leftrightarrow \mathbf{p}_2)}_{\Delta \tau}(t,t')$ term, which contains temporal shifts and momentum exchange.  

Particular cases of these conditions give the previous expressions derived in \cite{Maxwell2015} for monochromatic fields. This includes hyperbolic structures and spine-like fringes which are reflection symmetric upon the diagonal $p_{1\parallel}=p_{2\parallel}$ of the $p_{1\parallel}p_{2\parallel}$ plane. Nonetheless, because the half-cycle symmetry is broken, such features are no longer symmetric about $(p_{1\parallel},p_{2\parallel})\leftrightarrow (-p_{1\parallel},-p_{2\parallel})$ and a bright interference maximum along the antidiagonal $p_{1\parallel}=-p_{2\parallel}$  is also absent. The specific case explored in this work, namely a linearly polarized few-cycle pulse, can be re-written as a three-color field \cite{Habibovich2021}. Thus, many of the present studies can be modified to a continuous bichromatic field or a long pulse with two colors. Additionally, some conclusions can be anticipated for this latter case. For instance, the inter-cycle patterns are expected to be more prominent for a continuous wave or long enough pulse.  Furthermore, for linearly polarized fields of frequencies $r\omega$  and $s\omega$ such that $r+s$ is even (odd), the half-cycle symmetry is retained (broken); thus, one expects that the interference fringe predicted to exist in the anti-diagonal will be present (absent). More systematic studies of how the field symmetries affect the two-electron interference building blocks and patterns would require the arguments and the formalism in \cite{Rook2022}. 

Furthermore, the bound-state prefactors $V_{\mathbf{p}_{2}e}$ and $V_{\mathbf{p}_{1}e,\mathbf{k}g}$ introduce additional phase shifts and momentum biases. The prefactor $V_{\mathbf{p}_{2}e}$ associated with the ionization of the second electron determines the shapes of the distributions, and  $V_{\mathbf{p}_{1}e,\mathbf{k}g}$ influences more their centers and widths. However, although the nodes are blurred for the latter prefactor, the field dependence embedded in it means that it will get skewed for a pulse due to the unequal cycles and gradients. Here, we provide an analysis of how both prefactors can be mapped to the $p_{1\parallel}p_{2\parallel}$ plane. In this work, we use argon as a target, for which there are six excitation pathways to $s$, $p$, and $d$  states. This facilitates a comparison with our previous work. 

Specifically for a pulse, it is crucial to identify dominant events, whose interference one must study. In our previous paper \cite{Faria2012}, we have studied this dominance using qualitative arguments. Here, we propose a parameter that brings together three key factors influencing an event's dominance: the ionization probabilities of the first and second electrons, which can roughly be associated with the minimal values of $(\mathrm{Im}[t''])^{-1}$ and $(\mathrm{Im}[t])^{-1}$, and the momentum region for which rescattering of the first electron has a classical counterpart. The dominance parameter is a good indicator of what should be discarded, and of what physically happens in the pulse. 

However, it has a series of limitations, which boil down to a single number not being sufficient to quantify dominance. The prevalence of a particular event depends strongly on the momentum region. Therefore, if they are too discrepant, one cannot use $\mathcal{D}(p_i,o_j)$ to predict where the single main maximum lies. Instead, it indicates which event dominates in the \textit{same} region. Furthermore, the gradient of  $\mathrm{Im}[t]$ is asymmetrical with regard to its minimum. Away from the minimum, imaginary parts of the times associated with different events follow each other closely.  This is particularly true for high-lying bound states, for which these gradients are steep. The results in Sec.~\ref{sec:dominance} show that the dominance parameter is not reliable in these cases. 

Another issue is that it does not allow us to compare an event leading to a bright signal in a small momentum region with another, less bright probability density occupying a larger region. A question remains on how to determine what contributes more to the final map. Finally, the prefactors also skew the momentum distributions and sometimes mask the dominance. Still, if $\mathcal{D}(p_i,o_j)$ is too low for a specific event, such as those at the trailing edges of the pulse, the event can be discarded without compromising the key results. 

For the two specific CEPs employed in this work, we have identified at most three dominant events in agreement with the qualitative findings in \cite{Faria2012}. Applying the equations and criteria in Sec.~\ref{sec:interfcondition}, we found 15 pairwise phase differences, which are given in Table \ref{tab:phasetab}. These phase differences were then used as a guide to disentangle the interference patterns obtained by subtracting the fully coherent from the fully incoherent single-channel distributions.  The complexity of the problem makes it sometimes impractical to obtain analytic expressions in all cases. Nonetheless, one may still observe an approximate hierarchy, as far as the interference types are concerned. 

Overall, the interference due to exchange-only processes is more pronounced and behaves as predicted, with a bright maximum along the main diagonal and hyperbolic patterns. We also observe that the bright fringe along the anti-diagonal and the symmetry upon exchanging the signals of both parallel momenta are absent. This agrees with the analytical predictions of Sec.~\ref{sec:interfcondition}, which states that these features are only present for fields with half-cycle symmetry. The interference effects due to temporal shifts reinforce exchange-only effects for the most part, except for the wing-shaped patterns in the second and fourth quadrants. They also lead to ring-shaped patterns, as predicted for temporal shifts of a full cycle, which resemble those in \cite{Wang2012}.   
The phase differences associated with electron exchange and temporal shift have the subtlest effect of all. Unfortunately, it is difficult to find a hierarchy of the building blocks that constitute these types of interference. Key challenges are to determine in which momentum regions the building blocks are more or less significant and how this might change with the pulse parameters given their strong dependence on the pulse symmetry. This will be of extreme importance in a more realistic scenario, in which the beam profile and geometric phases must be considered. Furthermore, we have also identified patterns associated with three or more interfering pathways, which are difficult to disentangle. 

Regardless, the exchange-only patterns are expected to be robust as many of them do not depend on the field. This also explains their robustness against focal averaging observed in \cite{Maxwell2016}. This suggests that these patterns would be the most prone to survive in a realistic setting. To some extent, this is backed by a comparison with existing experiments \cite{Kubel2014}, in which traces of hyperbolic features and fringes along the diagonals $p_{1\parallel}=\pm p_{2\parallel}$ are present. Because the experiments had CEP-averaged pulses, we were able to apply the simplified model developed for monochromatic fields and compensated for the different frequency bandwidths that exist in few-cycle pulses by employing ad-hoc phases and amplitudes. Thereby, inter-channel interference was also considered.  

However, one may in principle find ways to enhance the other types of interference by manipulating the field parameters.  This can be achieved, for instance, by adapting the knowledge obtained for controlling one-electron PMDs in tailored fields to RESI \cite{Xie2017,Hoang2017,Eicke2020,Habibovich2021,Sun2022} It is also plausible that inter-cycle interference was observed in a TDSE calculation, in form of ring-shaped structures \cite{Parker2006}. However, a direct comparison with the present results is not possible, as those calculations were performed for a parameter range in which EI is dominant. Interference between the EI and RESI pathway has also been reported in \cite{Baier2006,Baier2007,Eckhardt2010}. 

Finally, several issues have been left out of this article. First, there is evidence, from our previous work, that inter-channel interference can lead to striking results and also survive focal averaging. So far, this type of interference has been studied in a simplified way, employing coherent superpositions of excitation channels whose phases and weighting were chosen in a partly ad-hoc way.  To tackle this problem more rigorously, it will be necessary to seek an alternative strategy to the analytic derivations performed here as intra-channel interference is much more influenced by the bound-state geometry. Thereby, key questions are how to establish whether specific channels are comparable and if the fringes resulting from this type of interference have high contrast.    

Second, a realistic model will require incorporating the residual binding potential in the electrons' continuum propagation. Our studies of ultrafast photoelectron holography, using a fully Coulomb-distorted path-integral method in a single electron context \cite{Lai2015a,Maxwell2017,Faria2020}, provide us with some insight into what to expect for RESI. 

For the first electron, the SFA with a single act of rescattering will be a good approximation for the actual dynamics, and, in order to account for Coulomb corrections, one may need at most a phase shift in the action. A recent comparison shows that the rescattered SFA is a good approximation for backscattered electron trajectories, which are those relevant to the first electron \cite{Rook2024}. These studies indicate that the kinematic constraints such as rescattering ridges remain practically unchanged. There are, however, small changes close to the ionization threshold, and shifts in the interference patterns arising from the two types (long and short) scattering returning orbits in a pair. These shifts are very small for the shortest, most relevant orbit pair, and the corresponding interference patterns are washed out upon transverse momentum integration \cite{Shaaran2012}. 

The dynamics of the second electron are expected to be more influenced by the Coulomb potential, due to its kinetic energy in the continuum being lower.  A key issue is that, along the field polarization, the electron can no longer escape with vanishing velocity due to the influence of the Coulomb potential \cite{Rodriguez2023}. Thus, there would be a suppression at the axes $p_{n\parallel}$ in the RESI distributions stemming from this effect. Furthermore, the longer direct orbit, which is degenerate for the SFA, splits into two field-dressed Kepler hyperbolae if the Coulomb potential is present \cite{Lai2015a}. One of them can still be considered ``direct", while the other will exhibit a hybrid character and will interact more with the core. Holographic patterns stemming from these three types of orbits show that their interference is strongly influenced by the binding potential \cite{Maxwell2017,Maxwell2017a}. Nonetheless, as these effects are angle-dependent, it is hard to predict what will survive upon transverse momentum integration.  Because both electrons' ionization times will change due to the presence of the Coulomb potential, we expect that two-electron interference effects involving time delays will be more sensitive to its presence than those resulting from electron exchange only. The Coulomb potential could also break some of the symmetries that arise from the field. 

Third, in NSDI there is the matter of final-state electron-electron repulsion. This effect has been incorporated for the electron-impact (EI) ionization pathway in \cite{Faria2004,Faria2004b} and affected the electron-momentum distributions. However, for electron-impact ionization, the second electron is freed without time delay, which justifies this repulsion being relevant. In contrast, for RESI, the second electron leaves at a later time, so that final-state electron-electron repulsion is expected to be much weaker. 

Fourth, there is focal averaging \cite{Kopold2002} and the effect of Gouy/Maslov phases, which will potentially produce shifts in the interference fringes and introduce momentum biases. Focal averaging has been considered in \cite{Maxwell2016} and Gouy/Maslov phases have been studied in the context of photoelectron holography \cite{Brennecke2020,Werby2021,Carlsen2024,rook2024influence}. A comparison with experiments will require taking these matters into account and will be the topic of future work. 

\acknowledgements
Discussions with Andrew Maxwell, Bradley Augstein, Tobin Holtmann, Gergely Eory, Abraham Jacob, and Lewis Ling are gratefully acknowledged. This work was partly funded by grant No.\ EP/J019143/1, from the UK Engineering and Physical Sciences Research Council (EPSRC) and by UCL.
\appendix*
\section{Prefactors}

In this appendix, we provide a summary of how the prefactors influence the electron momentum distributions when integrated over the momentum components perpendicular to the driving field polarization.  This is important in both understanding the shapes of the two-electron correlated momentum distributions, and in locating the phase shifts present in the quantum interference plots. They were calculated assuming that all bound states involved are hydrogenic and include all normalization and phases necessary for computing coherent superpositions. For details we refer to our previous publications \cite{Shaaran2010,Shaaran2010a,Maxwell2015}. 

\subsection{Expressions}

Here we give the general expressions for the prefactors.
The excitation prefactor reads
\begin{equation}
\begin{aligned}
    V_{p_1 e, k g}= & \sum_{L=\left|l_e-l_g\right|}^{l_e+l_g} \sum_{M=-L}^L(-i)^L A_1 Y_L^M\left(\theta_q, \phi_q\right) \\
& \times \frac{\left(\left\langle l_g, l_e, 0,0 \mid L, 0\right\rangle\left\langle l g, l e, m_g,-m_e \mid L, M\right\rangle\right.}{\sqrt{(2 L+1)}} I_r 
\label{eq:angmom}
\end{aligned}
\end{equation}
where 
\begin{equation}
\begin{aligned}
I_r= & \sum_{k_g=0}^{b_{n_g l_g}} \sum_{k_e=0}^{b_{n_e l_e}} \frac{(-1)^{k_g+k_e} 2^{a_1-1-2 L} \xi^{L-a_1} \Xi_{l_g k_g}^{n_g} \Xi_{l_e k_e}^{n_g} \Gamma\left(a_1\right)}{k_{g} ! k_{e} !\left(b_{n_g l_g}-k_{g}\right) !\left(b_{n_e l_e}-k_{e}\right) ! \Gamma\left(\frac{3}{2}+L\right)} \\
& \times d_{l_g k_g}^{n_g} d_{l_e k_e}^{n_e}\frac{q}{\xi}^L{ }_2 F_1\left(\frac{1}{2} a_1, \frac{1}{2}\left(a_1+1\right) ; \frac{3}{2}+L ;-\frac{q^2}{\xi^2}\right)
\end{aligned}
\end{equation}

and 
$$
\begin{aligned}
A_1 & =(-1)^{m_e} C_{n_g l_g} C_{n_e l_e} \quad \frac{V_{12}(\boldsymbol{q})}{\sqrt{2 \pi}} \sqrt{\left(2 l_g+1\right)\left(2 l_e+1\right)}, \\
C_{n l} & =\sqrt{\frac{(n-l-1) !}{2 n(n+l) !}}, \quad \Xi_{l k}^n=\left(\sqrt{2 E_n}\right)^{\frac{3}{2}+l+k}, \\
d_{l k}^n & =\frac{(n+l) !}{(2 l+k+1) !}, \quad \xi=\sqrt{2 E_{n_g}}+\sqrt{2 E_{n_e}}, \\
a_1 & =3+k_g+k_e+l_g+l_e+L, \quad b_{n l}=n-l-1 .
\end{aligned}
$$,
and

The ionization prefactor $V_{p_{2e}}$ is given by:
\begin{equation}
\begin{aligned}
V_{p_2, e}= & A_2 \sum_{k=0}^{b_{n_e l_e}}(-1)^k \frac{2^k\left(\sqrt{2 E_{n_e}}\right)^{-\frac{1}{2}-l_e} p_2^{l_e}}{\left(b_{n_e l_e}-k\right) ! k !} d_{l_e k}^{n_e} \\
& \times \frac{\bar{\Gamma}\left(a_2\right)}{\Gamma\left(\frac{3}{2}+l_e\right)}{ }_2 F_1\left(\frac{1}{2} a_2, \frac{1}{2}\left(a_2+1\right) ; \frac{3}{2}+l_e ;-\frac{p_2^2}{2 E_{n_e}}\right),
\end{aligned}
\end{equation}

where
$$
\begin{aligned}
A_2 & =2(-i)^{l_e} C_{n_e l_e} Y_{l_e}^{m_e}\left(\theta_{p_2}, \phi_{p_2}\right), \\
a_2 & =2+k+2 l_e .
\end{aligned}
$$

\subsection{Excitation prefactor shifts and mapping}

\begin{figure}
    \centering
\includegraphics[width=\columnwidth]{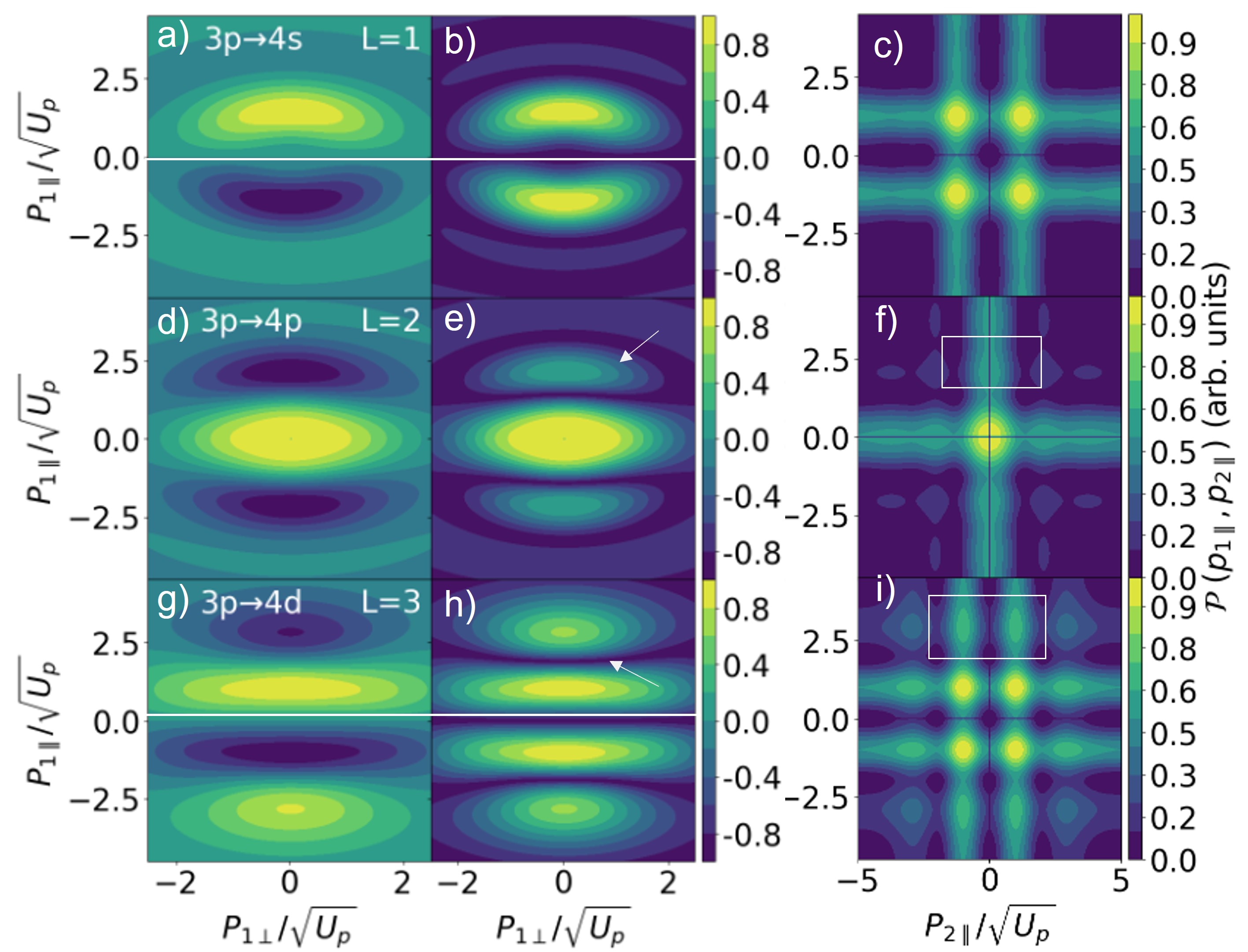}
    \caption{The ``unshifted" excitation prefactor and corresponding mapping calculated using the short orbit and the most dominant pair $p_4$ for the $3p \rightarrow 4s$ (panels (a)-(c)), $3p \rightarrow 4p$ (panels (d)-(f)) and $3p \rightarrow 4d$ (panels (g)-(i)) transitions. The left column shows the imaginary part of the prefactor for odd orbital quantum number $L$ (panels (a), (g)) and the real part for even $L$ in (d). The same pulse parameters have been employed as in Fig.~\ref{fig:cep65singlechannel}, with CEP $65^{\circ}$. Arrows in panels (e) and (h) indicate extra maxima present due to the higher orbital quantum number $L$ and white boxes in (f) and (i) indicate how these maxima translate to the mapping.}
    \label{fig:unshiftedpf1}
\end{figure}

\begin{figure}
    \centering
\includegraphics[width=\columnwidth]{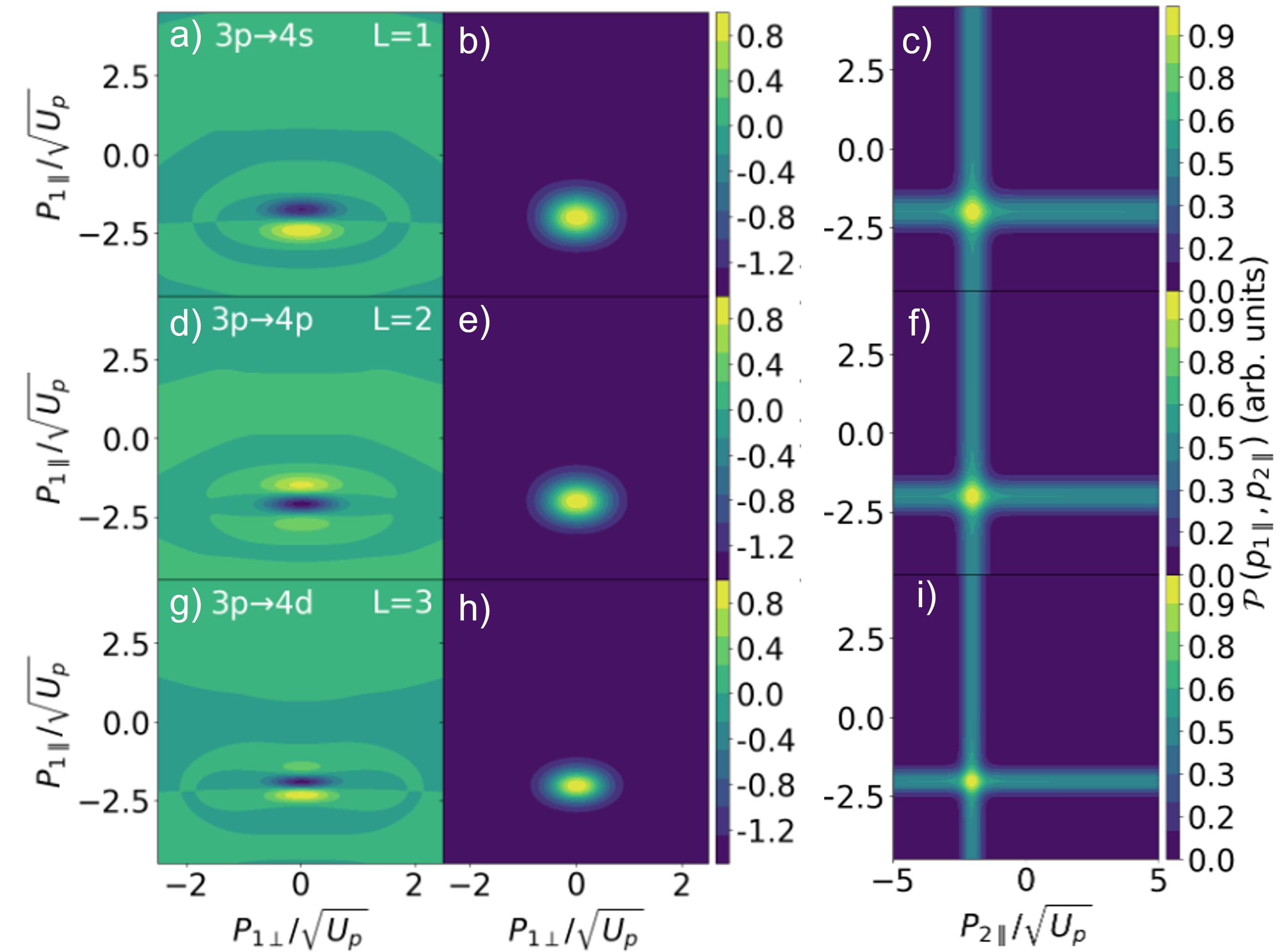}
    \caption{The ``shifted" excitation prefactor and corresponding mapping calculated using the short orbit and the most dominant pair $p_4$ for the $3p \rightarrow 4s$ (panels (a)-(c)), $3p \rightarrow 4p$ (panels (d)-(f)) and $3p \rightarrow 4d$ (panels (g)-(i)) transitions. The left column shows the imaginary part of the prefactor for odd orbital quantum number $L$ (panels (a), (g)) and the real part for even $L$ in (d). The same pulse parameters have been employed as in Fig.~\ref{fig:cep65singlechannel}, with CEP $65^{\circ}$.}
    \label{fig:shiftedpf1}
\end{figure}

\begin{figure}
\includegraphics[width=\columnwidth]{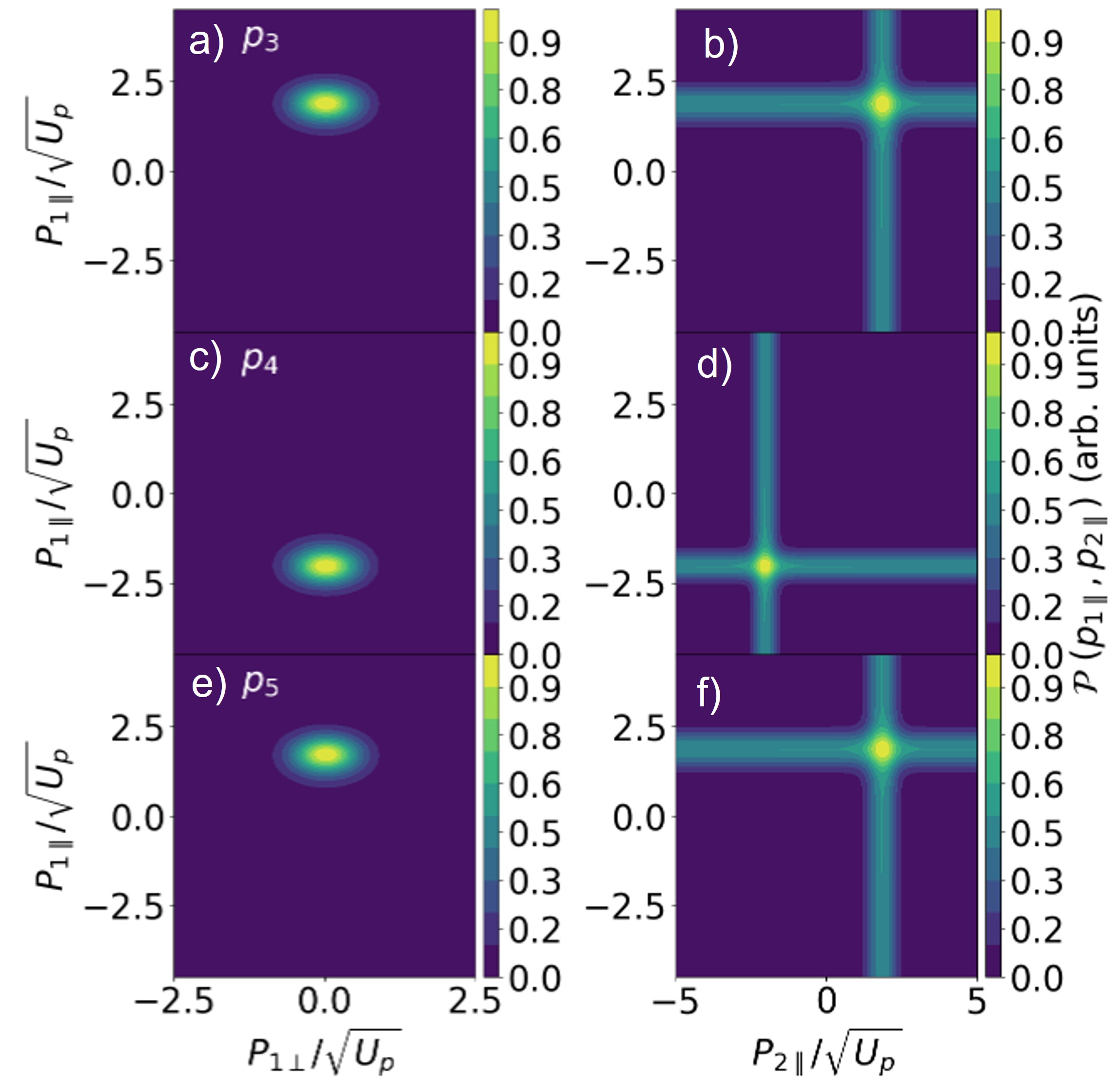}
    \caption{The ``shifted" excitation prefactor [ (a)-(c)] and corresponding mappings (panels (d)-(f)) calculated using the short orbit for the $3p \rightarrow 4d$ channel using the three most dominant events. The same pulse parameters have been employed as in Fig.~\ref{fig:cep65singlechannel}, with CEP $65^{\circ}$.}
    \label{fig:shiftedpf1eventwise}
\end{figure}

Eq.~\ref{eq:Vp1ekg} shows a dependence of the excitation prefactor on $q=\sqrt{\mathbf{q}^2}$, where $\mathbf{q}=\mathbf{p}_1-\mathbf{k}$, where $\mathbf{k}$ represents the intermediate momentum of the first electron. Also, the shape of this prefactor is strongly influenced by the orbital angular momentum $L$ resulting from the sum over the angular momenta $l_e$ and $l_g$ of the second electron's ground and excited states in Eq.~\eqref{eq:angmom}. The magnetic quantum number $m=0$ is used throughout for simplicity, although the expressions are general.  

A general "unshifted" form of this prefactor was calculated by setting $\mathbf{q}=\mathbf{p}_1$. Fig.~\ref{fig:unshiftedpf1}(b), (e) and (h) show the absolute value of this unshifted prefactor for three values of orbital angular momentum $L$. With odd $L$, there are nodes along $p_{1\parallel}=0$. These prefactors are also purely imaginary [panels (a) and (g)] due to the properties of the spherical harmonics and Clebsch-Gordon coefficients. They have opposite and alternating phases above and below the central node. With even $L$, this node at  $p_{1\parallel}=0$ is replaced with a maximum, and the prefactor becomes real [panel (e)]. The nodes are now at around $\pm\sqrt{U_p}$. There are still alternating phases above and below the nodes. It should be noted that the unshifted prefactor is not orbit-dependent i.e. remains the same for both the long and short orbits. 

The locations of the maxima and nodes strongly translate to the corresponding mappings of the excitation prefactor itself. For the $s$ and $d$ states, there is suppression along the $p_{1\parallel}$ axis and brightest maxima at $\pm\sqrt{U_p}$. The $p$ state leads to brightest maxima at the origin forming a cross-shape. With larger L, there are two additional maxima indicated by arrows in Fig.~\ref{fig:unshiftedpf1}(e) and (h), leading to secondary maxima at their corresponding locations indicated by the boxes in panels (f) and (i). 

The effect of the radial nodes is lost when we shift the prefactor. Thus, they are neglected in this discussion. It is difficult to predict the effect of shifting the momentum $\mathbf{p}_1$ by $\mathbf{k}$, not least because the intermediate momentum has an imaginary component. The shifted prefactor is also now orbit-dependent and therefore shapes for only the non-divergent solution are shown here. For further details on the divergence of orbits and the uniform approximation, see \cite{Faria2003,Shaaran2010}.

The imaginary parts of the shifted prefactor, with the short orbit and taking the dominant $p_4$ pair, are shown in Fig.~\ref{fig:shiftedpf1}(a) and (g). For odd $L$, the central node has now shifted to $-2\sqrt{U_p}$. The phases above and below the node still alternate but have swapped sign, possibly due to $\mathbf{k}$ being complex which introduces extra phase shifts. For even $L$, there now exists an imaginary component for the prefactor with a node at $-2\sqrt{U_p}$ coming from the $\mathbf{k}$ shift as with the odd $L$ case and the central maximum in the real part of the prefactor shifts to $-2\sqrt{U_p}$ [panel (d)]. The real parts of the shifted prefactor for all $L$ have some stretching distortion towards lower values of $p_{1\parallel}$. This is because the real part of $\mathbf{k}$ dominates and increases linearly with the x-axis with an average of around $\sqrt{U_p}$ so prefactors are also shifted by around this amount. 

The prefactor has also significantly narrowed in the momentum plane, so much so that the absolute value of the shifted prefactor loses its characteristic geometry-dependent shape when mapped, since the nodes have been washed out. The mappings therefore lead to a bright maximum around $-2\sqrt{U_p}$. The position of this maximum is dependent on where the event in question is located in the pulse, since $\mathbf{k}$ depends on the ionization and rescattering times of the first electron. Fig.~\ref{fig:shiftedpf1eventwise} shows how the maximum of the shifted prefactor, and thus its mapping changes with event for a given channel.

\section{Detangling interference due to $\alpha^{(A^2)}(t',t'')$}
\label{appendix:interf}
\begin{figure}[]
     \centering
    \includegraphics[width=\columnwidth]{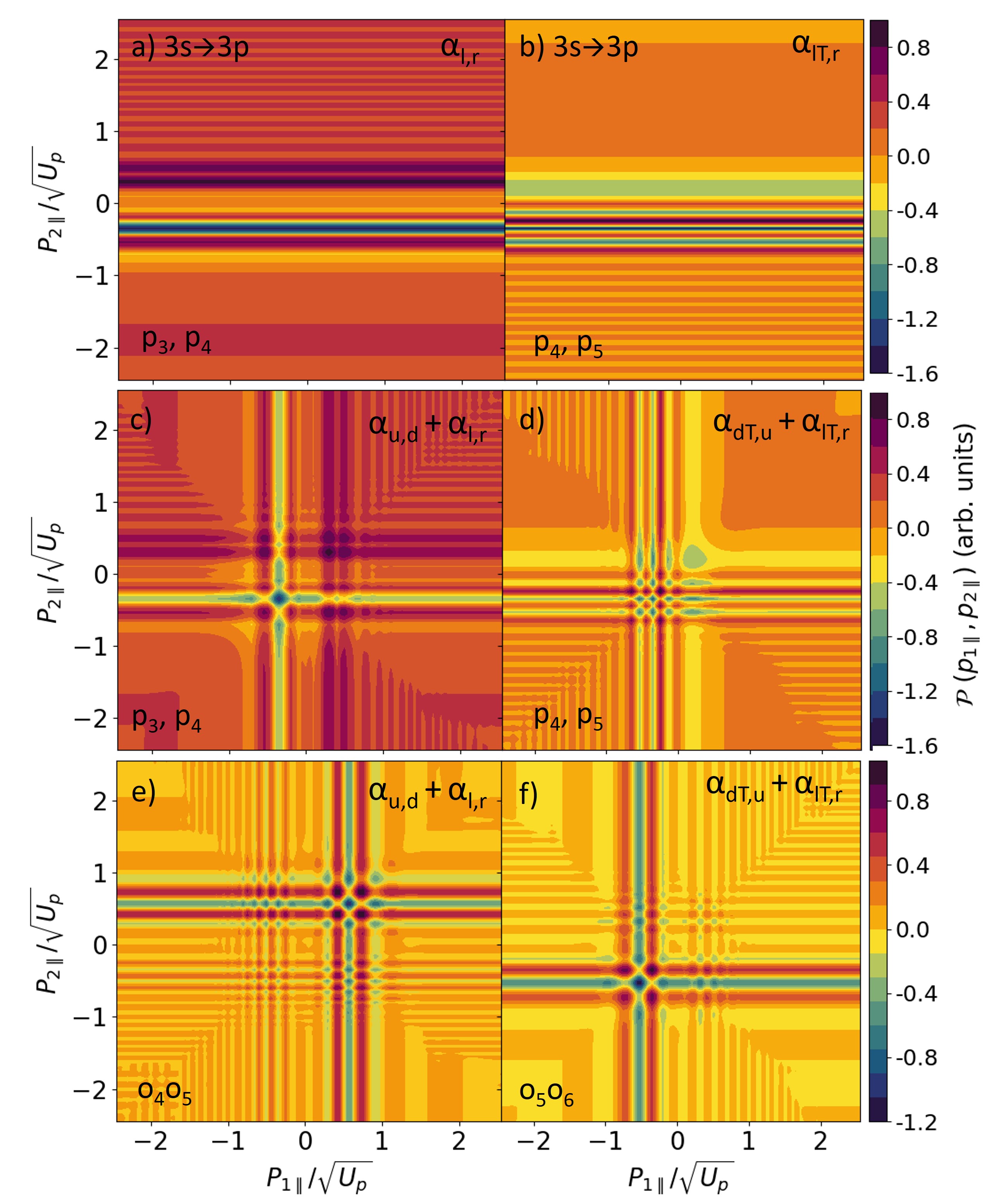}
    \caption{Pure temporal-shift interference patterns separated for the first electron (panels (a)-(d)) and the second electron (panels (e)-(f)) shown unsymmetrized (panels (a)-(b)) and symmetrized (panels (c)-(f)), computed for pairwise combinations of the most dominant events that separated by half a cycle for the $3s \rightarrow 3p$ transition and CEP $65^{\circ}$. All other pulse parameters employed are the same as in Fig.~\ref{fig:cep65singlechannel}.
    \label{fig:eventelectronbreakdown}}
\end{figure}

Here, we consider the interference due to temporal shifts for each electron separately in order to detangle effects arising from $\alpha^{(A^2)}(t',t'')$ explicitly. We employ this to determine whether the chequerboard patterns are due to quantum interference or incoherent symmetrization effects. This can be done by computing $\alpha_{l, r}$, $\alpha_{lT, r}$, $\alpha_{lT, l}$ and their transposed counterparts with just the pairs (first electron), or just the orbits (second electron). These momentum distributions with the first electron will contain all four of the phase differences associated with the temporal shifts. However, those with the second electron will be missing the interference resulting from  $\alpha^{(A^2)}(t',t'')$ since this is dependent only on the times for the first electron. In Fig.~\ref{fig:eventelectronbreakdown}, the interferences coming from the $p_3,p_4$ and $p_4,p_5$ interfering pairs for the first electron are shown, unsymmetrized [panels (a), (b)] and symmetrized [panels (c), (d)].  The symmetrized interference coming from $o_4,o_5$ and $o_5,o_6$ interfering orbits are shown in panels (e) and (f). The differences of (c) and (e), and (d) and (f) are taken to find the contribution of interferences from $\alpha^{(A^2)}(t',t'')$, for the two sets of events. 

Fig.~\ref{fig:eventelectronbreakdown}(a), (b) show alternating horizontal stripes for the unsymmetrized case, which when symmetrized lead to a cross with small chequerboards at the intersection of the horizontal and vertical stripes, in the quadrant associated with the more dominant of the pair of interfering pairs (panels (c) and (d)) for the first electron. Similarly, chequerboards in the quadrants associated with the more dominant of the two orbits for the second electron can be seen in panels (e) and (f). When the difference between the electron one only and electron two only distributions are found, these chequerboards overlap and cause more intricate and complex patterns.

\end{document}